\documentstyle[editedvolume]{crckapb} 

\input epsf.sty

\newcommand{\bms}{\bf}		% `bold math sans serif', for matrices

\newcommand{\be}{\begin{equation}}
\newcommand{\ee}{\end{equation}}
\newcommand{\ba}{\begin{eqnarray}}
\newcommand{\ea}{\end{eqnarray}}
\newcommand{\aap}[2]{{\it Astron.\ \& Astrophys.}, {\bf #1}, #2}
\newcommand{\aj}[2]{{\it A.\ J.}, {\bf #1}, #2}
\newcommand{\araa}[2]{{\it Ann.\ Rev.\ Astr.\ Ap.}, {\bf #1}, #2}
\newcommand{\apj}[2]{{\it Ap.\ J.}, {\bf #1}, #2}
\newcommand{\apjs}[2]{{\it Ap.\ J.\ Supplement}, {\bf #1}, #2}
\newcommand{\ass}[2]{{\it Ap.\ Space Sci.}, {\bf #1}, #2}
\newcommand{\mn}[2]{{\it MNRAS}, {\bf #1}, #2}
\newcommand{\nature}[2]{{\it Nature}, {\bf #1}, #2}
\newcommand{\pasp}[2]{{\it Pub.\ Astr.\ Soc.\ Pacific}, {\bf #1}, #2}
\newcommand{\phrept}[2]{{\it Physics Reports}, {\bf #1}, #2}
\newcommand{\etal}{{\it et al.\/}}

\newcommand{\smoothdelta}{\tilde\delta}
\newcommand{\smoothk}{\tilde k}
\newcommand{\smoothP}{\tilde P}
\newcommand{\coleP}{\tilde P}

\newcommand{\rhobar}{\bar\rho}
\newcommand{\deltarho}{\delta}
\newcommand{\nbar}{\bar n}

\newcommand{\FTa}{\hat a}
\newcommand{\FTalpha}{\hat\alpha}
\newcommand{\FTb}{\hat b}
\newcommand{\FTdelta}{\hat\delta}
\newcommand{\FTdeltarho}{\hat{\deltarho}}
\newcommand{\FTf}{\hat f}
\newcommand{\FTw}{\hat w}
\newcommand{\FTxi}{\hat \xi}
\newcommand{\FTbS}{\hat {\bms S}}
\newcommand{\FTR}{\hat R}

\newcommand{\el}{\ell}

\newcommand{\k}{{\mbox{\boldmath$k$}}}
\newcommand{\bl}{{\mbox{\boldmath$l$}}}
\newcommand{\n}{{\mbox{\boldmath$n$}}}
\newcommand{\r}{{\mbox{\boldmath$r$}}}
\newcommand{\s}{{\mbox{\boldmath$s$}}}
\newcommand{\bv}{{\mbox{\boldmath$v$}}}
\newcommand{\w}{{\mbox{\boldmath$w$}}}
\newcommand{\x}{{\mbox{\boldmath$x$}}}
\newcommand{\z}{{\mbox{\boldmath$z$}}}
\newcommand{\K}{{\mbox{\boldmath$K$}}}
\newcommand{\bL}{{\mbox{\boldmath$L$}}}
\newcommand{\R}{{\mbox{\boldmath$R$}}}
\newcommand{\bmiA}{{\mbox{\boldmath$A$}}}
\newcommand{\bmia}{{\mbox{\boldmath$a$}}}
\newcommand{\bmib}{{\mbox{\boldmath$b$}}}

\newcommand{\one}{{\bms 1}}

\newcommand{\bPsi}{{\bms\Psi}}
\newcommand{\bdelta}{\mbox{\boldmath$\delta$}}
\newcommand{\bdeltarho}{\mbox{\boldmath$\deltarho$}}
\newcommand{\bnabla}{\mbox{\boldmath$\nabla$}}

\newcommand{\bxi}{\mbox{\boldmath$\xi$}}

\newcommand{\bC}{{\bms C}}
\newcommand{\bN}{{\bms N}}
\newcommand{\bS}{{\bms S}}

\newcommand{\LG}{{\rm LG}}
\newcommand{\Mpc}{{\rm Mpc}}

\newcommand{\kms}{{\rm km} \, {\rm s}^{-1}}
\newcommand{\obs}{{\rm obs}}
\newcommand{\nn}{\nonumber \\}
\newcommand{\Tr}{{\rm Tr}}

\newcommand{\ga}{\raisebox{-0.5ex}[1.5ex][0ex]{
                \begin{array}[b]{@{}c@{\;}} > \\
                [-1.7ex] \sim \end{array}}}
\newcommand{\la}{\raisebox{-0.5ex}[1.5ex][0ex]{
                \begin{array}[b]{@{}c@{\;}} < \\
                [-1.7ex] \sim \end{array}}}
\newcommand{\para}{{\mbox{\mathsurround=0pt\raisebox{0.2ex}[1.2ex][0ex]{
                  \hspace{-.35em}$\scriptscriptstyle/$
                  \hspace{-.5em}$\scriptscriptstyle/$}}}}
\begin{opening}
\title{LINEAR REDSHIFT DISTORTIONS: A REVIEW}

\author{A.\ J.\ S.\ HAMILTON}
\institute{JILA \& Dept.\ of Astrophysical and Planetary Sciences\\
           Box 440, U.\ Colorado, Boulder, CO 80309, USA;\\
           Andrew.Hamilton@colorado.edu;\\
           http:/\hspace{-2pt}/casa.colorado.edu/$\sim$ajsh}
\end{opening}

\runningtitle{LINEAR REDSHIFT DISTORTIONS}

\begin{document}

% The \begin{document} command comes after the \end{opening}
% command.

\noindent
Invited review to appear in
Hamilton, D. (ed.)
{\it Ringberg Workshop on Large-Scale Structure},
held at Ringberg Castle, Germany, 23--28 September 1996,
Kluwer Academic, Dordrecht.
\vspace{8mm}

\begin{abstract}
Redshift maps of galaxies in the Universe are distorted by the peculiar
velocities of galaxies along the line of sight.
The amplitude of the distortions on large, linear scales
yields a measurement of the linear redshift distortion parameter,
which is $\beta \approx \Omega_0^{0.6}/b$
in standard cosmology with cosmological density $\Omega_0$
and light-to-mass bias $b$.
All measurements of $\beta$ from
linear redshift distortions published up to mid 1997 are reviewed.
The average and standard deviation of the reported values
is $\beta_{\rm optical} = 0.52 \pm 0.26$
for optically selected galaxies,
and $\beta_{\it IRAS} = 0.77 \pm 0.22$
for {\it IRAS\/} selected galaxies.
The implied relative bias is $b_{\rm optical}/b_{\it IRAS} \approx 1.5$.
If optical galaxies are unbiased, then
\raisebox{0ex}[2ex][0ex]{$\Omega_0 = 0.33^{+0.32}_{-0.22} \,$},
while if {\it IRAS\/} galaxies are unbiased, then
\raisebox{0ex}[2ex][0ex]{$\Omega_0 = 0.63^{+0.35}_{-0.27} \,$}.
\end{abstract}

%\begin{sloppypar}

%\setcounter{tocdepth}{2}
%\tableofcontents

\section{Introduction}
\label{intro}
\setcounter{equation}{0}

The organisers of this thoroughly enjoyable workshop
asked me to write a review of redshift distortions
aimed primarily at graduate students and others who are not familiar with
the field.
The review aims at fairly thorough coverage
(up to mid 1997)
within a rather limited scope:
the subject of redshift distortions in the large scale, linear regime.
The review does not attempt to cover the large body of work
involving the direct measurement of peculiar velocities.
The latter has been the subject of recent comprehensive reviews by
Strauss \& Willick (1995),
and by Dekel (1994).
Both of those reviews included sections on redshift distortions.
Nor does the present review cover nonlinear redshift distortions,
except insofar as they affect linear redshift distortions.
For an entry to the literature on nonlinear redshift distortions, try
Davis, Miller \& White (1997).

Hubble's (1929) law states that the recession velocity $cz$ of a galaxy
is proportional to its distance $d$
\be
  cz = H_0 d
  \ ,
\ee
with constant of proportionality the Hubble constant $H_0$
(the subscript $0$ signifies its present day value).
The recession velocity $cz$ of a galaxy can be measured
from the redshift $z$ of its spectrum
($c$ is the speed of light),
a great deal more easily and accurately than its true distance $d$.
This has been a primary motivation for redshift surveys
(see e.g.\ Strauss 1997
% Geller 1997
for a recent review),
which map the Universe in 3 dimensions
using the recession velocity $cz$ of each galaxy as a measure of its distance.

Hubble's law is not perfect, however.
Galaxies have peculiar velocities $\bv$ relative to the general
Hubble expansion.
Thus it is necessary in general to distinguish between
a galaxy's {\bf redshift distance} $s$
(conveniently expressed in velocity units)
\be
  s \equiv cz
\ee
and its true distance $r$
(also conveniently expressed in velocity units)
\be
  r \equiv H_0 d
  \  .
\ee
The redshift distance $s$ of a galaxy differs from the true distance $r$
by its peculiar velocity $v \equiv \hat\r . \bv$ along the line of sight:
\be
\label{srv}
  s = r + v
  \ .
\ee

The peculiar velocities of galaxies thus cause them to appear displaced
along the line of sight in redshift space.
These displacements lead to {\bf redshift distortions}
in the pattern of clustering of galaxies in redshift space.
Although such distortions complicate the interpretation of redshift maps
as positional maps,
they have the tremendous advantage of bearing information about
the dynamics of galaxies.
In particular, the amplitude of distortions on large scales yields a
measure of the {\bf linear redshift distortion parameter} $\beta$,
which is related to the cosmological density $\Omega_0$,
the present day ratio of the matter density of the Universe
to the critical density required to close it, by
(see \S\ref{continuity})
\be
\label{beta}
  \beta = {f(\Omega_0) \over b} \approx {\Omega_0^{0.6} \over b}
\ee
in standard pressureless Friedmann cosmology with light-to-mass bias $b$.
The goal of much of the current work on redshift distortions
is to measure the linear distortion parameter $\beta$,
and perhaps, pious hope,
if bias can be quantified through nonlinear effects or otherwise,
to determine the cosmological density $\Omega_0$ itself.

\subsection{Plan}
\label{plan}

The plan of this paper is as follows.

Section~\ref{look} attempts to convey visually what redshift distortions
look like and why,
both \S\ref{schematic} schematically,
and \S\ref{observation} observationally.

Section~\ref{correlation} is about power spectra.
Sections~\ref{real} and \ref{redshift} contains definitions,
needed for subsequent reference,
of correlation functions and power spectra in real and redshift space.
Section~\ref{hilbert} advertises the delights of Hilbert space.
Section~\ref{power} explains why power spectra are better.

Section~\ref{theory}
presents the theory of redshift distortions in the linear regime.
The first part, \S\ref{continuity},
explores what $\beta$, the linear redshift distortion parameter,
really means.
The second part, \S\ref{operator},
derives the linear redshift distortion operator,
which transforms real space into redshift space,
for fluctuations in the linear regime.
The third and fourth parts, \S\S\ref{LG} and \ref{selfn},
cover two small but important details, the peculiar motion of us, the observers,
and the difference between the selection functions in real and redshift space.

Section~\ref{methods}
describes the three different types of method that have been
used to date to measure $\beta$ from linear redshift distortions:
\S\ref{redtoreal}, the ratio of real to redshift angle-averaged power;
\S\ref{quadrupole},
the ratio of quadrupole-to-monopole harmonics of the redshift power spectrum;
and \S\ref{ML}, the maximum likelihood approach.

Section~\ref{example}
interjects an example of measuring $\beta$ from linear redshift distortions,
partly as an illustration, partly to bring out the difference
between optical and {\it IRAS\/}-selected galaxies, and partly to demonstrate
the importance of nonlinearities.

Section~\ref{translinear} describes the two methods that have been used
to date to deal with nonlinearity:
\S\ref{pairdispersion}, a model in which linear redshift distortions
are modulated by a random velocity dispersion;
and \S\ref{zeldovich}, the Zel'dovich approximation.

Section~\ref{measurement}
compiles measurements of $\beta$
from linear redshift distortions, complete up to the first half of 1997.
Section~\ref{compilation}
summarizes the measurements of $\beta$ in a Table,
and gives the average and standard deviation of the measurements,
which results are the ones quoted in the abstract.
Section~\ref{individual} offers commentary on all the individual measurements.
I apologize to authors whose work has been inadvertently omitted,
or inadequately portrayed.

Finally, \S\ref{cosmological} discourses briefly on cosmological redshift
distortions, which arise from differences in the geometry
of the Universe perceptible at high enough redshift.

The greater part of this review is just that, a review;
but there are some new things here and there.
The expression for the linear redshift distortion operator $\bS^{s\,\LG}$
for the practical case where
(a) the redshift overdensity is measured in the Local Group frame,
and (b) the selection function is measured in redshift space,
also in the Local Group frame,
appears here explicitly for the first time, equation~(\ref{SsLG}).
The analysis of the Stromlo-APM survey reported in \S\ref{example}
has not been published elsewhere.

\section{What Redshift Distortions Look Like}
\label{look}
\setcounter{equation}{0}

\subsection{Schematically}
\label{schematic}

\begin{figure}
\vbox to62mm{\rule{0pt}{62mm}}
\begin{center}
\leavevmode
\includegraphics{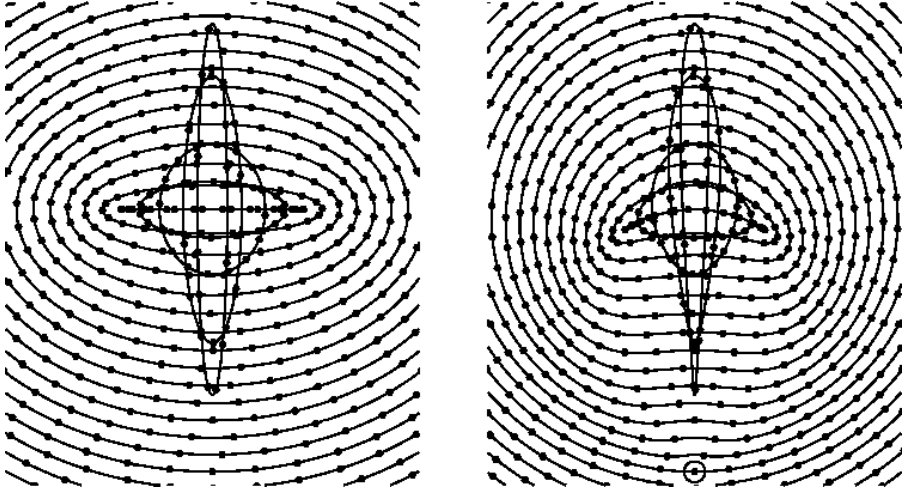}
\end{center}
  \caption[1]{
A spherical overdensity
appears distorted by peculiar velocities
when observed in redshift space.
On large (linear) scales
the overdensity appears squashed along the line of sight,
while on small (nonlinear) scales fingers-of-god appear.
At left, the overdensity is far from the observer
(who is looking upward from somewhere way below the bottom of the diagram),
and the distortions are effectively plane-parallel.
At right, the overdensity is near the observer (large dot),
and the large scale distortions appear kidney-shaped,
while the finger-of-god is sharpened on the end pointing at the observer.
The observer shares the infall motion towards the overdensity.
A similar diagram appears in Kaiser (1987).
\label{fogs}
}
\end{figure}

\begin{figure}
\begin{center}
\leavevmode
\epsfxsize=3in \epsfbox{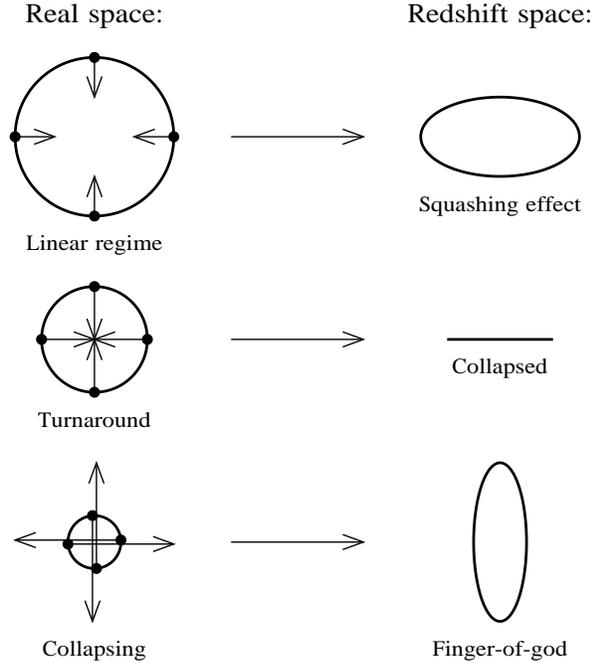}
\end{center}
  \caption[1]{
Detail of how peculiar velocities lead to the redshift distortions
illustrated in Figure~\protect\ref{fogs}.
The dots are `galaxies' undergoing infall towards a spherical overdensity,
and the arrows represent their peculiar velocities.
At large scales,
the peculiar velocity of an infalling shell is small compared to its radius,
and the shell appears squashed.
At smaller scales, not only is the radius of a shell smaller,
but also its peculiar infall velocity tends to be larger.
The shell that is just at turnaround,
its peculiar velocity just cancelling the general Hubble expansion,
appears collapsed to a single velocity in redshift space.
At yet smaller scales,
shells that are collapsing in proper coordinates
appear inside out in redshift space.
The combination of collapsing shells
with previously collapsed, virialized shells,
gives rise to fingers-of-god.
\label{caustic}
}
\end{figure}

Figure~\ref{fogs} illustrates how a spherical overdensity appears
distorted by peculiar velocities along the line of sight,
when observed in redshift space.
The initial spherical overdensity perturbation here was taken to be
a power law with radius, $\delta \propto r^{-1}$,
located in an expanding Universe with critical mean density, $\Omega = 1$.
The free-fall gravitational collapse of such a spherical pressureless
overdensity can be computed analytically
(Peebles 1980, \S18).
The dots (galaxies) started out uniformly distributed in the initial conditions,
being uniformly placed around a series of uniformly spaced concentric shells.
Thus the density of dots in Figure~\ref{fogs} indicates the density of
galaxies in the collapsing overdensity, as observed in redshift space.
Figure~\ref{fogs} omits shells that have collapsed to less than
half their radius at turnaround,
which shells may be expected to scatter off previously collapsed shells,
and to virialize.

Figure~\ref{caustic} shows how peculiar velocities produce
the pattern illustrated in Figure~\ref{fogs}.
On large scales, peculiar infall towards the overdensity causes it to appear
squashed along the line of sight.
The squashing increases to smaller scales down to the point of turnaround,
where the peculiar infall velocity exactly cancels the general Hubble expansion.
In the turnaround shell,
the near and far parts of the shell appear collapsed to a single radial
velocity in redshift space.
At smaller scales,
shells that have turned around and are collapsing in real
space appear turned `inside out' in redshift space.
At even smaller scales (not shown in Figures~\ref{fogs} or \ref{caustic}),
collapsed shells are expected to virialize.
The combination of collapsing and virialized regions of galaxy clusters
gives rise to {\bf fingers-of-god}.

\subsection{Observationally}
\label{observation}

Fingers-of-god are well-known features of redshift surveys.
Prominent examples are the fingers-of-god in the Coma cluster
(de Lapparent, Geller \& Huchra 1986)
and the Perseus cluster
(Wegner, Haynes \& Giovanelli 1993, Figs.~7--10).

The envelope of the finger-of-god in Figure~\ref{fogs} forms a caustic,
a surface of infinite density (but finite mass).
Such caustics are not obviously seen in real fingers-of-god
(Reg\H{o}s \& Geller 1989);
presumably the caustics are smeared out by subclustering.
The structure of the well-studied Coma cluster, for example,
is quite complicated
(Colless \& Dunn 1996).

Visually, the large scale squashing effect is more subtle to discern
in real data.
Are prominent transverse structures such as the Great Wall
(Ramella, Geller \& Huchra 1992)
enhanced by redshift distortions?
Probably yes, at some level
(e.g.  Praton, Melott \& McKee 1997).
However,
Dell'Antonio, Geller \& Bothun (1996)
conclude from their analysis of the peculiar velocity field of the Great Wall
that any infall is small, $\la 150 \, \kms$.

\begin{figure}
\vbox to75mm{\rule{0pt}{75mm}}
\includegraphics{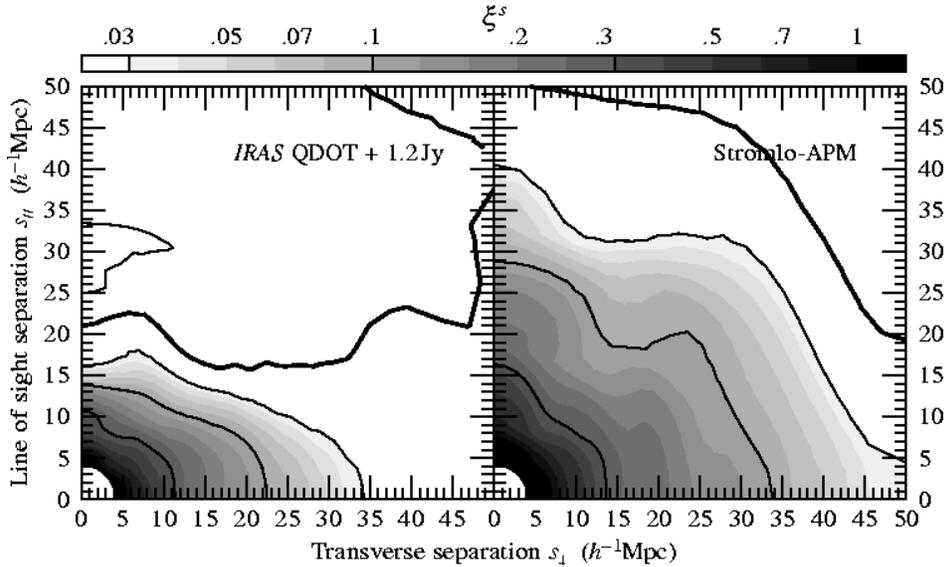}
  \caption[1]{
Contour plots of the redshift space two-point correlation function
$\xi^{s}$ as a function of separations $s_\para$ and $s_\perp$
parallel and perpendicular to the line of sight
in: (left) the {\it IRAS\/} QDOT and 1.2~Jy redshift surveys,
merged over the angular region of the sky common to both surveys;
and (right) the optical Stromlo-APM survey.
In each case the region within $25 h^{-1} {\Mpc}$ of the Milky Way
has been excluded, so as to eliminate bias from the local overdensity.
A near minimum variance pair weighting has been applied.
The thick contour signifies $\xi^{s} = 0$,
and other contours are logarithmically spaced
at intervals of 0.5~dex above $10^{-1.5}$
(the left panel also shows one negative contour, at $-10^{-1.5}$).
Shading is graduated at intervals of 0.1~dex above $10^{-1.5}$.
The correlation function here has been smoothed over pair separation
\protect\raisebox{0ex}[2ex][0ex]{$s = ( s_\perp^2 + s_\para^2 )^{1/2}$}
with a tophat window of width $0.2$~dex,
and over angles
\protect\raisebox{0ex}[2ex][0ex]{$\theta = \tan^{-1} ( s_\perp / s_\para )$}
to the line of sight with a Gaussian window
with a 1$\sigma$ width of $10^\circ$.
\label{xilconts}
}
\end{figure}

The large scale squashing effect can however be detected statistically,
from the distortion of the redshift space correlation function $\xi^{s}$,
or of its Fourier transform the redshift space power spectrum $P^{s}$.
These quantities are defined formally in the next section,
\S\ref{correlation}.
Physically,
the redshift correlation function $\xi^{s}(s_\para, s_\perp)$
is the mean fractional excess of galaxy neighbours of a galaxy
at separations $s_\para$ and $s_\perp$
parallel and perpendicular to the line of sight,
a definition that suffices to understand Figure~\ref{xilconts}.

Figure~\ref{xilconts} shows redshift space correlation functions
$\xi^{s}(s_\para, s_\perp)$
measured from two sets of redshift surveys,
one (QDOT + 1.2~Jy) selected in the infrared,
the other (Stromlo-APM) in the optical.
The infrared survey is a merger of two
{\it Infrared Astronomical Satellite\/} ({\it IRAS\/})
redshift surveys, the QDOT survey, and the 1.2~Jy survey.
The revised QDOT survey
(Lawrence \etal\ 1997, in preparation)
is a redshift survey of a 1-in-6 subset of galaxies brighter than 0.6~Jy
from the {\it IRAS\/} Point Source Catalog (PSC).
It contains 2376 galaxies over 9.29365 steradians,
covering most of the sky above galactic latitude $|b| > 10^\circ$.
The {\it IRAS\/} 1.2~Jy survey
(Fisher \etal\ 1995a; Strauss \etal\ 1992a)
is a complete redshift survey of galaxies
brighter than 1.2~Jy from the same {\it IRAS\/} PSC.
It contains 5321 galaxies over 11.02577 steradians,
covering most of the sky above galactic latitude $|b| > 5^\circ$.
For Figure~\ref{xilconts},
the surveys were merged over the part of the sky common to both surveys,
yielding 5752 galaxies over 9.26740 steradians.
Removing the region closer than $25 \, h^{-1} \Mpc$,
to avoid local bias
(Hamilton \& Culhane 1996, \S2.1)
left 4826 galaxies.
The optical survey is the Stromlo-APM survey
(Loveday \etal\ 1996b),
which is a redshift survey of a 1-in-20 subset of galaxies
brighter $b_J = 17.15$
from the Automatic Plate Measuring (APM) survey
(Maddox \etal\ 1990a,b, 1996).
The Stromlo-APM redshift survey contains 1790 galaxies over 1.32467 steradians
centred roughly around the South Galactic Pole.
Eliminating the region closer than $25 \, h^{-1} \Mpc$
left 1725 galaxies.

To ensure the validity of the plane-parallel approximation,
the redshift correlation function shown in Figure~\ref{xilconts}
was computed including only pairs closer than $50^\circ$ on the sky,
the line of sight to each pair being defined as the bisector
(the angular midpoint on the sky) of the pair.
A near minimum variance pair weighting was applied
as described by Hamilton (1993b, \S5, eqs.~[60] \& [61]).
Further details of the procedure used here to calculate
$\xi^{s}(s_\para, s_\perp)$
are given in the latter paper.

Figure~\ref{xilconts} shows clearly the expected large scale squashing effect,
while fingers-of-god show up as an enhancement of
the redshift correlation function along the line-of-sight axis.
Besides these two expected effects,
two other features are apparent.
Firstly, the redshift correlation function of the optically selected galaxies
is roughly a factor of 2 higher than that of the {\it IRAS\/} selected galaxies.
Secondly, the fingers-of-god are longer and more prominent in the
optical galaxies.
Both these features can be attributed to the fact that
{\it IRAS\/} galaxies, which are dusty, gas-rich spirals,
avoid the centres of rich clusters of galaxies,
where ellipticals rule.

A contour plot of the redshift space power spectrum from an $N$-body simulation
can be found in Cole, Fisher \& Weinberg (1994, Fig.~1),
while Bromley, Warren \& Zurek (1997, Fig.~1)
show plots of the redshift space power spectrum
as a function of $\theta = \tan^{-1}(k_\perp/k_\para)$
from two large (17 million particle) high resolution simulations.

\section{Correlation Functions and Power Spectra}
\label{correlation}
\setcounter{equation}{0}

Sections~\ref{real} and \ref{redshift} collect standard definitions
of correlation functions and power spectra,
in real and redshift space respectively,
needed elsewhere in this review.
Compared to theoretical correlation functions,
defined in \S\S\ref{realdef} and \ref{redshiftdef},
observed correlation functions contain additional noise,
the dominant contribution to which is often taken to be
Poisson sampling noise, also known as shot noise,
discussed in \S\S\ref{shot} and \ref{redshot}.
I adhere throughout this review to what has become the standard convention
in the field of galaxy clustering for defining Fourier transforms,
notwithstanding the extraneous factors of $2\pi$ that result.

Section~\ref{hilbert} points out how Hilbert space
provides a compact, powerful notation.

Section~\ref{power} explains what makes power spectra special.

\subsection{Real Space}
\label{real}

A fundamental proposition
is that the three dimensional distribution of matter in the Universe
constitutes a statistically homogeneous and isotropic random density field.
This proposition, a reflection of the proposition that the Universe
at large is homogeneous and isotropic
--- Milne's 1933 Cosmological Principle (see Peebles 1980, \S3B) ---
is powerfully evidenced by the isotropy of the Cosmic Microwave Background
(Bennett \etal\ 1996; G\'orski 1997).

The statistical properties of such a field are completely determined
by its irreducible moments,
or correlation functions as they are commonly called
in the discipline of large scale structure
(Peebles 1980, Ch.~III).
The first two irreducible moments are the mean, a constant,
and the covariance, or $2$-point correlation function,
a function of separation.

In the case of a multivariate Gaussian random field,
for which by definition the third and all higher correlation functions vanish,
the mean and $2$-point correlation function
completely specify the statistical properties of the field.
If fluctuations generated in the early Universe
were the result of a superposition of many independent random processes,
as for example quantum fluctuations during inflation,
then the Central Limit Theorem guarantees that the fluctuations will
form a multivariate Gaussian.
Fluctuations on large, linear scales today would then also be Gaussian.
Available observational evidence is consistent with fluctuations
being Gaussian on linear scales
(Stirling \& Peacock 1996;
Kogut \etal\ 1996).

Once density fluctuations grow large, they cannot remain Gaussian,
since density must remain positive.
Thus on small, nonlinear scales the third and higher order correlation
functions must be non-vanishing.
Nonetheless the covariance, the $2$-point correlation function,
is well-defined in the nonlinear regime,
and remains a statistic of fundamental importance.

\subsubsection{The True Density Field of the Universe}
\label{realdef}

Let $\rho(\r)$ denote the density of matter at (real, not redshift)
position $\r$ in the Universe,
and let $\rhobar$, a constant in space, denote the mean density.
The density $\rho(\r)$ may be regarded as the density of mass,
or of particles, or of galaxies,
or of some particular set of objects one is interested in.
Whatever the case,
the density $\rho(\r)$ signifies the true density,
not the observed density of objects that may be recorded in a real survey
(which is considered in \S\S\ref{shot} and \ref{redshift} following),
and $\r$ is the true position, not the redshift position.
The true {\bf overdensity} $\deltarho(\r)$ at position $\r$ is defined by
\be
\label{deltarho}
  \deltarho(\r) \equiv {\rho(\r) - \rhobar \over \rhobar}
  \ .
\ee

It is to be noted that although $\deltarho(\r)$ is defined to be the
`true' overdensity, it is nevertheless a hypothetical quantity,
to be distinguished from a quantity,
such as the redshift, angular position, or flux of a galaxy,
which is directly observed.
This is the usual situation in statistics,
where one imagines a hypothetical `true' population from which
observed data are drawn.
The properties of statistical homogeneity and isotropy attributed to
the `true' density field are likewise theoretical hypotheses,
whose validity can be tested against observed data,
though never proven absolutely.
The distinction between observed data and theoretical models thereof
appears starkly in the maximum likelihood procedure, \S\ref{ML}.

The most basic statistic that can be constructed from the overdensity
(the mean having been subtracted off)
is its variance, its second irreducible moment,
the {\bf correlation function}
$\xi(r_{12})$
(also known as the 2-point correlation function, or 2-point function,
or covariance function, or autocovariance function),
\be
\label{xi}
  \xi(r_{12}) \equiv
    \langle \deltarho(\r_1) \hspace{1pt} \deltarho(\r_2) \rangle
  \ .
\ee
Equation~(\ref{xi}) states that
the expectation value of the product of overdensities at a pair of
randomly positioned points separated by $r_{12}$ in the Universe
is $\xi(r_{12})$.
More physically,
the correlation function $\xi(r_{12})$ is
the mean overdensity of neighbours around a random particle (galaxy)
(Peebles 1980, \S31).
The assumption that the density field is statistically homogeneous and
isotropic means that the correlation function $\xi(r_{12})$ is a function
only of the scalar separation $r_{12} \equiv |\r_1 - \r_2|$
of the points $\r_1$ and $\r_2$,
not of their overall location or orientation.

The Fourier transform of the overdensity
defines the Fourier modes $\FTdeltarho(\k)$ at wavevector $\k$
(hats denote Fourier transforms throughout this review)
\be
\label{dk}
  \FTdeltarho(\k) \equiv \int e^{i \k.\r} \deltarho(\r) \, d^3 r
  \ , \quad
  \deltarho(\r) = \int e^{- i \k.\r} \FTdeltarho(\k) \, d^3 k/(2\pi)^3
  \ .
\ee
The {\bf power spectrum} is by definition
the covariance of Fourier modes,
which from the definitions~(\ref{dk}) and (\ref{xi})
is equal to the Fourier transform of the correlation function
\be
\label{dkdk}
  \langle \FTdeltarho(\k_1) \hspace{1pt} \FTdeltarho(\k_2) \rangle
  =
  \int e^{i \k_1.\r_1 + i \k_2.\r_2} \xi(r_{12}) \, d^3 r_1 d^3 r_2
  \ .
\ee
Since the correlation function $\xi(r_{12})$ is a function only of separation,
equation~(\ref{dkdk}) reduces to\footnote{
It is also fine to define the power spectrum as the covariance of modes
with one of the modes taken to be the complex conjugate,
in which case
\protect\raisebox{0ex}[2ex][0ex]{$
\langle \FTdeltarho(\k_1) \FTdeltarho^\ast(\k_2) \rangle =
(2\pi)^3 \discretionary{}{}{} \delta_D(\k_1 - \k_2)
\discretionary{}{}{} P(k_1)$}.
The equivalence of the two definitions is made clear in
\S\protect\ref{hilbert}.
The advantage of the symmetric choice~(\protect\ref{dkP})
becomes more apparent when dealing with higher order correlation functions,
such as the 3-point function
\protect\raisebox{0ex}[2ex][0ex]{$
\langle \FTdeltarho(\k_1) \FTdeltarho(\k_2) \FTdeltarho(\k_3) \rangle$}.
}
\be
\label{dkP}
  \langle \FTdeltarho(\k_1) \hspace{1pt} \FTdeltarho(\k_2) \rangle
  =
  (2\pi)^3 \delta_D(\k_1 + \k_2) P(k_1)
  \ ,
\ee
where $P(k)$ is also called the power spectrum,
\be
\label{P}
  P(k) \equiv \int e^{i \k.\r} \xi(r) \, d^3 r
  \ , \quad
  \xi(r) = \int e^{- i \k.\r} P(k) \, d^3 k/(2\pi)^3
  \ .
\ee
The `momentum-conserving' 3-dimensional Dirac delta function
$\delta_D(\k_1 + \k_2)$
in equation~(\ref{dkP})
%(whose integral is one, $\int \delta_D(\k) d^3 k = 1$,
%integrated over any region containing the origin $\k = 0$)
expresses the assumed translation invariance, i.e.\ statistical homogeneity,
of clustering,
while the fact that $P(k)$
is a function only of the absolute value $k$ of the wavevector $\k$
expresses statistical isotropy.

\begin{sloppypar}
It is useful also to give here results in spherical transform space,
in which the density field is expanded in spherical harmonics
about the observer.
Let $\deltarho_{\el m}(r)$ denote spherical modes in real space
\be
\label{drlm}
  \deltarho_{\el m}(r) \equiv \int Y_{\el m}(\hat\r) \deltarho(\r) \, do_\r
  \ , \quad
  \deltarho(\r) = \sum_{\el m} Y^\ast_{\el m}(\hat\r) \deltarho_{\el m}(r)
\ee
($do_\r$ denotes an interval of solid angle in real space)
and similarly let $\FTdeltarho_{\el m}(k)$
denote spherical modes in Fourier space
\be
\label{dklm}
  \FTdeltarho_{\el m}(k) \equiv \int Y_{\el m}(\hat\k) \FTdeltarho(\k) \, do_\k
  \ , \quad
  \FTdeltarho(\k) = \sum_{\el m} Y^\ast_{\el m}(\hat\k) \FTdeltarho_{\el m}(k)
\ee
($do_\k$ denotes an interval of solid angle in Fourier space)
where $Y_{\el m}$ are the usual orthonormal spherical harmonics.
The spherical transforms in real and Fourier space are related by
\ba
\label{dkrlm}
  \FTdeltarho_{\el m}(k) &=&
    i^\el 4\pi \int_0^\infty \! j_\el(kr) \deltarho_{\el m}(r) \, r^2 dr
  \ ,
  \nn
  \deltarho_{\el m}(r) &=&
    i^{-\el} 4\pi \int_0^\infty \! j_\el(kr) \FTdeltarho_{\el m}(k)
    \, k^2 dk/(2\pi)^3
\ea
where $j_\el(kr)$ are spherical Bessel functions.
The reality conditions $\deltarho^\ast(\r) = \deltarho(\r)$, hence
$\FTdeltarho^\ast(\k) = \FTdeltarho(-\k)$,
along with the usual properties
$Y^\ast_{\el m} = (-)^m Y_{\el,-m}$
and
$Y_{\el m}(-\hat\k) = (-)^\el Y_{\el m}(\hat\k)$
of the spherical harmonics,
imply
\be
  \deltarho^\ast_{\el m}(r) = (-)^m \deltarho_{\el,-m}(r)
  \ , \quad
  \FTdeltarho^\ast_{\el m}(k) = (-)^{\el+m} \FTdeltarho_{\el,-m}(k)
  \ .
\ee
The correlation function of spherical modes in real space is
\be
\label{ddrlm}
  \langle \deltarho_{\el_1 m_1}(r_1) \hspace{1pt}
    \deltarho_{\el_2 m_2}(r_2) \rangle =
    (-)^{m_1} \delta^K_{\el_1\el_2} \delta^K_{m_1, -m_2}
    \xi_\el(r_1,r_2)
\ee
where $\delta^K$ is the Kronecker delta,
and the reduced correlation function $\xi_\el(r_1,r_2)$ is related to
the power spectrum $P(k)$ by
\be
\label{xilP}
  \xi_\el(r_1,r_2) =
    (4\pi)^2 \int_0^\infty \! j_\el(kr_1) j_\el(kr_2) P(k) \, k^2 dk/(2\pi)^3
  \ .
\ee
The extraneous minus signs in~(\ref{ddrlm}),
and also in the next equation~(\ref{ddklm}),
disappear if one takes the complex conjugate on one of the overdensities,
as in $\langle \deltarho \hspace{1pt} \deltarho^\ast \rangle$
in place of $\langle \deltarho \hspace{1pt} \deltarho \rangle$;
see \S\ref{hilbert} for clarification of this point.
The correlation function of spherical modes in Fourier space is
\be
\label{ddklm}
  \langle \FTdeltarho_{\el_1 m_1}(k_1) \hspace{1pt}
    \FTdeltarho_{\el_2 m_2}(k_2) \rangle =
    (-)^{\el_1+m_1} \delta^K_{\el_1 \el_2} \delta^K_{m_1, -m_2}
    (2\pi)^3 \delta_D(k_1\!-\!k_2) k_1^{-2} P(k_1)
\ee
where $\delta_D(k_1-k_2)$ is the 1-dimensional Dirac delta-function,
satisfying
%$\int \delta_D(k) \discretionary{}{}{} d k = 1$
$\int \delta_D(k) d k = 1$
integrated over any interval containing the origin $k = 0$.
As explained in \S\ref{hilbert},
the expression in front of $P(k_1)$
on the right hand side of equation~(\ref{ddklm})
is just the unit matrix in $k\el m$-space, equation~(\ref{unitklm}).
\end{sloppypar}

\subsubsection{The Observed Density Field and Shot Noise}
\label{shot}

Real surveys probe only a portion of the density field of the Universe,
and what they do survey is liable to be an imperfect representation
of the true density field.
This subsection discusses what is often assumed to be the dominant
source of noise in a survey, which is Poisson sampling noise,
often also called shot noise.
It is assumed in this subsection that the data lie in real space,
not redshift space.
Corresponding results in redshift space are given in \S\ref{redshot}.

Typically, a galaxy survey does not include all galaxies in a region of space,
but only, say, those brighter than some flux limit.
Thus a survey generally blends a finer sampling of dim galaxies nearby
with a sparser sampling of luminous galaxies farther away.
A survey is characterized by its {\bf selection function} $\nbar(\r)$,
which is the expected mean number of galaxies at position $\r$
given the selection criteria (e.g.\ the flux limit) of the survey.

The selection function $\nbar(\r)$ of a survey must be measured from it.
The problem of measuring the selection function of a survey is reviewed by
Binggeli, Sandage \& Tammann (1988),
and is discussed further in \S\ref{selfn}.
Some additional comments on the measurement of the selection function
appear in the maximum likelihood section, \S\ref{ML},
in the paragraph following equation~(\ref{Csij}).

In order to combine heterogeneously sampled regions,
the (testable) hypothesis is commonly made that the galaxies observed
are drawn randomly from a hypothetical continuous underlying population
of galaxies.
The observed galaxies then form a {\bf Poisson process} on the underlying
population,
and the selection function $\nbar(\r)$ is interpreted as specifying
the probability (in units of number of galaxies per unit volume)
of including a galaxy at position $\r$ into the survey.

Let $n(\r)$ denote the observed number density of galaxies at
(real, not redshifted) position $\r$ in a survey.
The number density $n(\r)$ is a sum of delta functions,
since galaxies come as discrete units.
The {\bf observed galaxy overdensity} $\delta_\obs(\r)$ is then defined by
\be
\label{delta}
  \delta_\obs(\r) \equiv {n(\r) - \nbar(\r) \over \nbar(\r)}
  \ .
\ee
In the Poisson process model,
the observed galaxy overdensity $\delta_\obs(\r)$
provides a discretized but unbiased estimate of the true overdensity
$\deltarho(\r)$, equation~(\ref{deltarho}).
The subscript $\obs$ is retained throughout this section~\ref{correlation}
to distinguish the estimate $\delta_\obs$ from the hypothesized true value
$\delta$,
but is dropped from most of this review
to avoid an overly ponderous notation
(correctly, estimated values not only of $\delta$ but also of all other
measured quantities should be distinguished from their `true' values).
It should be clear from the context whether what is meant is
the theoretical `true' value of a quantity,
or an observational estimate thereof
(the theoretical sections~\ref{theory}, \ref{methods}, and \ref{translinear}
refer largely to theoretical quantities,
while the observational sections~\ref{example} and \ref{measurement}
report mainly estimates thereof).

In the Poisson process model,
the expectation value $C(\r_1,\r_2)$ of the covariance of observed
overdensities is a sum of the true correlation function $\xi(r_{12})$
with a {\bf Poisson sampling noise}, or {\bf shot noise}, term:
\be
\label{C}
  \langle \delta_\obs(\r_1) \hspace{1pt} \delta_\obs(\r_2) \rangle \equiv
    C(\r_1,\r_2) =
    \xi(r_{12}) + \delta_D(\r_1-\r_2) [\nbar(\r_1)]^{-1}
  \ .
\ee
The form
$\delta_D(\r_1-\r_2) [\nbar(\r_1)]^{-1}$
of the Poisson sampling term can be derived from the following argument.
First, note that
$\int_V n(\r_1) n(\r_2) d^3 r_1 d^3 r_2
= \int_V n(\r_1) d^3 r_1 = 0$ or 1
when the integration is over an infinitesimal volume $V$,
which either does not ($0$) or does ($1$) contain a galaxy.
It follows that the expectation value of the product of densities
in the same infinitesimal volume element is
$\langle n(\r_1) n(\r_2) \rangle =
\delta_D(\r_1-\r_2) \langle n(\r_1) \rangle =
\delta_D(\r_1-\r_2) \nbar(\r_1)$,
whence
$\langle \delta_\obs(\r_1) \delta_\obs(\r_2) \rangle =
\delta_D(\r_1-\r_2) [\nbar(\r_1)]^{-1}$,
which is the shot noise term of equation~(\ref{C}) as claimed.
The Poisson sampling term reflects the fact that the probability
of finding yourself as a neighbour at zero separation is unity.
The shot noise becomes infinite in regions outside a survey
where the selection function is zero, which makes sense.

Equation~(\ref{C}) shows that,
in the Poisson process model,
the expectation value of the survey covariance
is equal to the true correlation function at any finite separation
\be
\label{Cn}
  C(\r_1,\r_2)_{\r_1 \neq \r_2} = \xi(r_{12})
  \ .
\ee
Thus any average
$\langle \delta_\obs(\r_1) \delta_\obs(\r_2) \rangle$
of products of pairs of overdensities at any finite separation
$r_{12} \equiv |\r_1-\r_2| \neq 0$,
weighted in any arbitrary a priori fashion,
provides an unbiased estimate of the true correlation function $\xi(r_{12})$.

In estimating the correlation function from a survey,
the shot noise contribution can be eliminated by excluding from the computation
all self-pairs of galaxies (pairs consisting of a galaxy and itself).
Alternatively, it may be convenient to include self-pairs,
and to subtract off the shot noise as a separate step.

\subsection{Redshift Space}
\label{redshift}

Redshift space quantities
(distinguished in this review by a superscript $s$)
are defined analogously to real space quantities.
Let $n^{s}(\s)$ denote the observed number density of galaxies
at redshift position $\s$ in a redshift survey
(the observer is at the origin $\s = 0$).

A slightly subtle point in the definition of overdensity $\delta^{s}(\s)$
in redshift space arises because,
as emphasized by Fisher, Scharf \& Lahav (1994),
the apparent brightness of a galaxy depends on its true distance $r$,
not its redshift distance $s$,
so that the selection function of a flux-limited redshift survey
is correctly a function $\nbar(\r)$ in real space, not redshift space.
This suggests that one might
define the observed overdensity in redshift space by
$\delta^{s}_\obs(\s) \equiv [n^{s}(\s) - \nbar(\r)] / \nbar(\r)$.
This is possible,
but it requires knowing not only the true selection function
$\nbar(\r)$ in real space, but also the true distance $r$ to each galaxy,
which involves carrying out a full reconstruction of the deredshifted density
field
(Yahil \etal\ 1991;
Fisher \etal\ 1995b;
Webster, Lahav \& Fisher 1997),
whereupon one might as well work with the overdensity $\delta_\obs(\r)$
in real space.

There remain two alternative possibilities for defining the redshift
overdensity.
The first option is to define the observed redshift space galaxy overdensity
$\delta^{s}_\obs(\s)$ at redshift position $\s$
relative to the real space selection function $\nbar(\s)$
evaluated at the same redshift position $\s$
\be
\label{deltas}
  \delta^{s}_\obs(\s) \equiv {n^{s}(\s) - \nbar(\s) \over \nbar(\s)}
  \ .
\ee
This is the definition adopted by Kaiser (1987),
and also adopted in \S\ref{operator}.

A drawback of the definition~(\ref{deltas}) is that it can be tricky to
estimate the real space selection function $\nbar(\s)$ from a redshift survey,
whereas it is relatively straightforward to estimate the redshift space
selection function $\nbar^{s}(\s)$ (see footnote\footnote{
Hamilton \& Culhane (1996) claim that the real and redshift selection
functions agree to linear order, but this is false.
See the paragraph just before \S\ref{turnermethod}.
}).
Thus a second possibility is to define the observed redshift space overdensity
$\delta_\obs^{ss}(\s)$
[with a double $ss$ superscript to distinguish it from $\delta_\obs^{s}(\s)$]
relative to the redshift space selection function $\nbar^{s}(\s)$
\be
\label{deltass}
  \delta^{ss}_\obs(\s) \equiv {n^{s}(\s) - \nbar^{s}(\s) \over \nbar^{s}(\s)}
  \ .
\ee

The relation between the real and redshift space selection functions
$\nbar(\s)$ and $\nbar^{s}(\s)$,
and the consequent relation between the redshift space overdensities
$\delta^{s}(\s)$ and $\delta^{ss}(\s)$,
is discussed in \S\ref{selfn}.

\subsubsection{The True Density Field in Redshift Space}
\label{redshiftdef}

As in real space, in redshift space
it is necessary to distinguish between the observed redshift overdensity
$\delta^{s}_\obs(\s)$, equation~(\ref{deltas}),
and the hypothetical `true' redshift overdensity
$\deltarho^{s}(\s)$
whose existence is predicted theoretically.
For fluctuations in the linear regime,
the theoretical prediction is that the true redshift overdensity
$\deltarho^{s}(\s)$
is related to the true unredshifted overdensity $\deltarho(\r)$
by equation~(\ref{ds}), derived in \S\ref{operator}.
Similarly, corresponding to the observed redshift overdensity
$\delta^{ss}_\obs(\s)$, equation~(\ref{deltass}),
is a hypothetical `true' redshift overdensity
$\deltarho^{ss}(\s)$
predicted by equation~(\ref{dss}).
Corresponding predictions for the redshift overdensities
$\delta^{s\,\LG}$ and $\delta^{ss\,\LG}$ measured in the Local Group
frame instead of the Cosmic Microwave Background frame
are given by equations~(\ref{dsLG}) and (\ref{dssLG}).
The definitions in this subsection remain valid
for all the various flavours of the redshift overdensity:
just change the superscripts appropriately.

Unlike the true overdensity $\deltarho(\r)$ in real space,
which is defined independent of the selection function of the survey,
the true overdensity $\deltarho^{s}(\s)$ in redshift space
depends on the selection function.
This is fine, except that an ambiguity arises for regions outside the
survey, where the selection function is zero.
Specifically, the true redshift overdensity $\delta^{s}(\s)$
depends on the selection function $\nbar(\r)$ through the quantity
$\alpha(\r) \equiv \partial\ln r^2 \nbar(\r)/\partial\ln r$
in the redshift distortion operator~(\ref{S}).
The quantity $\alpha(\r)$ is zero divided by zero outside the survey,
where the selection function is zero, so is ambiguous.

Two comments can be made about this ambiguity in the definition of
the true redshift overdensity $\delta^{s}(\s)$ outside the survey.
Firstly,
one can choose to resolve the ambiguity in whatever manner is convenient,
and this choice has no effect whatsoever on the comparison between
theory and observation.
This is because the ambiguity in $\alpha(\r)$ occurs only at points
outside the survey, whereas comparison between theory and observation
is made only at points inside the survey.
A convenient choice in a flux-limited survey,
where the selection function $\nbar(r)$, hence $\alpha(r)$,
is a function only of depth $r$, not of direction $\hat\r$ within the angular
boundaries of the survey, is to take $\alpha(r)$ to have the same value
outside the angular boundaries as within.
The advantage of this choice is that it preserves the angular symmetry
of the `true' redshift correlation function about the observer.
Beyond the radial extent of the survey, $\alpha(r)$ could be set to zero,
or any other convenient choice.

The second comment is that
the ambiguity disappears in the plane-parallel, or distant observer, limit,
where the $\alpha$ term in the distortion operator goes to zero.
In the plane-parallel limit,
the `true' redshift space overdensity $\deltarho^{s}(\s)$
becomes independent of the selection function,
and the `true' plane-parallel redshift space correlation function
and power spectrum are correspondingly independent of the selection function.
This is the context in which the concept of a redshift space power spectrum
is in any case most often considered (\S\ref{planepar}).

Suppose hereafter then the true redshift space overdensity $\deltarho^{s}(\s)$
has been defined through all space, the ambiguity outside the boundaries
of the survey being resolved if possible as described in the previous
two paragraphs.
The {\bf redshift space correlation function}
$\xi^{s}(s_{12},s_1,s_2)$
is defined by
\be
\label{xisdd}
  \xi^{s}(s_{12},s_1,s_2) \equiv
    \langle \deltarho^{s}(\s_1) \hspace{1pt} \deltarho^{s}(\s_2) \rangle
  \ .
\ee
Redshift distortions partially destroy the symmetry
enjoyed by the unredshifted correlation function $\xi(r_{12})$,
so that the redshift correlation function
$\xi^{s}(s_{12}, s_1, s_2)$
is a function of the redshift distances $s_1$ and $s_2$
as well as the separation $s_{12}$ of a pair of galaxies.
Redshift distortions do however preserve the rotational symmetry
of the correlation function about the position of the observer\footnote{
\label{rot}
More correctly, the redshift correlation function has orientation symmetry
provided that the selection function is independent of direction
over the parts of the sky surveyed (which need not be the whole sky),
as in a redshift survey with a uniform flux limit.
A selection function that is variable with direction over the sky
destroys orientation symmetry about the observer,
through the
$\alpha(\r)$
term in the redshift distortion operator~(\protect\ref{S}).
}.

If the angle between the positions $\s_1$ and $\s_2$ of the galaxy pair
is small enough,
then the line-of-sight redshift distortions are effectively plane-parallel,
as illustrated in the left panel of Figure~\ref{fogs},
and the redshift correlation function $\xi^{s}$ reduces to a function
only of the components $s_\para$ and $s_\perp$
of the pair separation $\s_{12}$
respectively parallel and perpendicular to the line of sight $\z$
\be
  \xi^{s}(s_{12}, s_1, s_2) \approx
    \xi^{s}(s_\para, s_\perp)
  \ .
\ee

The Fourier transform of the true redshift space overdensity
defines the redshift Fourier modes $\FTdeltarho^{s}(\k)$
(here the necessity for $\deltarho^{s}(\s)$ to be defined everywhere,
including outside the survey, is apparent)
\be
  \FTdelta^{s}(\k) \equiv \int e^{i\k.\s} \delta^{s}(\s) \, d^3 s
  \ , \quad
  \delta^{s}(\s) = \int e^{-i\k.\s} \FTdelta^{s}(\k) \, d^3 k/(2\pi)^3
  \ .
\ee
The covariance of redshift Fourier modes defines the
{\bf redshift space power spectrum},
which is equal to the Fourier transform of the redshift correlation function
\be
\label{dkdks}
  \langle \FTdelta^{s}(\k_1) \hspace{1pt} \FTdelta^{s}(\k_2) \rangle
  =
  \int e^{i \k_1.\s_1 + i \k_2.\s_2}
    \xi^{s}(s_{12}, s_1, s_2) \, d^3 s_1 d^3 s_2
  \ .
\ee
The line-of-sight redshift distortions
destroy statistical homogeneity,
so that the redshift power spectrum is no longer a diagonal matrix.
The residual orientation symmetry about the observer\footnote{
\label{rotk}
Again,
orientation symmetry is strictly true only if the selection function is
uniform within the angular boundaries of the survey,
because of the $\alpha(\r)$ term
in the redshift distortion operator~(\protect\ref{S}).
}
implies that the redshift power spectrum
is a function only of scalar combinations of its arguments
\be
  \langle \FTdelta^{s}(\k_1) \hspace{1pt} \FTdelta^{s}(\k_2) \rangle
  =
  \FTxi^{s}(|\k_1 + \k_2|, k_1, k_2)
  \ .
\ee

In the plane-parallel approximation,
however,
redshift distortions do preserve statistical homogeneity,
and in that case the redshift power spectrum is again a diagonal matrix
\be
\label{dskdskP}
  \langle \FTdelta^{s}(\k_1) \hspace{1pt} \FTdelta^{s}(\k_2) \rangle
  \approx
  (2\pi)^3 \delta_D(\k_1 + \k_2) P^{s}(k_{1 \para}, k_{1 \perp})
  \ ,
\ee
where $k_{1 \para}$ and $k_{1 \perp}$ are the components of the wavevector
$\k_1$ respectively parallel and perpendicular to the line of sight $\z$, and
\be
  P^{s}(k_\para, k_\perp) \equiv
    \int e^{i \k.\s} \xi^{s}(s_\para, s_\perp) \, d^3 s
  \ .
\ee

Spherical modes $\delta^{s}_{\el m}(s)$ and $\FTdelta^{s}_{\el m}(k)$
in redshift space are defined analogously to equations~(\ref{drlm})
and (\ref{dklm}).
Redshift distortions preserve rotational symmetry about the observer
(but see footnote~\ref{rot}),
so the redshift correlation functions remain diagonal with respect
to the angular indices $\el m$.
Equations~(\ref{drlm})--(\ref{ddrlm}) take the same form in redshift space,
aside from the addition of $s$ superscripts.
The relation~(\ref{xilP}) is modified to
\be
\label{xislP}
  \xi^{s}_\el(s_1,s_2) =
    (4\pi)^2 \int_0^\infty \! j_\el(k_1 s_1) j_\el(k_2 s_2)
    \FTxi^{s}_\el(k_1,k_2) \, k_1^2 d k_1 k_2^2 d k_2/(2\pi)^6
\ee
and the correlation function of spherical modes in Fourier space becomes
\be
\label{ddsklm}
  \langle \FTdelta^{s}_{\el_1 m_1}(k_1) \hspace{1pt}
    \FTdelta^{s}_{\el_2 m_2}(k_2) \rangle
  =
    (-)^{\el_1+m_1} \delta^K_{\el_1 \el_2} \delta^K_{m_1, -m_2}
    \FTxi^{s}_\el(k_1,k_2)
  \ .
\ee

\subsubsection{The Observed Density Field and Shot Noise in Redshift Space}
\label{redshot}

As in real space,
in redshift space the observed galaxy density is subject to Poisson sampling
noise, or shot noise.
In the Poisson process model described in \S\ref{shot},
the expectation value
$C^{s}(\s_1,\s_2)$
of the covariance of observed redshift space
overdensities $\delta^{s}_\obs(\s)$ defined by~(\ref{deltas})
is a sum of the true redshift space correlation function
$\xi^{s}(s_{12}, s_1, s_2)$, equation~(\ref{xis}),
with a Poisson sampling noise term
\be
\label{Cs}
  \langle \delta^{s}_\obs(\s_1) \hspace{1pt}
    \delta^{s}_\obs(\s_2) \rangle \equiv
    C^{s}(\s_1,\s_2) =
    \xi^{s}(s_{12}, s_1, s_2) + \delta_D(\s_1-\s_2) [\nbar(\s_1)]^{-1}
\ee
which may be compared to the corresponding equation~(\ref{C}) in real space.
The coefficient
$[\nbar(\s_1)]^{-1}$ of the Dirac delta-function in the Poisson sampling term
is valid for a flux-limited redshift survey;
the coefficient could conceivably be different if there were some criterion
other than a flux limit for selecting galaxies into the redshift survey.
The form of the Poisson sampling term can be derived by an argument
similar to the one in real space which followed equation~(\ref{C}).
An important part of the argument in redshift space is that
the expectation value of the product of redshift densities in the same
infinitesimal volume element is
$\langle n^{s}(\s_1) n^{s}(\s_2) \rangle =
\delta_D(\s_1-\s_2) \langle n^{s}(\s_1) \rangle =
\delta_D(\s_1-\s_2) \nbar(\s_1)$,
the last step of which comes from the fact that,
in a flux-limited redshift survey, the ensemble-averaged expectation value
of the redshift density at a point is equal to the real (not redshift)
selection function, $\langle n^{s}(\s) \rangle = \nbar(\s)$.

Similarly,
the expectation value
$C^{ss}(\s_1,\s_2)$
of the covariance of observed redshift space
overdensities $\delta^{ss}_\obs(\s)$ defined by~(\ref{deltass}) is
a sum of the corresponding true redshift space correlation function
$\xi^{ss}(s_{12}, s_1, s_2)$, equation~(\ref{xiss}),
with a Poisson sampling noise term
\be
\label{Css}
  \langle \delta^{ss}_\obs(\s_1) \hspace{1pt}
    \delta^{ss}_\obs(\s_2) \rangle \equiv
    C^{ss}(\s_1,\s_2) =
    \xi^{ss}(s_{12}, s_1, s_2)
    + \delta_D(\s_1-\s_2) {\nbar(\s_1) \over [\nbar^{s}(\s_1)]^2}
  \ .
\ee
Again, the coefficient $\nbar(\s_1)/[\nbar^{s}(\s_1)]^2$
of the Poisson sampling term is valid
for a flux-limited redshift survey.
The form of the coefficient follows from the same argument as that
following equation~(\ref{Cs}).

Equation~(\ref{Cs}) shows that, in the Poisson process model,
the covariance of observed redshift space overdensities at any
finite separation $\s_1 \neq \s_2$ provides an unbiased estimate of
the true redshift correlation function,
\be
\label{Csn}
  C^{s}(\s_1,\s_2)_{\s_1 \neq \s_2} = \xi^{s}(s_{12}, s_1, s_2)
  \ ,
\ee
similar to the corresponding result~(\ref{Cn}) in real space.
Similarly, from equation~(\ref{Css}),
$C^{ss}(\s_1,\s_2)_{\s_1 \neq \s_2} = \xi^{ss}(s_{12}, s_1, s_2)$.
As in real space,
the shot noise contribution to the survey covariance in redshift space can be
removed by excluding all self-pairs, consisting of a galaxy and itself.

\subsection{Hilbert Space}
\label{hilbert}

It can be confusing to keep track of
the mess of $2\pi$'s and other factors,
and the juxtaposition of discrete and continuous quantities,
in equations such as~(\ref{ddklm}).
Hilbert space offers a compact, unifying formalism
that takes care of the bookkeeping and provides much insight.
It is not the place of this subsection to present a complete account of Hilbert
space;
rather, the goal is to demonstrate how concepts familiar from
quantum mechanics also prove powerful in the present statistical context.

The overdensity can be regarded as a vector $\bdeltarho$
in an infinite-dimensional vector space, Hilbert space.
This vector has a meaning independent of the basis,
i.e.\ complete set of linearly independent functions,
with respect to which it is expanded.
Thus for example the overdensity vector $\bdeltarho$ has components
$\deltarho_\r$ [$= \deltarho(\r)$] when expanded with respect to a basis
of delta functions in real space,
and it has components $\deltarho_\k$ [$= \FTdeltarho(\k)$, but now the hat
is dropped from $\deltarho_\k$ in recognition of the fact
$\deltarho_\r$ and $\deltarho_\k$ are really the same vector]
when expanded with respect to a basis of Fourier modes $e^{-i\k.\r}$;
but from a Hilbert space point of view the vector $\bdeltarho$ remains
unchanged:
only the `coordinate system', the basis of expansion functions, changed.
This is in precise analogy to ordinary finite-dimensional vectors,
which have a meaning independent of the particular coordinate system
used to locate them.
Similarly, the covariance matrix
\be
\label{bxi}
  \bxi = \langle \bdeltarho \hspace{1pt} \bdeltarho \rangle
\ee
(not an inner product!)
can be regarded as a matrix in Hilbert space,
with, for example, components $\xi_{\r_1 \r_2}$ in real space,
or $\xi_{\k_1 \k_2}$ in Fourier space.

The quintessential property of Hilbert space is the existence of an inner
product, or scalar product, of any two vectors ${\bmia}$ and ${\bmib}$,
which maps the two vectors into a scalar, a (complex-valued, in general) number.
This scalar value is independent of any basis with respect to which the
vectors may be expressed.
Various compact notations exist for writing the scalar product:
\be
\label{ab}
  \langle a \hspace{1pt} | \hspace{1pt} b \rangle =
  \Tr \, {\bmia}^\dagger {\bmib} =
  a^i b_i
  \ .
\ee
The first expression in~(\ref{ab}) is the Dirac bra-ket notation,
in which the ket $| \hspace{1pt} a \rangle$ denotes a vector
and the bra $\langle a \hspace{1pt} |$
its Hermitian conjugate, the complex conjugate transpose of the vector
(the angle brackets $\langle \, \rangle$ here are not to be confused
with statistical averages; in the Dirac notation, a vertical line
always separates the bra and the ket).
The middle expression is matrix notation,
in which the vector ${\bmia}^\dagger$ denotes the Hermitian
conjugate of the vector ${\bmia}$, and $\Tr$ signifies the trace.
The final expression of~(\ref{ab})
is the inner product in component form:
the quantities $a_i$ are the components of the vector ${\bmia}$ with respect
to some orthonormal basis of functions
(see eqs.~[\ref{psipsi}], [\ref{ai}]),
while the quantities $a^i = a^\ast_i$ with raised index
denote their Hermitian conjugates,
and there is implicit summation over the indices $i$
(the similarity to tensor notation in general relativity is not coincidental).
It is to be understood that the index $i$ may be either discrete or continuous,
or a combination of both, and that in the continuous case summation
signifies integration with respect to some defined measure of $i$;
examples of the latter are seen immediately below.
Other variants of the inner product notation also occur,
such as the omission of the trace sign $\Tr$ in the third expression,
if there is no ambiguity,
or the insertion of an explicit summation or integration over $i$
in the last expression.
In practice it may be typographically convenient to keep all indices lowered,
it being understood that when doubled indices occur,
one of the pair is implicitly raised, signifying the Hermitian conjugate.

In a statistically homogeneous 3-dimensional random field,
equal volume elements $d^3 r$ have equal statistical `weight',
so it is logical to define the inner product of two
(complex-valued, in general) functions
$a(\r)$ and $b(\r)$ as an integral over $d^3 r$.
With respect to the various bases encountered in \S\ref{real},
the inner product of $a(\r)$ and $b(\r)$ can be written
\ba
\label{aibi}
  \langle a \hspace{1pt} | \hspace{1pt} b \rangle &\!\!\!=\!\!\!&
  \int a^\ast(\r) b(\r) \, d^3 r =
  \int \FTa^\ast(\k) \FTb(\k) \, d^3 k/(2\pi)^3
  \\
  \nonumber
  &\!\!\!=\!\!\!&
  \int_0^\infty \sum_{\el m} a^\ast_{\el m}(r) b_{\el m}(r) \, r^2 dr =
  \int_0^\infty \sum_{\el m} \FTa^\ast_{\el m}(k) \FTb_{\el m}(k)
    \, k^2 dk/(2\pi)^3
  \ .
\ea
The equality between the real and Fourier integrals,
the second and third expressions in equation~(\ref{aibi}),
is Parseval's theorem, an apparent triviality in the Hilbert formalism.
Note that if the inner product is complex-valued,
then the order of the inner product matters:
the inner product in the opposite order is its complex conjugate,
$\langle b \hspace{1pt} | \hspace{1pt} a \rangle =
\langle a \hspace{1pt} | \hspace{1pt} b \rangle^\ast$.

A basis of functions $\psi_i(\r)$
is said to be orthonormal if their inner products constitute the unit matrix
(see eq.~[\ref{unitmatrix}])
\be
\label{psipsi}
  \int \psi_i(\r) \psi^j(\r) \, d^3 r = 1_i^j
  \ , \quad
  \sum_i \psi^i(\r) \psi_i(\r') = \delta_D(\r-\r')
\ee
in which either equality implies the other,
given that $\psi_i(\r)$ constitute a basis, a complete linearly independent set.
In index notation, equations~(\ref{psipsi}) are
\be
  \psi_i^{\ \r} \psi^j_{\ \r} = 1_i^j
  \ , \quad
  \psi^i_{\ \r} \psi_i^{\ \r'} = 1_\r^{\r'}
  \ .
\ee
The orthonormal functions $\psi_i(\r_j)$ can be regarded as the rows
of an orthonormal matrix $\Psi_{ij}$, satisfying
$\bPsi\bPsi^\dagger = \bPsi^\dagger\bPsi = \one$.
The components $a_i$ of a vector ${\bmia}$
with respect to the orthonormal basis $\psi_i(\r)$ are given by
\be
\label{ai}
  a_i = \int \psi_i(\r) a(\r) \, d^3 r
  \ , \quad
  a(\r) = \sum_i \psi^i(\r) a_i
  \ .
\ee
In index notation, these equations are
\be
  a_i = \psi_i^{\ \r} a_\r
  \ , \quad
  a_\r = \psi^i_{\ \r} a_i
  \ .
\ee

The rule that raising a lowered index $i$
(or lowering a raised index $i$)
means take the complex conjugate works fine for vectors,
but the rule becomes less easy to interpret for a matrix,
where there is more than one index.
In practical calculations it is easier to apply an equivalent set of rules,
valid for matrices that are real-valued in real space
(as is usually the case for matrices of physical interest, such as
the covariance matrix $\langle \bdelta \hspace{1pt} \bdelta \rangle$,
eq.~[\ref{bxi}],
or the linear redshift distortion operator $\bS$, eq.~[\ref{S}]).
The rules, which depend on the basis,
are that raising (or lowering) an index $i$:
\begin{itemize}
\item[$\circ$]
in real space does nothing;
\item[$\circ$]
in Fourier space transforms $\k_i \rightarrow -\k_i$;
\item[$\circ$]
in $r\el m$-space transforms $m_i \rightarrow -m_i$
and multiplies by $(-)^{m_i}$;
\item[$\circ$]
in $k\el m$-space transforms $m_i \rightarrow -m_i$
and multiplies by $(-)^{\el_i+m_i}$.
\end{itemize}

The unit matrix $1_i^j$,
whether for discrete or continuous indices,
is defined by the property that
\be
\label{unitmatrix}
  1_i^j a_j = a_i
\ee
for any vector $a_i$.
From this definition together with the definition~(\ref{aibi}) of the inner
product, it is straightforward to determine that the unit matrix
in real space is
(the following equations [\ref{unitr}]--[\ref{unitklm}]
give both the unit matrix and its expression with both indices lowered)
\be
\label{unitr}
  1_\r^{\r'} = \delta_D(\r-\r')
  \ , \quad
  1_{\r\r'} = \delta_D(\r-\r')
  \ .
\ee
Similarly the unit matrix in Fourier space is
\be
  1_\k^{\k'} = (2\pi)^3 \delta_D(\k-\k')
  \ , \quad
  1_{\k\k'} = (2\pi)^3 \delta_D(\k+\k')
  \ .
\ee
The unredshifted power spectrum~(\ref{dkP}) is thus recognized
as being a diagonal matrix, with eigenvalues $P(k)$.
In spherical transform space with the radial index in real space,
the unit matrix is
(abbreviating $\mu = r\el m$)
\be
  1_\mu^{\mu'} =
    \delta^K_{\el\el'} \delta^K_{mm'} {\delta_D(r-r') \over r^2}
  \ , \quad
  1_{\mu\mu'} =
    (-)^m \delta^K_{\el\el'} \delta^K_{m,-m'} {\delta_D(r-r') \over r^2}
  \ .
\ee
In spherical transform space with the radial index in Fourier space,
the unit matrix is
(abbreviating $\mu = k\el m$)
\be
\label{unitklm}
  1_\mu^{\mu'} =
    \delta^K_{\el\el'} \delta^K_{mm'} {(2\pi)^3 \delta_D(k-k') \over k^2}
  \ , \ \ 
  1_{\mu\mu'} = (-)^{\el+m}
    \delta^K_{\el\el'} \delta^K_{m,-m'} {(2\pi)^3 \delta_D(k-k') \over k^2}
  \ .
\ee
Thus again the unredshifted power spectrum~(\ref{ddklm})
in $k\el m$-space is seen to be a diagonal matrix, with eigenvalues $P(k)$.

There is no need to go further here into the Hilbert space formalism,
which can be learned from any quantum mechanical textbook
(e.g.\ Landau \& Lifshitz 1958).
The advantage of the formalism is that it frees expressions,
such as the relation~(\ref{ds}) between overdensity in real and redshift space,
or the expression~(\ref{L}) for the Gaussian likelihood function,
from being tied to any particular basis, such as real space or Fourier space.
If such expressions are written in component form
(i.e.\ the vectors and matrices are written with indices),
then any index can be regarded as referring to any basis of functions,
possibly a different basis for each index, with the following two provisos:
(1) indices and the bases to which they refer should match between
the left and right hand sides of an equation;
(2) wherever paired indices occur, indicating an inner product,
they should be evaluated in the same (arbitrary) basis.
In evaluating an inner product, it suffices to bear in mind two rules.
First, one of a pair of indices $i$ should always be raised
(implicitly, if not typographically) indicating the Hermitian conjugate;
in the case of Fourier space for example,
this means using $-\k_i$ for one index and $+\k_i$ for the other
(at least for vectors and matrices that are real-valued in real space).
Second, if the paired index is continuous,
then the volume element of integration is consistently the same,
in accordance with the definition of the inner product in the basis used;
in Fourier space for example,
the volume element is $d^3 k_i/(2\pi)^3$, according to equation~(\ref{aibi}).

\subsection{Merit of the Power Spectrum}
\label{power}

Blackman \& Tukey (1959, \S B.3)
in their classic text on power spectra state
``The autocovariance function is of little use except as a basis
for measuring the power spectrum''.
The power spectrum of large scale structure in the Universe was apparently
first measured by
Yu \& Peebles (1969) and Peebles (1973).

The fundamental advantage of the (true, unredshifted) power spectrum is that
the Fourier modes of a statistically homogeneous random field are uncorrelated:
their covariance matrix
$\langle \FTdeltarho(\k_1) \FTdeltarho(\k_2) \rangle$ is a diagonal matrix,
equation~(\ref{dkP}).
Modes that are uncorrelated are said to be
{\bf statistically orthogonal}.
The statistical orthogonality of Fourier modes
does not imply that they are independent in general,
but it does imply that they are independent in the important case of a
multivariate Gaussian field, for which by definition the 3-point
and all higher correlation functions are zero.

Note that the advantage of the power spectrum over (for example)
the spatial correlation function is not that it contains any more information
--- the power spectrum and the correlation function have
precisely the same information content ---
but rather that the power spectrum parcels the information into
uncorrelated chunks.

The statistical orthogonality of the (true, unredshifted) Fourier modes
is intimately associated with the assumed
translation invariance of the statistical properties of the density field.
As is familiar from quantum mechanics,
the translation operator, the generator of an infinitesimal translation,
is $- i \partial/\partial\r$, which is also the momentum operator $\k$
(Landau \& Lifshitz 1958, \S15).
The eigenfunctions of the translation operator $- i \partial/\partial\r$
are just the Fourier modes $\FTdeltarho(\k)$, with eigenvalues $\k$.
The operator of an overall translation is
$-i \partial/\partial\R \equiv \sum -i \partial/\partial\r$,
which is the same as the total momentum operator $\K \equiv \sum \k$.
The sum here is over all relevant coordinates,
e.g.\ the 2 coordinates of the 2-point correlation function.
Statistical homogeneity means that the statistical properties,
the correlation functions, of the density field
are unchanged by an overall translation,
$\partial\xi/\partial\R = 0$, or equivalently $\K\xi = 0$.
Thus the correlation functions have zero total momentum, $\K = \sum\k = 0$,
as in equation~(\ref{dkP}).

Similarly,
the statistical orthogonality of (true, unredshifted) spherical modes
with respect to angular indices $\el m$,
equations~(\ref{ddrlm}) and (\ref{ddklm}),
is intimately associated with statistical isotropy about the observer
(or indeed about any other point).
The operator of an infinitesimal rotation is the angular momentum operator
$\bl$, whose eigenmodes are the spherical harmonics $Y_{\el m}$.
Statistical isotropy (about some point)
implies that the correlation functions have zero total angular momentum
(about that point), $\bL = \sum\bl = 0$,
as in equations~(\ref{ddrlm}) and (\ref{ddklm}).

Redshift distortions destroy translation invariance,
but they preserve isotropy about the observer
(but see footnote~\ref{rot}),
and hence also preserve the statistical orthogonality of spherical modes
with respect to the angular indices $\el m$.
Thus, as emphasized by Fisher, Scharf \& Lahav (1994)
and Heavens \& Taylor (1995),
the advantage of the power spectrum can best be preserved in redshift space
by working in spherical harmonics.

The statistical orthogonality of Fourier modes
windowed through any finite survey
(i.e., if the overdensity is multiplied by the,
possibly weighted, selection function of the survey)
is destroyed by the fact that the survey is not translation invariant,
especially at wavelengths approaching the size of the survey.
Similarly, the statistical orthogonality of spherical modes
windowed through a non-all-sky survey
is destroyed by the break down of rotational symmetry.
However, the desirability of measuring quantities that are uncorrelated
remains.
Ways to construct statistically orthogonal modes in real surveys
are described by
Vogeley \& Szalay (1996),
Tegmark, Taylor \& Heavens (1997),
and Tegmark \etal\ (1997),
and ways to decorrelate power spectra are described by
Hamilton (1997),
and Tegmark \etal\ (1997).

\section{Theory of Linear Redshift Distortions}
\label{theory}
\setcounter{equation}{0}

The theory of linear redshift distortions was greatly clarified in
a fundamental paper by Kaiser (1987).
Although the existence of large scale redshift distortions
and the possibility of using them to measure the cosmological density $\Omega_0$
had been recognized previously (Sargent \& Turner 1977; Peebles 1980, \S76B),
the complete, correct equations
describing linear redshift distortions were first derived by Kaiser.
%(See Fisher 1995 for connection between Kaiser \& Peebles?).

The linear distortion equations are presented in \S\ref{operator}.
First, however, it is useful to discuss
what the linear distortion parameter $\beta$,
which is the quantity actually measurable from linear redshift distortions,
and the focus of much of the fuss,
actually means.

\subsection{The Linearized Continuity Equation}
\label{continuity}

I vividly recall a conversation with Jerry Ostriker (in 1991, I believe)
in which I was excitedly explaining how one could measure the cosmological
density $\Omega_0$ from redshift distortions in the linear regime.
Jerry, a fan of explosive cosmologies
(Ostriker \& Cowie 1981),
was unimpressed.
He pointed out that redshift distortions are no more than
a consequence of the continuity equation,
that the continuity equation is of widespread applicability,
and that what redshift distortions measure is not $\Omega_0$
but rather (exact quote)
``a dimensionless number of no particular significance''.

\begin{figure}
\begin{center}
\leavevmode
\epsfxsize=4in \epsfbox{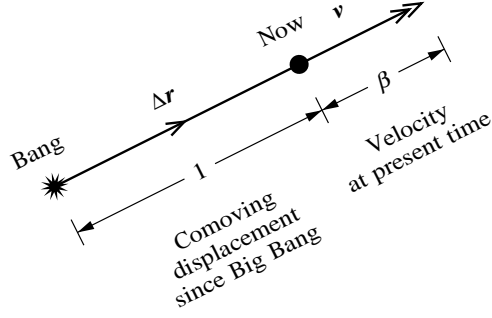}
\end{center}
  \caption[1]{
{\it Problem:}
From the linearized continuity equation~(\protect\ref{cty}) show that,
in the linear regime,
the peculiar velocity $\bv$ of a particle (galaxy) at the present time
is related to its comoving displacement $\Delta \r$
(measured in velocity units) since the Big Bang by
\protect\(
  \bv = \beta \Delta \r
  \ .
\protect\)
In deriving this result you will need to assume that the density field
is initially uniform,
and that the velocity is the gradient of a potential.
Amongst other things, you should find that
in the linear regime particles move in straight lines.
\label{move}
}
\end{figure}

Jerry's view may be on the blueshifted side of the spectrum of perspectives,
but his point is well taken.
Physically,
to grow an overdense region,
galaxies have to move in the general direction of the overdensity.
More precisely,
{\em the value of $\beta$ that is measured from redshift distortions
in the linear regime is the value that solves the linearized
continuity equation}
\be
\label{cty}
  \beta \delta + \bnabla . \bv = 0
  \ .
\ee
Actually, an additional proposition enters here,
which is that the peculiar velocity field $\bv$ is irrotational
(has zero vorticity), as predicted by gravitational growth theory.
These matters are explained below.
The reader who wishes to gain insight into the problem
is invited to solve the problem illustrated in Figure~\ref{move}.

When dealing with dynamics in cosmology,
it is convenient to work in a {\bf comoving coordinate system},
which expands with the general Hubble expansion of the Universe.
To maintain contact with observational reality,
it is also useful, as in \S\ref{intro},
to express comoving distances in velocity units evaluated at the present day,
using the present day values $H_0$ of the Hubble constant
and $a_0$ of the cosmic scale factor
to relate comoving distance and velocity.
Thus let $\r$ denote the comoving (unredshifted) position of a galaxy,
measured in velocity units at the present time
($\r$ is related to Peebles' 1980 \S7
comoving distance $\x$ by $\r = H_0 a_0 \x$).
For example, the comoving distance to the Coma cluster
is $r \approx 7000 \, \kms$, as inferred from its measured redshift;
by definition of comoving coordinates,
this comoving distance does not change (much) as the Universe expands.

The proper {\bf peculiar velocity} $\bv$ of a galaxy is its proper velocity
in the comoving frame
\be
  \bv \equiv {a \, d\r \over H_0 a_0 dt} = {d\r \over d\tau}
\ee
where $t$ is proper time,
$\tau$ is the dimensionless {\bf conformal time}
defined by $d\tau \equiv H_0 a_0 dt/a$,
and $a(\tau)$ is the cosmic scale factor.
One of the advantages of working with conformal time is that
objects (e.g.\ photons) moving at the speed of light $c$
have peculiar velocities $d\r/d\tau = c$ at all times.

At this point
it is necessary to distinguish carefully between matter,
which satisfies a continuity equation,
and galaxies, which, thanks to galaxy formation and later merging, may not.
The continuity, Euler, and Poisson equations
for cold (pressureless) matter (subscripted $M$)
in a perturbed Friedmann-Robertson-Walker Universe
are, expressed in the comoving coordinate system
(Hui \& Bertschinger 1996),
\be
\label{ctym}
  {\partial\delta_M \over \partial\tau} + \bnabla . (1 + \delta_M) \bv_M = 0
\ee
\be
\label{euler}
  {\partial a \bv_M \over a \, \partial\tau} + \bv_M . \bnabla \bv_M =
    - \bnabla \phi
\ee
\be
\label{poisson}
  \nabla^2 \phi = {4\pi G \bar\rho_M a^2 \delta_M \over H_0^2 a_0^2}
    = {3 \Omega_M H^2 a^2 \delta_M \over 2 H_0^2 a_0^2}
\ee
where $\bnabla \equiv \partial/\partial\r$ is the comoving gradient,
$\bar\rho_M \propto a^{-3}$ is the proper mean matter density,
and the cosmological matter density $\Omega_M$ and Hubble constant $H$
both evolve in time.
In the linear regime, where $| \delta_M | \ll 1$,
the continuity and Euler equations reduce to
\be
\label{ctymlin}
  {\partial\delta_M \over \partial\tau} + \bnabla . \bv_M = 0
\ee
\be
\label{eulerlin}
  {\partial a \bv_M \over a \, \partial\tau} = - \bnabla \phi
  \ .
\ee
If the peculiar velocity is decomposed (as can always be done)
into its gradient (irrotational, or longitudinal, or scalar) and
curl (rotational, or transverse, or vector) parts,
\be
  \bv_M = \bv_M^G + \bv_M^C
  \ ,
\ee
in which the longitudinal part $\bv_M^G = \bnabla\psi$
is the gradient of some scalar potential $\psi$,
while the transverse part $\bv_M^C = \bnabla \times {\bmiA}$
is the curl of some vector potential ${\bmiA}$,
then the linearized Euler equation~(\ref{eulerlin}) becomes
\be
\label{vGC}
  {\partial a \bv_M^G \over a \, \partial\tau} = - \bnabla \phi
  \ , \quad
  {\partial a \bv_M^C \over a \, \partial\tau} = 0
  \ .
\ee
The second of equations~(\ref{vGC}) shows that the curl part of the velocity
decays as $\bv_M^C \propto a^{-1}$ as the Universe expands,
so should be sensibly equal to zero in the absence of a mechanism
to generate vorticity.
If only gravity operates, then the peculiar velocity in the linear regime
should be pure gradient.

When the linearized continuity~(\ref{ctymlin}),
linearized Euler~(\ref{eulerlin}),
and Poisson~(\ref{poisson}) equations
are combined, the result is a second order linear differential equation
for the overdensity $\delta_M$,
which, when further combined with the unperturbed solution for
the evolution $a(\tau)$ of the cosmic scale factor,
leads to growing and decaying solutions
(Peebles 1980, \S\S10--13; Padmanabhan 1993, \S4.5)
\be
\label{D}
  \delta_M(\r,\tau) \propto D(\tau)
\ee
which evolve in time without change of shape.
The interesting solution is the unstable, growing solution,
and $D(\tau)$ below is taken to be the linear growth factor of the growing mode.
The linearized continuity equation~(\ref{ctymlin}) can then be written
\be
\label{ctyf}
  {H a f \over H_0 a_0} \delta_M + \bnabla . \bv_M = 0
\ee
where $f$ is the {\bf dimensionless linear growth rate} of the growing mode
\be
  f \equiv {H_0 a_0 \over H a} {d \ln D \over d\tau}
    = {d \ln D \over d \ln a}
  \ .
\ee
In the standard pressureless matter-dominated cosmology,
the dimensionless growth rate $f$
is an analytic function of $\Omega_M$
with a celebrated approximation as a power law\footnote{
Occasionally a small fuss is made about the exponent of the power law in this
approximation (cf.\ Lightman \& Schechter 1990).
A sensible person would use
the exact analytic expression for $f(\Omega_M)$ when reporting
actual results.}
(Peebles 1980, eq. [14.8])
\be
\label{f}
  f(\Omega_M) \approx \Omega_M^{0.6}
  \ .
\ee
Lahav \etal\ (1991) give an approximation in the case where there is
a cosmological constant\footnote{
Einstein's cosmological constant $\Lambda$,
which could arise from a non-zero vacuum density $\rho_{\rm vac}$,
is related to $\Omega_\Lambda$ by
$\Lambda =  8\pi G\rho_{\rm vac} = 3 H^2 \Omega_\Lambda \,$.}
so that the total cosmological density
is a sum of matter and cosmological constant parts,
$\Omega = \Omega_M + \Omega_\Lambda$:
\be
\label{flambda}
  f(\Omega_M, \Omega_\Lambda)
  \approx \Omega_M^{0.6}
    + {\Omega_\Lambda \over 70} \left( 1 + {\Omega_M \over 2} \right)
  \ .
\ee
Evidently the growth rate $f$ depends mainly on the matter density $\Omega_M$,
and is only weakly dependent on the cosmological constant.

Equation~(\ref{ctyf}) is the continuity equation for the matter;
to go from this to the continuity equation~(\ref{cty}) for galaxies
involves one further issue --- bias.

For more than a decade it has been apparent that galaxies may not
constitute an unbiased tracer of the underlying matter density.
If galaxies were an unbiased tracer of the matter, they would,
by definition, satisfy $\delta = \delta_M$
(see footnote\footnote{
Since galaxies come as discrete units,
one should understand this equation in a probabilistic sense.
Galaxies are unbiased if the probability
of finding a galaxy in a volume element $dV$ at position $\r$ is
proportional to the amount of matter $\rho_M(\r) dV$
in that volume element.
The discretized galaxy density $n(\r)$ is then said to be a
Poisson process, with mean proportional to $\rho_M(\r)$.
The statistical properties of such a discretized distribution,
i.e.\ its 2-point and higher correlation functions,
coincide with those of the underlying matter density field aside from
the addition of delta-function discreteness terms at zero separation
(cf.\ \S\ref{shot}).
}).
That some bias exists, at least in some galaxy populations,
follows from the fact that galaxies selected in different ways
have correlation functions with different amplitudes
(e.g.\ Peacock \& Dodds 1994; Oliver \etal\ 1996; Peacock 1997).
For example,
the correlation function of optical galaxies is approximately a factor of 2
larger than that of {\it IRAS\/} galaxies,
a fact evident from Figure~\ref{xilconts},
which demonstrates that optical galaxies and {\it IRAS\/} galaxies
cannot both be unbiased tracers
(Oliver \etal\ 1996).

The simplest model of bias postulates that the galaxy overdensity $\delta$
is linearly biased by a constant factor,
the {\bf linear bias factor} $b$,
relative to the underlying mass density $\delta_M$, so that
\be
\label{b}
  \delta = b \delta_M
  \ ,
\ee
whereas galaxy velocities faithfully follow the velocity of matter
\be
\label{vvm}
  \bv = \bv_M
  \ .
\ee
The linear biasing model~(\ref{b}) was originally motivated by
{\bf threshold biasing},
in which galaxies are supposed to form only where the matter density
exceeds a certain threshold
(Kaiser 1984),
and by
{\bf peaks biasing}
where galaxies form at the peaks of the matter density
(Bardeen \etal\ 1986).
These models predict that,
at least in some regimes,
the galaxy correlation function
$\xi(r_{12}) \equiv \langle \delta(\r_1) \delta(\r_2) \rangle$
should be amplified over the matter correlation function
$\xi_M(r_{12}) \equiv \langle \delta_M(\r_1) \delta_M(\r_2) \rangle$
by an approximately constant factor
\be
  \xi(r_{12}) \approx b^2 \xi_M(r_{12})
  \ .
\ee
What is true observationally is that the correlation functions of
galaxies selected in different ways
differ mainly in their amplitudes,
but only weakly in their shapes
(Peacock \& Dodds 1994; Peacock 1997).
Ultimately,
it remains unclear how nature chooses to bias galaxies in reality.

The linearized continuity equation~(\ref{ctyf}) for the matter,
evaluated at the present time,
together with the linear bias model~(\ref{b}),
yield the linearized continuity equation~(\ref{cty}) for galaxies,
where the dimensionless quantity $\beta$ is related to the
present day value $f_0$ of the linear growth rate $f$,
equation~(\ref{f}) or (\ref{flambda}),
and the bias factor $b$, equation~(\ref{b}), by
\be
\label{betaf}
  \beta = {f_0 \over b}
  \ .
\ee
With the standard formula~(\ref{f}) for $f$,
this becomes the oft-quoted relation~(\ref{beta}).

If the matter peculiar velocity $\bv_M$ is curl-free in the linear regime
(see the argument following eq.~[\ref{vGC}]),
and the galaxy velocity is unbiased, equation~(\ref{vvm}),
then the galaxy peculiar velocity field $\bv$ is also curl-free
in the linear regime.
The linearized galaxy continuity equation~(\ref{cty})
then implies that the galaxy peculiar velocity $\bv$ is related to the
galaxy overdensity $\delta$ at the present day by
\be
\label{bv}
  \bv = - \beta \bnabla \nabla^{-2} \delta
\ee
where $\nabla^{-2}$ is the inverse Laplacian.
All measurements of $\beta$ in the linear regime
are in effect measurements of the ratio of the peculiar velocity $\bv$
to the galaxy overdensity $\delta$,
the (testable) relation~(\ref{bv}) being assumed to hold true.
This is true not only in the case of large scale redshift distortions,
but also for measurements of $\beta$
from direct comparison of overdensities and peculiar velocities
(Strauss \& Willick 1995),
or from comparison of the peculiar motion of the Local Group of galaxies
to the dipole anisotropy of galaxies over the sky
(Strauss \etal\ 1992b).

Two final points about bias are worth making.
Firstly,
the linear bias model~(\ref{b}) must break down in the nonlinear regime
(if $b > 1$, as generally expected),
since otherwise equation~(\ref{b}) would predict
that regions empty of matter, $\delta_M = -1$,
would have negative galaxy density, which is impossible.
Various models of nonlinear biasing can be contemplated
(Fry \& Gazta\~naga 1993;
Mann, Peacock \& Heavens 1997).

Secondly,
if a linear bias $b_i$ is established
so $\delta(\r,t_i) = b_i \delta_M(\r,t_i)$
at some initial time $t_i$,
after which galaxies satisfy continuity
(being neither created nor destroyed),
so that
$[1 + \delta(\r)]/[1 + \delta_M(\r)] \propto n(\r)/\rho_M(\r)$
remains constant in Lagrangian elements after the initial time $t_i$,
then the bias factor $b = \delta/\delta_M$ evolves.
During linear growth, $| \delta_M | \ll 1$,
the bias decreases in time while remaining constant in space,
\be
\label{bt}
  b(t) - 1 =
    {D(t_i) \over D(t)} (b_i - 1)
  \ ,
\ee
where $D(t)/D(t_i) = \delta_M(\r,t)/\delta_M(\r,t_i)$
is the linear growth factor.
Equation~(\ref{bt}) shows that the action of continuity
is to drive the bias factor closer to
(but not necessarily all the way to) unity,
as fluctuations grow.
If continuity persists into the nonlinear regime,
then the bias factor becomes a function also of position,
\be
\label{brt}
  b(\r,t) - 1 =
    \delta_M(\r,t_i) \left( {1 \over \delta_M(\r,t)} + 1 \right)
    (b_i - 1)
\ee
saturating
at $b = 1$ in underdense regions (as $\delta_M(\r,t) \rightarrow -1$),
and at $b = 1 + \delta_M(\r,t_i) (b_i - 1)$
in overdense regions (as $\delta_M(\r,t) \rightarrow \infty$).
In the nonlinear regime, streams from several different
initial points may overlap at the same final point, so correctly
equation~(\ref{brt}) should be averaged over streams.

The model described in the previous paragraph is of course simplistic.
The point is that the bias factor $b$ probably evolves,
and that there is a mechanism --- continuity ---
that tends to drive the bias factor closer to unity,
that is, to make galaxies less biased as time goes on.

\subsection{The Linear Redshift Distortion Operator}
\label{operator}

In the linear regime,
the overdensity $\delta^{s}$ in redshift space is related
to the overdensity $\delta$ in real space
by a {\bf linear redshift distortion operator} $\bS$,
\be
\label{ds}
  \delta^{s} = \bS \, \delta
  \ .
\ee
Conceptually, it is helpful to recognize that
{\em the linear redshift distortion operator $\bS$ is just a linear operator}
(a matrix in Hilbert space),
much like the linear operators one encounters in quantum mechanics.
In real (as opposed to Fourier, say) space,
the distortion operator, derived immediately below,
is an integro-differential operator
(the inverse Laplacian $\nabla^{-2}$ is the integral part)
\be
\label{S}
  \bS = 1 + \beta \left(
{\partial^2 \over \partial r^2} + {\alpha(\r) \partial \over r \partial r}
\right) \nabla^{-2}
\ee
where $\alpha(\r)$ is the logarithmic derivative of $r^2$ times the
real space selection function $\nbar(\r)$,
\be
\label{alpha}
  \alpha(\r) \equiv {\partial \ln r^2 \nbar(\r) \over \partial \ln r}
  \ .
\ee
The redshift distortion operator~(\ref{S})
is valid in the frame of reference of stationary observers,
those at rest with respect to the Cosmic Microwave Background (CMB),
and provided that the redshift overdensity $\delta^{s}(\r)$ is defined
relative to the real space selection function $\nbar(\r)$,
as in equation~(\ref{deltas}).
The modifications to the linear redshift distortion operator $\bS$
for the cases where the redshift overdensity is defined in the Local Group
frame, and/or relative to the selection function $\nbar^{s}(\r)$
measured in redshift space,
are given by equations~(\ref{SLG}), (\ref{Ss}), and (\ref{SsLG}).

The linear redshift distortion equations were first derived by
Kaiser (1987, \S2).
The derivation below follows most closely that of
Hamilton \& Culhane (1996, \S2).

The starting point of the derivation of the linear redshift distortion
operator~(\ref{S}) is a conservation equation for galaxies,
which expresses the fact that peculiar velocities displace galaxies
along the line of sight, but they do not make galaxies appear or disappear
(this conservation equation is not the same as the dynamical continuity
equation discussed in \S\ref{continuity})
\be
\label{nsn}
  n^{s}(\s) \, d^3 s = n(\r) \, d^3 r
  \ .
\ee
In terms of the redshift space overdensity $\delta^{s}(\s)$
defined by equation~(\ref{deltas}),
the galaxy conservation equation~(\ref{nsn}) is
\be
\label{nsnd}
  \nbar(\s) \bigl[ 1 + \delta^{s}(\s) \bigr] \, s^2 d s =
  \nbar(\r) \bigl[ 1 + \delta(\r) \bigr] \, r^2 dr
  \ .
\ee
With the relation between redshift position $\s$ and real position $\r$
\be
  \s = \r + v \hat\r
  \ ,
\ee
equation~(\ref{nsnd}) rearranges to
\be
\label{ds1}
  1 + \delta^{s}(\s) = {r^2 \nbar(\r) \over (r + v)^2 \nbar(\r+v\hat\r)}
    \left( 1 + {\partial v \over \partial r} \right)^{-1}
    \bigl[ 1 + \delta(\r) \bigr]
  \ .
\ee
So far equation~(\ref{ds1}) is exact, valid in the linear and nonlinear regimes,
at least until orbit-crossing occurs.

The next step is to linearize equation~(\ref{ds1}).
In addition to the linear theory assumption
$| \delta(\r) | \ll 1$,
which also implies
$| \partial v/\partial r | \ll 1$ if the velocity $v$ is given by linear
theory, equation~(\ref{v}),
it is necessary to assume that peculiar velocities $v$ of galaxies
are small compared to their distances $r$ from the observer
(actually, this assumption can be dropped,
if the redshift space overdensity $\delta^{s}(\r)$
in the CMB frame is constructed according to equation~[\ref{dsLGp}];
see \S\ref{knownvLG}),
\be
\label{vllr}
  | v | \ll r
  \ .
\ee
To linear order equation~(\ref{ds1}) then reduces to
(note in particular that $\delta^{s}(\s) = \delta^{s}(\r)$ to linear order)
\be
\label{dsv}
  \delta^{s}(\r) = \delta(\r)
    - \left( {\partial \over \partial r} + {\alpha(\r) \over r} \right) v
\ee
where $\alpha(\r)$ is defined by~(\ref{alpha}).
In linear theory, the peculiar velocity $\bv$ is given by equation~(\ref{bv}),
so the line-of-sight component $v$ of the peculiar velocity
in the CMB frame is
\be
\label{v}
  v = - \beta {\partial \over \partial r} \nabla^{-2} \delta
  \ .
\ee
\begin{sloppypar}
Inserting this formula into equation~(\ref{dsv})
yields the distortion equation~(\ref{ds})
with the redshift distortion operator $\bS$ given by equation~(\ref{S}),
as was to be proven.
\end{sloppypar}

It follows from equation~(\ref{ds}) that the redshift correlation function
$\xi^{s}(r_{12}, \discretionary{}{}{} r_1,r_2)$,
equation~(\ref{xisdd}),
in the linear regime is related to its
unredshifted counterpart $\xi(r_{12})$ by
\ba
\label{xis}
\lefteqn{\ \ 
  \xi^{s}(r_{12},r_1,r_2) \: = \:
    \bS_1 \bS_2 \, \xi(r_{12})
} && \\
  \nonumber
  && =
    \left[ 1 + \beta \left(
    {\partial^2 \over \partial r_1^2}
    + {\alpha(\r_1) \partial \over r_1 \partial r_1}
    \right) \nabla_1^{-2} \right]
    \! \left[ 1 + \beta \left(
    {\partial^2 \over \partial r_2^2}
    + {\alpha(\r_2) \partial \over r_2 \partial r_2}
    \right) \nabla_2^{-2} \right]
    \xi(r_{12})
  \ . \ 
\ea
This equation is correct for observers who are located randomly in the Universe,
who are at rest with respect to the CMB,
and who define the redshift overdensity relative to the real space
selection function $\nbar(\r)$, as in equation~(\ref{deltas}).
As discussed in \S\ref{redshot},
in a real redshift survey,
where the overdensity field is sampled by discrete galaxies,
the correlation function of the observed overdensity field
contains an additional shot noise term, equation~(\ref{Cs}).

We, sitting on a galaxy, the Milky Way, are not at a random location,
but reside in an overdense region of the Universe, as galaxies are wont to do.
To guard against `local bias',
it would a reasonable precaution to excise at least part of the local region
when analysing a galaxy survey.

Furthermore, we are not at rest in the CMB.
What to do about that is discussed in \S\ref{LG}.

Finally, it is much more straightforward to measure the selection function
$\nbar^{s}(\r)$ in redshift space than in real space.
What to do about that is discussed in \S\ref{selfn}.

\subsubsection{The Linear Plane-Parallel Redshift Distortion Operator}
\label{planepar}

\begin{figure}
\epsfbox{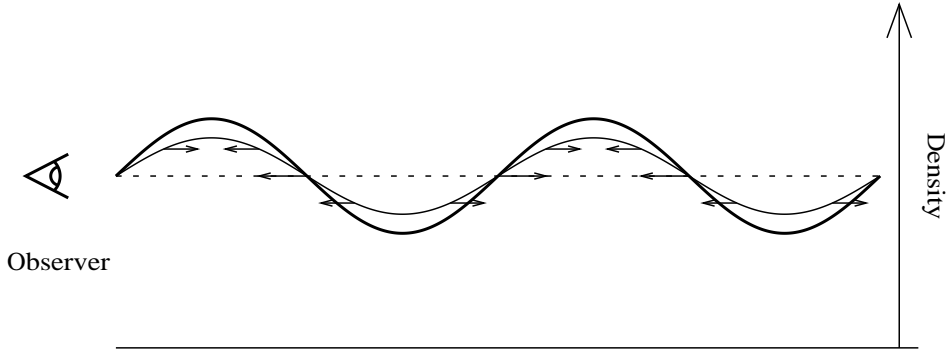}
  \caption[1]{
A wave of amplitude $\delta(\k)$ in real space (thin line)
appears as a wave with enhanced amplitude $\delta^{s}(\k)$ in redshift space
(thick line) because of peculiar velocities (arrows).
If the wavevector $\k$ is along the line of sight,
then the amplification factor is $1 + \beta$.
More generally,
a wave that is angled to the line of sight appears amplified in redshift
space by a factor
\protect\raisebox{0ex}[2ex][0ex]{$1 + \beta \mu_\k^2$},
equation~(\protect\ref{dsk}), where
\protect\raisebox{0ex}[2ex][0ex]{$\mu_\k \equiv \hat\z . \hat\k$}
is the cosine of the angle between the
wavevector $\k$ and the line of sight $\z$.
\label{wave}
}
\end{figure}

In the plane-parallel, or distant observer, limit,
the linear distortion operator~(\ref{S}) reduces to
(the superscript $p$ denotes plane-parallel)
\be
\label{Sp}
  \bS^p = 1 + \beta {\partial^2 \over \partial z^2} \nabla^{-2}
\ee
where $z$ is distance along the line of sight.
In Fourier space,
$( \partial / \partial z )^2 \nabla^{-2} = k_z^2 / k^2 = \mu_{\k}^2$,
where $\mu_\k \equiv \hat\z . \hat\k$ is the cosine of the angle between
the wavevector $\k$ and the line of sight $\z$.
Thus in Fourier space the plane-parallel distortion operator
reduces to a diagonal operator
\be
\label{Spk}
  \bS^p = 1 + \beta \mu_{\k}^2
  \ ,
\ee
so that,
as illustrated in Figure~\ref{wave},
a Fourier mode $\FTdelta^{s}(\k)$ in redshift space is simply
equal to the unredshifted mode $\FTdelta(\k)$
amplified by a factor $1 + \beta \mu_{\k}^2$
\be
\label{dsk}
  \FTdelta^{s}(\k) = ( 1 + \beta \mu_{\k}^2 ) \FTdelta(\k)
  \ .
\ee
It follows from equation~(\ref{dsk}) that,
in the plane-parallel approximation,
the redshift space power spectrum $P^{s}(\k)$ is amplified by
$( 1 + \beta \mu_{\k}^2 )^2$
over its unredshifted counterpart $P(k)$
\be
\label{Kaiser}
  P^{s}(\k) = (1 + \beta \mu_{\k}^2)^2 P(k)
  \ ,
\ee
an elegant formula first pointed out by Kaiser (1987, eq.~[3.5]).

The simplicity of formula~(\ref{Kaiser})
arises from the fact that the plane-parallel distortion operator
is diagonal in Fourier space,
which is intimately associated with the fact that
plane-parallel redshift distortions preserve translation invariance.
The translation operator $- i \bnabla$ commutes with the
plane-parallel distortion operator,
and therefore the eigenmodes of the translation operator
$- i \bnabla$,
which are precisely the Fourier modes $\FTdelta(\k)$,
are also eigenmodes of the plane-parallel distortion operator.

Translation invariance means that redshifted Fourier modes remain
statistically orthogonal in the plane-parallel approximation,
equation~(\ref{dskdskP}),
preserving the advantage of the unredshifted power spectrum
(\S\ref{power}).

\subsubsection{The Linear Radial Redshift Distortion Operator}

Unfortunately,
the full distortion operator~(\ref{S}),
which I call here the radial redshift distortion operator\footnote{
Hamilton \& Culhane (1996) call it the spherical distortion operator.
}
to distinguish it from the plane-parallel distortion operator~(\ref{Sp})
or (\ref{Spk}),
proves to be quite a bit more complicated to deal with than
the plane-parallel operator.
The radial operator destroys translation symmetry,
so that Fourier modes are no longer eigenmodes of the distortion operator,
and the redshift space power spectrum
$\langle \delta^{s}(\k_1) \delta^{s}(\k_2) \rangle$
is no longer a diagonal matrix.
The linear radial distortion operator in Fourier space is an integral operator,
a continuous matrix,
\ba
\label{Sk}
\lefteqn{\ \
  \FTbS(\k, \k') =
    \int e^{i \k.\r} \bS(\r) e^{- i \k'.\r} d^3 r
} && \\
  \nonumber
  && =
  (2\pi)^3 \delta_D(\k-\k')
    \left( 1 + \beta {\partial^2 \over \partial k^2}
    k^2 \nabla_\k^{-2} k^{-2} \right)
    - \beta \FTalpha(\k-\k') {\partial \over \partial k'}
    k' \nabla_{\k'}^{-2} k'^{-2}
  \ \ 
\ea
where $\bnabla_\k \equiv \partial/\partial\k$,
and $\FTalpha(\k) \equiv \int \alpha(\r) e^{i \k.\r} d^3 r$ is the Fourier
transform of $\alpha(\r)$.
The distortion operator~(\ref{Sk})
yields the redshifted Fourier modes $\FTdelta^{s}(\k)$
when acting on the unredshifted Fourier modes $\FTdelta(\k)$
\be
  \FTdelta^{s}(\k) = \int \FTbS(\k, \k') \FTdelta(\k') \, d^3 k'/(2\pi)^3
  \ .
\ee
The form of the linear radial distortion operator in Fourier space
is discussed by Zaroubi \& Hoffman (1996).

The radial redshift distortion operator does however
preserve angular symmetry about the observer\footnote{
Correctly, the distortion operator preserves angular symmetry
to the extent that the selection function is independent of direction,
so that the quantity $\alpha(\r)$ in the distortion operator~(\ref{S})
is a function $\alpha(r)$ only of depth $r$,
as is true in a survey that is uniformly flux-limited
within its angular boundaries.
Over unsurveyed regions of the sky (where $\alpha(\r)$ is zero divided by zero),
the distortion operator can be taken to have the same form
as over the surveyed part, as discussed in \S\protect\ref{redshiftdef}.
To avoid confusion,
it should be understood that it is the `true' redshift power spectrum
that possesses orientation symmetry.
When the power spectrum is windowed through a non-all-sky survey,
then the lack of angular symmetry of the window
destroys the angular symmetry of the windowed redshift power spectrum.
},
so that at least some of the merits of the power spectrum
can be retained by working in spherical harmonics.
In particular,
spherical harmonic modes remain statistically orthogonal with respect
to the angular indices in redshift space (\S\ref{power}),
a point emphasized and taken advantage of first by
Fisher, Scharf \&  Lahav (1994),
and then by Heavens \& Taylor (1995).
Mathematically, the distortion operator $\bS$ commutes with the angular momentum
operator $\bL$, so that the eigenmodes of the angular momentum operator,
which are the spherical harmonics $Y_{\el m}$,
are also eigenmodes of the distortion operator.

Hamilton \& Culhane (1996, \S4)
point out that the linear radial distortion operator~(\ref{S})
is nearly scale-free, to the extent that the quantity $\alpha(\r)$
is approximately constant.
If follows that the distortion operator commutes approximately with
the operator $\partial/\partial\ln r$,
the logarithmic derivative with respect to depth $r$,
or equivalently with the operator
$\partial/\partial\ln k = - \partial/\partial\ln r - 3$,
the logarithmic derivative with respect to wavenumber $k$.
A complete set of commuting Hermitian operators for the spherical distortion
operator $\bS$ is then, to the extent that $\alpha(\r)$ is constant,
\be
\label{op}
  - i \left( {\partial \over \partial \ln r} + {3 \over 2} \right) =
    i \left( {\partial \over \partial \ln k} + {3 \over 2} \right)
 \ ,\ \ 
 L^2
 \ ,\ \ 
 L_z
 \ .
\ee
The operators $L^2$ and $L_z$
are the square and $z$-component
(along some arbitrary axis) of the angular momentum operator
$\bL = i \, \r \times \partial/\partial\r =
i \, \k \times \partial/\partial \k$,
which is the same operator in real and Fourier space.
The radial eigenfunctions
are radial waves in logarithmic depth $r$ or wavenumber $k$,
while the angular eigenfunctions are the usual orthonormal
spherical harmonics $Y_{\el m}$.
Thus the eigenfunctions of the commuting set (\ref{op})
are spherical waves with radial parts in logarithmic real or Fourier space.

There are two drawbacks to the basis of logarithmic spherical waves
proposed by Hamilton \& Culhane.
First,
$\alpha(\r)$,
the logarithmic derivative of $r^2 \nbar(\r)$,
equation~(\ref{alpha}),
is not really constant, although it is slowly varying,
and it is typically nearly constant in the nearer part of a redshift survey
where the selection function is found empirically to approximate a power law.
The second problem
is that it is not enough that the modes should be eigenfunctions of the
distortion operator.
It is also desired that the modes be statistically orthogonal, or nearly so.
Now the modes would be statistically orthogonal if the power spectrum
were, like the distortion operator, scale-free,
i.e.\ a power law $P(k) \propto k^n$,
which may not be a bad approximation in reality.
However, in any real survey,
Poisson sampling noise, \S\S\ref{shot} and \ref{redshot},
destroys the scale-free symmetry,
except in the case that the power spectrum itself is that of Poisson noise,
$P(k) =$ constant.
It remains to be seen if this idea can be exploited.

The radial redshift distortion operator~(\ref{S}) is not Hermitian,
unlike the plane-parallel distortion operator~(\ref{Sp}), which is Hermitian.
Amongst other things, this means that the eigenfunctions of the
radial distortion operator do not all have real eigenvalues.
Presumably this means that redshift distortions cause modes close to the
observer to undergo a `phase shift' in passing from real to redshift space.
The Hermitian conjugate $\bS^\dagger$ of the distortion operator is
\be
\label{Sdag}
  \bS^\dagger =
    1 + \beta \nabla^{-2} r^{-2} {\partial \over \partial r}
    \left( {\partial \over \partial r} - {\alpha(\r) \over r} \right) r^2
  \ .
\ee
Equation~(\ref{Sdag}) is valid in the CMB frame.
The Hermitian conjugate of the redshift distortion operator in the Local Group
frame is given by equation~(\ref{SLGdag}).

Tegmark \& Bromley (1995) derive an expression for the inverse $\bS^{-1}$,
the Green's function,
of the distortion operator for the particular case $\alpha = 2$,
corresponding to a volume-limited sample.

\subsection{Peculiar Velocity of the Local Group}
\label{LG}

The redshift distortion operator~(\ref{S}) is valid from the point of
view of observers at rest with respect to the CMB.
The Milky Way however is not stationary, but moving.
The linear part of the motion of the Milky Way is the
peculiar velocity of the {\bf Local Group} (LG) of galaxies.
The essential feature of the Local Group,
which contains about 30 galaxies including our own,
is that it is the local region of space, about $1 \, \Mpc$ in radius,
that has turned around from the general Hubble expansion of the Universe,
and is beginning to fall together for the first time.

In the LG frame, the redshift distance $\s^\LG$
of a galaxy at true position $\r$ relative to the observer is
\be
\label{sLG}
  s^\LG = r + v - \hat\r . \bv^\LG
\ee
which differs from the redshift distance $s = r + v$ in the CMB frame
by the component of $- \bv^\LG$ along the direction $\hat\r$ to the galaxy.
From conservation of galaxies,
$n^\LG(\s^\LG) \, d^3 s^\LG = n(\s) \, d^3 s$,
one concludes that,
at distances much greater than the LG peculiar velocity,
$r \gg v^\LG$,
the redshift space overdensity $\delta^{s\LG}(\r)$ observed in the LG frame
is related to
the redshift space overdensity $\delta^{s}(\r)$ in the CMB frame
by
\be
\label{dsLGa}
  \delta^{s\LG}(\r) = \delta^{s}(\r) + \alpha(\r) {\hat\r . \bv^\LG \over r}
  \ .
\ee
The difference term is a dipole directed along the LG motion $\bv^\LG$,
modulated by the function $\alpha(\r)$, equation~(\ref{alpha}).

Since the LG shares in the linear motion of the nearby region,
there is no dipole distortion of the nearby region in the LG frame.
A dipole distortion appears in the CMB frame because of the streaming
of galaxies past the stationary observer.
Galaxies streaming toward the observer appear compressed to a smaller volume
in redshift space by the spherical convergence of the volume element,
and conversely galaxies downstream appear decompressed.

Two general strategies to deal with the redshift distortion produced
(or rather removed) by the motion of the LG are described below.
A third strategy is to ignore the problem,
which is what the plane-parallel approximation does.

\subsubsection{The Distortion Operator in the Local Group Frame}
\label{operatorLG}

The first strategy,
used by Fisher, Scharf \& Lahav (1994),
is to work entirely in the LG frame,
and to write the LG peculiar velocity as its value in linear theory,
which is given by the usual formula~(\ref{bv}) for peculiar velocity,
evaluated at the position of the LG, the origin, $\r = 0$:
\be
\label{vLG}
  \bv^\LG = - \beta \left. {\partial \over \partial\r} \right|_{\r = 0}
    \!\!\! \nabla^{-2} \delta
  \ .
\ee
An equivalent expression is
\be
\label{vLGp}
  \bv^\LG = \beta \int {\hat\r \delta(\r) \over 4\pi r^2} \, d^3 r
\ee
which expresses the usual fact that peculiar velocity in linear theory
is proportional to peculiar gravity,
which is proportional to the integrated dipole overdensity about
the observer.

The value~(\ref{vLG}), inserted in equation~(\ref{dsLGa}),
combined with the redshift distortion equation $\delta^{s} = \bS \delta$
in the CMB frame,
equation~(\ref{ds}),
yields the distortion equation for the redshift overdensity $\delta^{s\LG}$
in the LG frame
\be
\label{dsLG}
  \delta^{s\LG} = \bS^\LG \, \delta
\ee
where $\bS^\LG$ is the linear redshift distortion operator~(\ref{S})
modified to the LG frame
\be
\label{SLG}
  \bS^\LG = 1 + \beta \left[
    {\partial^2 \over \partial r^2} + {\alpha(\r) \hat\r \over r} \, .
    \left( {\partial \over \partial\r}
    - \left. {\partial \over \partial\r} \right|_{\r = 0} \right)
    \right] \nabla^{-2}
  \ .
\ee
The first $\partial/\partial\r$ in the $\alpha$ term is the gradient
evaluated at the position $\r$,
while $\left. \partial/\partial\r \right|_{\r = 0}$ is the gradient
evaluated at the position of the LG.
The assumption of linear theory here automatically guarantees that
the peculiar velocities of galaxies in the LG frame are small
compared to their distances,
\be
\label{vllrLG}
  | v - \hat\r . \bv^\LG | \ll r
  \ ,
\ee
unlike the corresponding condition~(\ref{vllr}) in the CMB frame,
which had to be incorporated as an additional assumption.

The redshift distortion operator $\bS^\LG$ in the LG frame,
equation~(\ref{SLG}),
can also be derived directly along the same lines as the derivation
in \S\ref{operator}
of the distortion operator $\bS$ in the CMB frame, equation~(\ref{S}).
The main difference is to replace
$v \rightarrow v - \hat\r . \bv^\LG$
throughout the derivation;
for example, equation~(\ref{dsv}) in the CMB frame
changes, in the LG frame, to
\be
\label{dsvLG}
  \delta^{s}(\r) = \delta(\r)
    - \left( {\partial \over \partial r} + {\alpha(\r) \over r} \right)
    (v - \hat\r.\bv^\LG)
\ee
which is Kaiser's (1987) equation~(3.3).

The redshift correlation function $\xi^{s\LG}(r_{12},r_1,r_2)$
in the LG frame is related to the unredshifted correlation function
$\xi(r_{12})$ by
\be
  \xi^{s\LG}(r_{12},r_1,r_2) \: = \:
    \bS^\LG_1 \bS^\LG_2 \, \xi(r_{12})
\ee
which may be compared to the corresponding equation~(\ref{xis})
in the CMB frame.

The Hermitian conjugate of the linear distortion operator $\bS^\LG$
in the LG frame can be written in a variety of ways.
One of the simpler ways, which however does not reveal the subtraction
of the LG velocity as manifestly as does the expression~(\ref{SLG})
for the distortion operator $\bS^\LG$ itself, is
\be
\label{SLGdag}
  \bS^{\LG\dagger} =
    1 + \beta \left[
    \nabla^{-2} r^{-2} {\partial \over \partial r}
    \left( {\partial \over \partial r} - {\alpha(\r) \over r} \right) r^2
    - {\hat\r \over r^2} . \left. {\partial \over \partial\r} \right|_{\r = 0}
    \nabla^{-2} \alpha(\r) r \right]
  \ ,
\ee
in which the last term inside the square brackets on the right hand side
is the term arising from the motion of the LG.
This equation~(\ref{SLGdag})
may be compared to the corresponding equation~(\ref{Sdag}) for the Hermitian
conjugate $\bS^\dagger$ of the distortion operator in the CMB frame.
Equation~(\ref{SLGdag}) is used in equation~(\ref{wc}),
and the equation~(\ref{wcp}) following that one
gives a rearrangement that brings out the subtraction of the LG velocity
more clearly.

\subsubsection{Using the Known Value of the Local Group Motion}
\label{knownvLG}

The second strategy to deal with the motion of the LG
is to use its known value,
measured from the dipole anisotropy of the CMB radiation,
corrected for the motion of the solar system about the centre of the Milky Way,
and for the motion of the Milky Way within the Local Group
(Yahil, Tammann \& Sandage 1977).
The LG peculiar velocity is
(Kogut \etal\ 1993, Table 3; Lineweaver \etal\ 1996)
\be
  \bv^\LG =
    %552.2 \pm 5.5 \, \kms
    627 \pm 22 \, \kms
  \quad \mbox{towards} \quad
    %l = 266.\!\!^\circ 5 \pm 0^\circ.3 , \ 
    %b = 29.\!\!^\circ 1 \pm 0.\!\!^\circ.4
    l = 276^\circ \pm 3^\circ , \ b = 30^\circ \pm 3^\circ
\ee
in the constellation of Hydra.

Thus if the redshift overdensity $\delta^{s\LG}$ is measured in the LG frame,
then it can be corrected to yield the overdensity $\delta^{s}$ in the CMB frame
by subtracting a dipole term in accordance with equation~(\ref{dsLGa}),
\be
\label{dsLGp}
  \delta^{s}(\r) = \delta^{s\LG}(\r) - \alpha(\r) {\hat\r . \bv^\LG \over r}
  \ .
\ee
If $\alpha(r)$ is a function only of depth $r$, independent of direction,
then the adjustment is a pure dipole overdensity,
but in the general case the adjustment involves other harmonics also.
The redshift correlation function of the overdensity in the CMB frame
is given by equation~(\ref{xis}).

An alternative version of the second strategy
is to adjust the redshift distances from the LG frame to the CMB frame
using equation~(\ref{sLG}), $s = s^\LG + \hat\r . \bv^\LG$,
and to measure the overdensity $\delta^{s}$ in the resulting CMB frame.
However, this alternative version is not equivalent to the original
when the condition $| v | \ll r$ is violated.
Closer examination of the derivation of the distortion operator $\bS$,
equation~(\ref{S}), reveals that in effect it {\it assumes\/} that
any streaming motion past the observer induces a dipole
in the redshift overdensity.
It follows that the best procedure is to measure the overdensity in the frame
where the streaming motion vanishes, which is the LG frame, and then
to transform to the CMB frame by making a dipole correction to the overdensity,
as previously recommended, equation~(\ref{dsLGp}).

If the overdensity $\delta^{s}(\r)$ in the CMB frame is constructed
according to the recommended procedure ---
measure the overdensity in the LG frame and then correct
to the CMB frame by subtracting a dipole streaming term,
equation~(\ref{dsLGp}) ---
then the distortion equation~(\ref{ds}) in the CMB frame will remain
correct even without the condition $| v | \ll r$, equation~(\ref{vllr}).

Which strategy is better for dealing with the motion of the LG,
to work entirely in the LG frame as described in \S\ref{operatorLG},
or to make a dipole correction and to work in the CMB frame,
as proposed in this subsection?
It seems to be largely a matter of judgement or convenience.
The difference in information content is not great.
If one likes,
the difference can be isolated into the single mode with the specific
form $\alpha(\r) \hat\r.\bv^\LG/r$, equation~(\ref{dsLGa}),
by making all other modes orthogonal to it
(Tegmark \etal\ 1997).
In the case of an all-sky survey with a uniform flux limit,
the mode is a dipole mode.
For a non-all-sky survey with a uniform flux limit,
the $\alpha(\r) \hat\r.\bv^\LG/r$ mode is effectively a `cut' dipole mode,
and making observed modes orthogonal to this mode means making them orthogonal
to the cut dipole.

Note that the peculiar velocity of the LG in the second strategy
is not used in order to allow a measurement of $\beta$
from any direct comparison between peculiar velocity and density fields;
rather the LG velocity is used to introduce a dipole redshift distortion,
which combines statistically with all the other redshift distortions
to yield a statistical estimate of $\beta$.

\subsubsection{Rocket Effect}
\label{rocket}

The fact that the peculiar motion of the observer induces a dipole overdensity
in redshift space in the direction of motion leads to an effect called the
{\bf rocket effect},
first pointed out by Kaiser (1987, \S2; Kaiser \& Lahav 1988, p.~366).
This dipole needs to be distinguished carefully from the true dipole anisotropy
of the density around the observer,
which in the standard gravitational instability picture
is responsible for inducing the observer's peculiar velocity,
equation~(\ref{vLGp})
(Strauss \etal\ 1992b).

Consider for example an observer situated in a perfectly uniform
distribution of galaxies, all with zero peculiar velocities.
Suppose that the observer hypothesizes that he is actually
moving at some velocity $\bv$ relative to the background.
This motion will induce a spurious dipole in the direction of motion,
which the witless observer may interpret as the dipole pulling him.

A complementary problem arises if an observer attempts to correct the
dipole around him for the reflex motion induced on the observer
by that dipole.
Here there is a tendency to instability at large depths,
beyond the median depth of the survey, where $\alpha(\r)$ goes negative
(Nusser \& Davis 1994).

\subsection{Real and Redshift Space Selection Functions}
\label{selfn}

The distortion equation~(\ref{ds}) with the distortion operator $\bS$
given by~(\ref{S}) provides a valid prediction $\delta^{s}(\r)$
of the observed redshift overdensity $\delta^{s}_\obs(\r)$
provided that the redshift overdensity is defined relative to the real space
selection function $\nbar(\r)$, as in equation~(\ref{deltas}).
In the practical case however,
the selection function is likely to be estimated in redshift space.
This subsection examines how
the real and redshift space selection functions $\nbar(\r)$
and $\nbar^{s}(\r)$ are related
for the typical case of a flux-limited redshift survey.

Suppose that the redshift survey includes all galaxies brighter than some flux
limit $F_{\lim}$, and suppose for simplicity that this limit is uniform
over the angular extent of the survey.
It is convenient to characterize the luminosity of a galaxy of flux $F$
and true distance $r$ by the maximum distance $r_L$ to which the galaxy
could be moved and still remain above the flux limit:
\be
\label{rL}
  r_L \equiv r \left( {F \over F_{\lim}} \right)^{1/2}
  \ .
\ee
The limiting distance $r_L$ of a galaxy is just a measure of its luminosity $L$,
since $L = 4\pi r^2 F = 4\pi r_L^2 F_{\lim}$.
Define $\phi(r_L) d\ln r_L$
to be the luminosity function, the mean number density of galaxies in a
logarithmic interval $d\ln r_L$ of limiting distances
(which is the same, up to a factor,
as the mean number density of galaxies in a logarithmic interval of luminosity,
or in an interval of absolute magnitude).
The selection function $\nbar(r)$ at (true, not redshift) depth $r$
in this flux-limited redshift survey
is the number density of galaxies luminous enough to be seen to depth $r$,
which is the luminosity function integrated over $r_L > r$
\be
\label{nbar}
  \nbar(r) = \int_r^\infty \phi(r_L) \, d\ln r_L
  \ .
\ee

All modern methods of measuring the selection function
(Binggeli, Sand\-age \& Tammann 1988; Koranyi \& Strauss 1997)
seek to eliminate the (otherwise dominant) uncertainty arising from
density inhomogeneity by assuming that the luminosity function $\phi(r_L)$
is a universal function, independent of the local density,
and estimating the luminosity function
by counting relative numbers of bright and faint galaxies in identical volumes.
If bright and faint galaxies at any point share the same peculiar velocity,
as seems probable in the linear regime,
then the measurement of the selection function in redshift space
should remain unbiased by inhomogeneity.
A systematic difference
between the real and redshift space luminosity functions,
hence selection functions,
does however result from the fact that
radial peculiar velocities $v$
shift galaxies not only in space, $s = r + v$,
but also in luminosity,
since the apparent limiting distance $s_L$ of a galaxy in redshift space,
defined by
\be
\label{sL}
  s_L \equiv s \left( {F \over F_{\lim}} \right)^{1/2} =
    {s \hspace{1pt} r_L \over r}
  \ ,
\ee
is $s/r$ times its true limiting distance $r_L$, equation~(\ref{rL}).

Hamilton \& Culhane (1996) claim that the real and redshift selection
functions agree to linear order, but this is false.
What is true is that the expectation values of the real and redshift selection
functions, averaged over an ensemble of observers, agree to linear order,
essentially because the peculiar velocity averaged over an ensemble of
observers is zero everywhere, $\langle \bv(\r) \rangle = 0$
(a difference between ensemble-averaged real and redshift selection functions
arises at next order, since the ensemble average dispersion of peculiar
velocities is non-zero).
In the actual case, however, the peculiar velocity does not vanish,
and the actual real and redshift selections differ to linear order.

\subsubsection{Turner's Method}
\label{turnermethod}

As a typical procedure,
consider Turner's (1979) procedure for estimating the selection function.
It can be shown that Turner's method, when applied over a grid of depths $r$,
yields the same selection function as that of the
``Stepwise Maximum Likelihood'' method of
Efstathiou, Ellis \& Peterson (1988),
although Turner's method does not yield uncertainties.
Turner's procedure is to construct the selection function $\nbar(r)$ from
\be
\label{turner}
  {d \ln \nbar(r) \over d \ln r} =
    \left. {\partial N(r_L,r) \over N(r,r) \, \partial \ln r_L}
    \right|_{r_L = r}
\ee
where $N(r_L,r)$ is the observed number of galaxies brighter than $r_L$
and at the same time closer than $r$.
In practice, equation~(\ref{turner}) is solved on a grid of depths $r$,
in which case the equation becomes, for bins of logarithmic width $\Delta\ln r$,
\be
\label{turnergrid}
  {\Delta\ln \nbar(r) \over \Delta\ln r} =
    {N(r e^{\Delta\ln r}, r) - N(r , r) \over N(r,r)}
  \ .
\ee
As pointed out by Strauss, Yahil \& Davis (1991),
if observed fluxes are rounded into discrete bins
(for example, Zwicky magnitudes are rounded mostly to $0.1$ magnitudes),
then the grid in $r$ must be chosen carefully to avoid a discretization bias.
For a logarithmic spacing $\Delta\ln F$ of discretized fluxes,
the estimate of the selection function is unbiased if
the logarithmic spacing $\Delta\ln r$ of depths is taken to be an
integral multiple of $\Delta\ln(F^{1/2})$;
the estimate is most biased if $\Delta\ln r$ is taken to be half
of $\Delta\ln(F^{1/2})$.
As with all such methods,
Turner's procedure leaves the overall normalization
of the selection function undetermined.
The normalization must be fixed as a separate step.

Suppose that Turner's method~(\ref{turner}) is applied in redshift space,
so that the redshift space selection function $\nbar^{s}(s)$ is estimated from
\be
\label{turners}
  {d \ln \nbar^{s}(s) \over d \ln s} =
    \left. {\partial N^{s}(s_L,s) \over N^{s}(s,s) \, \partial \ln s_L}
    \right|_{s_L = s}
\ee
where $N^{s}(s_L,s)$ is the observed number of galaxies brighter
than $s_L$ and closer than $s$ in redshift space.
How is the redshift space selection function estimated by~(\ref{turners})
related to the real space selection function?
Let
$f(r_L,\r)$
denote the number density of galaxies in an interval $d\ln r_L \, d^3r$
of luminosity and position in real space, and similarly let
$f^{s}(s_L,\s)$
denote the number density of galaxies in an interval $d\ln s_L \, d^3s$
in redshift space.
Conservation of galaxies implies
\be
\label{nsnL}
  f^{s}(s_L,\s) \, d\ln s_L \, d^3 s = f(r_L,\r) \, d\ln r_L \, d^3 r
\ee
with $s = r + v$ and $s_L = r_L s/r$,
which generalizes equation~(\ref{nsn}).
The relation $s = r + v$ is valid in the CMB frame;
in the LG frame,
replace $v \rightarrow v - \hat\r.\bv^\LG$.
The assumption that the luminosity function $\phi(r_L)$ is a universal function,
independent of position, implies that the real space number density $f(r_L,\r)$
factorizes to
\be
\label{fr}
  f(r_L,\r) = \phi(r_L) \bigl[ 1 + \delta(\r) \bigr]
  \ .
\ee
Given this assumption~(\ref{fr}),
a derivation similar to that leading from~(\ref{nsn}) to (\ref{dsv}) shows that,
to linear order in the overdensity $\delta(\r)$ and peculiar velocity $v(\r)$,
equation~(\ref{nsnL}) implies
\be
\label{fs}
  f^{s}(r_L,\r) =
    \phi(r_L) \left[ 1 + \delta(\r)
    - \left( {\partial \over \partial r} + {\alpha_\phi(r_L) \over r} \right)
    v(\r) \right]
\ee
where the dimensionless quantity
$\alpha_\phi(r_L) \equiv d\ln[r_L^2\phi(r_L)]/d\ln r_L$
is similar but not identical to the quantity $\alpha(\r)$ defined by
equation~(\ref{alpha}).
By definition,
the number $N^{s}(s_L,s)$ of galaxies brighter than $s_L$ and closer than $s$ is
\be
\label{Ns}
  N^{s}(s_L,s) \equiv
    \int_0^s \!\!\! \int_{\!A} \! \int_{s_L}^\infty
    f^{s}(s'_L,\s') \, d\ln s'_L \, d^3s'
\ee
where the angular part of the spatial integral is over the solid angle $A$
of the survey.
A few-step calculation using equation~(\ref{fs}) in equation~(\ref{Ns})
leads to
\be
\label{Nss}
  {\partial N^{s}(r_L,r) \over N^{s}(r_L,r) \, \partial \ln r_L}
  =
  {d \ln \nbar(r_L) \over d \ln r_L} -
    {d^2 \ln \nbar(r_L) \over d \ln r_L^{\,\,2}}
    \left\langle {v \over r} \right\rangle_{\!\!r}
\ee
where $\nbar(r_L)$, the real space selection function,
is the integral~(\ref{nbar}) of the luminosity function,
and $\langle {v / r} \rangle_r$ is the volume-average of $v/r$ within depth $r$
over the solid angle $A$ of the survey
\be
\label{vr}
  \left\langle {v \over r} \right\rangle_{\!\!r} \equiv
    {3 \over A r^3} \int_0^r \!\!\! \int_{\!A} {v(\r') \over r'} \, d^3 r'
  \ .
\ee
Equation~(\ref{Nss}) shows that the
basic relation between the real and redshift space selection functions,
when the latter is estimated according to Turner's procedure~(\ref{turners}),
is, to linear order in the overdensity and peculiar velocity,
\be
\label{nbarnbars}
  {d \ln [\nbar(r)/\nbar^{s}(r)] \over d \ln r} =
    {d^2 \ln\nbar(r) \over d\ln r^{\,2}}
    \left\langle {v \over r} \right\rangle_{\!\!r}
  \ .
\ee

Integrating equation~(\ref{nbarnbars}) yields
\be
\label{lnnbarnbars}
  \ln \left( \nbar(r) \over \nbar^{s}(r) \right) =
    - \int_r^{r_{\max}}
    {d^2 \ln\nbar(r') \over d\ln r'^{\,2}}
    \left\langle {v \over r} \right\rangle_{\!\!r'}
    \, {d r' \over r'}
\ee
where the limit $r_{\max}$ of integration
is an arbitrary depth which can conveniently be taken
to be, say, the maximum depth of the survey.
Changing $r_{\max}$ changes 
$\ln [\nbar(r)/\nbar^{s}(r)]$
by an integration constant,
the arbitrariness of which reflects the fact that Turner's method leaves
the overall normalization of the selection function undetermined.
Changing the order of integration
transforms equation~(\ref{lnnbarnbars}) to
\be
\label{lnnbarnbarsv}
  \ln \left( \nbar(r) \over \nbar^{s}(r) \right) =
    - \int_0^{r_{\max}} \!\!\! \int_{\!A}
    a\bigl[\max(r,r')\bigr] \, {v(\r') \over r'} \, d^3 r'
\ee
where the integration is over the volume of the survey,
and the function $a(r)$ is
\be
\label{ar}
  a(r) \equiv
    {3 \over A} \int_r^{r_{\max}} {1 \over r'^3}
    {d^2 \ln\nbar(r') \over d\ln r'^{\,2}} \, {d r' \over r'}
  \ .
\ee

In practice it would normally be better to estimate the redshift selection
function $\nbar^{s\,\LG}(r)$ in the LG frame rather than in the CMB frame,
since the derivation of equation~(\ref{fs}) requires amongst other things
that $|v| \ll r$, which is not necessarily satisfied at small depths $r$,
whereas the corresponding condition $|v - \hat\r.\bv^\LG| \ll r$
in the LG frame is automatically satisfied, at least over linear scales.
Equations in the LG frame are obtained by replacing
$v \rightarrow v - \hat\r.\bv^\LG$ everywhere.
In the LG frame, equations~(\ref{lnnbarnbars}) and (\ref{lnnbarnbarsv}) become
\ba
  \!\!\!\!\!\ln \left( \nbar(r) \over \nbar^{s\,\LG}(r) \right)
  &\!\!\!=\!\!\!&
    - \int_r^{r_{\max}}
    {d^2 \ln\nbar(r') \over d\ln r'^{\,2}}
    \left( \left\langle {v \over r} \right\rangle_{\!\!r'}
    - \left\langle {\hat\r \over r} \right\rangle_{\!\!r'} . \bv^\LG \right)
    \, {d r' \over r'}
  \nn
\label{lnnbarnbarsLG}
  &\!\!\!=\!\!\!&
    - \int_0^{r_{\max}} \!\!\! \int_{\!A}
    a\bigl[\max(r,r')\bigr] \, {[ v(\r') - \hat\r'.\bv^\LG ] \over r'} \, d^3 r'
\ea
where
$\langle \hat\r/r \rangle_r \equiv
3/(A r^3) \int_0^r \!\! \int_A (\hat\r'/r') d^3 r'$
is defined to be the volume average of $\hat\r/r$ within depth $r$
over the solid angle $A$ of the survey,
similarly to the definition~(\ref{vr}) of $\langle v/r \rangle_r$.
Note that $\langle \hat\r/r \rangle_r = 0$ in the case of an all-sky survey.

Just as there were two strategies to deal with the LG motion,
described in \S\S\ref{operatorLG} and \ref{knownvLG},
so here there are two general strategies to deal with the
difference between real and redshift selection functions.
The two strategies are described in the next two subsections,
\S\S\ref{operatornbar} and \ref{fixnbar}.

\subsubsection{The Distortion Operator Relative to the Redshift Selection Function}
\label{operatornbar}

The first strategy is to define the redshift overdensity $\delta^{ss}(\s)$
relative to the selection function $\nbar^{s}(\s)$ in redshift space,
as in equation~(\ref{deltass}),
and to modify the linear redshift distortion operator $\bS$ accordingly.

It is straightforward to ascertain that
the overdensities $\delta^{ss}(\r)$ and $\delta^{s}(\r)$
defined respectively relative to the redshift and real space selection
functions, equations~(\ref{deltass}) and (\ref{deltas}),
are related to linear order by
\be
\label{dssnbar}
  \delta^{ss}(\r) = \delta^{s}(\r) + \ln\bigl[\nbar(\r)/\nbar^{s}(\r)\bigr]
  \ .
\ee
Equation~(\ref{lnnbarnbarsv}),
together with the usual linear expression~(\ref{v}) for the radial peculiar
velocity $v$, shows that,
in the case of Turner's (1979) method applied to a flux-limited redshift survey,
$\ln[\nbar(r)/\nbar^{s}(r)]$ can be written
\be
\label{nbarnbarsS}
  \ln\bigl[\nbar(r)/\nbar^{s}(r)\bigr] =
    \int \bS^{\nbar}(r,\r') \delta(\r') \, d^3 r'
\ee
(more compactly,
the right hand side of eq.~[\ref{nbarnbarsS}] is just $\bS^{\nbar} \delta$
in the notation of matrices and vectors in Hilbert space, \S\ref{hilbert})
where the linear operator $\bS^{\nbar}$ is
\be
\label{Snbar}
  \bS^{\nbar}(r,\r') =
    \beta \, a\bigl[\max(r,r')\bigr] \omega(\r')
    {\partial \over r' \partial r'} \nabla'^{-2}
\ee
with $\omega(\r') \equiv 1$ inside, and $0$ outside, the volume of the survey.

Thus the linear redshift distortion equation for the redshift overdensity
$\delta^{ss}$ is
\be
\label{dss}
  \delta^{ss} = \bS^{s} \delta
\ee
where the linear distortion operator $\bS^{s}$
is the sum of the linear distortion operator $\bS$, equation~(\ref{S}),
and an extra piece $\bS^{\nbar}$
given, for Turner's method, by equation~(\ref{Snbar}):
\be
  \bS^{s} = \bS + \bS^{\nbar}
  \ .
\ee
A complete expression for the distortion operator $\bS^{s}$ is
\be
\label{Ss}
  \bS^{s}(\r,\r') =
    \delta_D(\r-\r') \left( 1
    + \beta \, {\partial^2 \over \partial r^2} \nabla^{-2} \right)
    + \beta \, {\alpha^{s}(\r,\r') \over r'} {\partial \over \partial r'}
    \nabla'^{-2}
\ee
where $\alpha^{s}(\r,\r')$ is
\be
\label{alphas}
  \alpha^{s}(\r,\r') \equiv \delta_D(\r-\r') \alpha(r)
    + a\bigl[\max(r,r')\bigr] \omega(\r')
\ee
with $\alpha(r)$ and $a(r)$ given by~(\ref{alpha}) and (\ref{ar}), and again
$\omega(\r') \equiv 1$ inside, and $0$ outside, the volume of the survey.

In practice,
it makes more sense to apply this strategy in the LG frame
than in the CMB frame, for the reason given in the paragraph
containing equation~(\ref{lnnbarnbarsLG}).
In the LG frame,
equations~(\ref{dssnbar})--(\ref{Ss}) carry through,
but superscripted $\LG$ appropriately.
In the LG frame,
the overdensities $\delta^{ss\,\LG}$ and $\delta^{s\,\LG}$ are related by
\be
\label{dssnbarLG}
  \delta^{ss\,\LG}(\r) =
    \delta^{s\,\LG}(\r) + \ln\bigl[\nbar(\r)/\nbar^{s\,\LG}(\r)\bigr]
  \ .
\ee
In the case of Turner's method applied to a flux-limited survey,
equation~(\ref{lnnbarnbarsLG}) shows that
$\ln[\nbar(r)/\nbar^{s\,\LG}(r)] = 
\int \bS^{\nbar\,\LG}(r,\r') \delta(\r') \, d^3 r'$
where the operator $\bS^{\nbar\,\LG}$ is
\be
\label{SnbarLG}
  \bS^{\nbar\,\LG}(r,\r') =
    \beta \, {a[\max(r,r')] \omega(\r') \hat\r' \over r'} \, .
    \left( {\partial \over \partial\r'}
    - \left. {\partial \over \partial\r'} \right|_{\r'=0}
    \right) \nabla'^{-2}
  \ .
\ee
The distortion equation for the redshift overdensity $\delta^{ss\,\LG}$
in the LG frame is
\be
\label{dssLG}
  \delta^{ss\,\LG} = \bS^{s\,\LG} \delta
\ee
where the linear distortion operator $\bS^{s\,\LG}$
is the sum of the distortion operator $\bS^\LG$, equation~(\ref{SLG}),
and the extra piece $\bS^{\nbar\,\LG}$ given, for Turner's method,
by equation~(\ref{SnbarLG}):
\be
  \bS^{s\,\LG} = \bS^{\LG} + \bS^{\nbar\,\LG}
  \ .
\ee
A complete expression for the distortion operator $\bS^{s\,\LG}$ is
\ba
  \bS^{s\,\LG}(\r,\r') &=&
    \delta_D(\r-\r') \left( 1
    + \beta \, {\partial^2 \over \partial r^2} \nabla^{-2} \right)
  \nn
\label{SsLG}
  & & \mbox{}
    + \beta \, {\alpha^{s}(\r,\r') \hat\r' \over r'} \, .
    \left( {\partial \over \partial\r'}
    - \left. {\partial \over \partial\r'} \right|_{\r'=0}
    \right) \nabla'^{-2}
\ea
with $\alpha^{s}(\r,\r')$ given by equation~(\ref{alphas}).
In evaluating $\alpha^{s}(\r,\r')$,
the selection function $\nbar(r)$ which enters the definitions~(\ref{alpha})
of $\alpha(r)$ and (\ref{ar}) of $a(r)$ can be replaced
to linear order by the measured redshift selection function $\nbar^{s\,\LG}(r)$.

Equation~(\ref{SsLG}) is an important equation.
To recapitulate,
it gives the form of the linear redshift distortion operator $\bS^{s\,\LG}$
for the case of a flux-limited redshift survey
in which the redshift space overdensity $\delta^{ss\,\LG}(\r)$ is defined
(a) in the frame of reference of the LG,
and (b) relative to the redshift space selection function $\nbar^{s\,\LG}(r)$
measured using Turner's (1979) method, equation~(\ref{turners}),
again in the LG frame.
That is,
the distortion equation~(\ref{dssLG}) gives the theoretically predicted form of
the overdensity $\delta^{ss\,\LG}(\r)$ in the linear regime,
to be compared to the observed redshift overdensity $\delta^{ss\,\LG}_\obs(\r)$
defined by
\be
  \delta^{ss\,\LG}_\obs(\r) \equiv
    {n^{s\,\LG}(\r) - \nbar^{s\,\LG}(r) \over \nbar^{s\,\LG}(r)}
\ee
with $n^{s\,\LG}(\r)$ the observed galaxy density at redshift position $\r$
in the LG frame,
and $\nbar^{s\,\LG}(r)$ the selection function measured in redshift space
in the LG frame by Turner's method~(\ref{turners}).

The redshift correlation function $\xi^{ss}(r_{12},r_1,r_2)$
of the overdensity $\delta^{ss}(\r)$
in the linear regime is related to the unredshifted correlation function
$\xi(r_{12})$ by
\be
\label{xiss}
  \xi^{ss}(r_{12},r_1,r_2) = \bS^{s}_1 \bS^{s}_2 \, \xi(r_{12})
  \ .
\ee
Similarly the redshift correlation function $\xi^{ss\,\LG}(r_{12},r_1,r_2)$
of the overdensity $\delta^{ss\,\LG}(\r)$ in the LG frame is
\be
\label{xissLG}
  \xi^{ss\,\LG}(r_{12},r_1,r_2) = \bS^{s\,\LG}_1 \bS^{s\,\LG}_2 \, \xi(r_{12})
  \ .
\ee

\subsubsection{Correcting the Redshift to the Real Space Selection Function}
\label{fixnbar}

The second strategy
is to measure the redshift space selection function $\nbar^{s}(\r)$ as before,
but then to correct it to the real space selection function $\nbar(\r)$.

The correction is given to linear order by equation~(\ref{lnnbarnbars}),
for the case of a flux-limited redshift survey in which the redshift space
selection function is estimated by Turner's (1979) method,
equation~(\ref{turners}).
Actually it would normally be preferable to estimate the redshift selection
function $\nbar^{s\,\LG}(r)$ in the LG frame, and to correct this
to the real selection function using equation~(\ref{lnnbarnbarsLG}).
More precisely, the correction is
\be
\label{lnnbarnbarsLGLG}
  \ln \left( \nbar(r) \over \nbar^{s\,\LG}(r) \right) =
    - \int_r^{r_{\max}}
    {d^2 \ln\nbar^{s\,\LG}(r') \over d\ln r'^{\,2}}
    \left( \left\langle {v \over r} \right\rangle_{\!\!r'}
    - \left\langle {\hat\r \over r} \right\rangle_{\!\!r'} . \bv^\LG \right)
    \, {d r' \over r'}
\ee
which is the same as equation~(\ref{lnnbarnbarsLG}) except that
the redshift selection function $\nbar^{s\,\LG}(r')$ (the thing you have)
replaces the real selection function $\nbar(r')$ (the thing you want)
in the integrand on the right hand side of equation~(\ref{lnnbarnbarsLGLG}).
The replacement is valid to linear order because
$\nbar^{s\,\LG}(r)$ and $\nbar(r)$ agree to zeroth order.

The correction~(\ref{lnnbarnbarsLGLG}) involves modelling the volume-average
$\langle v/r \rangle_r \hspace{1pt}$, equation~(\ref{vr}),
of the radial peculiar velocity $v$ divided by depth $r$,
as a function of depth $r$ in the survey.
Note that $\langle v/r \rangle_r$ is an average of the actual $v/r$,
not an ensemble average.
One way to do this is to carry out a complete reconstruction of the
unredshifted density and velocity field
(Yahil \etal\ 1991;
Fisher \etal\ 1995b;
Webster, Lahav \& Fisher 1997).
Complete reconstruction seems extravagant merely for the purpose of
estimating $\langle v/r \rangle_r \hspace{1pt}$;
perhaps a simpler approximation could be devised.

It is to be noted that in an all-sky survey
$\langle v/r \rangle_r$
is negative if galaxies are on average converging towards the observer,
which occurs if the observer is in an overdense region of the Universe.
The LG lies in such an overdense region,
as is typical for galaxy-bound observers,
which is an example of local bias.

In a non-all-sky survey,
$\langle v/r \rangle_r$
depends in addition on the alignment of the survey with large scale flows.

However carefully $\langle v/r \rangle_r$ is modelled,
there is liable to be some uncertainty in it,
hence some uncertainty in the correction
$\ln[\nbar(r)/\nbar^{s\,\LG}(r)]$, equation~(\ref{lnnbarnbarsLGLG}).
The correction
$\ln[\nbar(r)/\nbar^{s\,\LG}(r)]$
to the selection function is essentially a correction to the redshift
overdensity, as in equation~(\ref{dssnbarLG}).
Measurements of the redshift space overdensity can be immunized against
uncertainty in the radial correction
by discarding from the statistical analysis the uncertain radial mode or modes
--- possibly all radial modes, to be conservative ---
and by making all other observed modes orthogonal to these radial modes
(Tegmark \etal\ 1997).
In the case of an all-sky survey with a uniform flux limit,
radial modes are monopole modes about the observer.
For a non-all-sky survey with a uniform flux limit,
radial modes are effectively `cut' monopole modes,
and observed modes can be immunized against radial uncertainty
by making them orthogonal to the cut monopole modes.

\section{Methods to Measure $\beta$}
\label{methods}
\setcounter{equation}{0}

To date, essentially three types of method have been used to measure
$\beta$ from linear redshift distortions:
\begin{enumerate}
\item
The ratio of angle-averaged redshift space to real space
power spectrum or correlation function;
\item
The ratio of quadrupole to monopole moments of the redshift space
power spectrum;
\item
A maximum likelihood approach
in which the data are the amplitudes of (many hundreds of) individual modes,
and the parameters are $\beta$ and the
power spectrum $P(k)$, parametrized in some way.
\end{enumerate}
These methods are discussed in turn below.

\subsection{Ratio of redshift space to real space angle-averaged power spectra}
\label{redtoreal}

This method was proposed in Kaiser's (1987) original paper.
It assumes the plane-parallel, or distant observer, approximation,
where structures are sufficiently far away that line-of-sight
peculiar velocities are effectively plane-parallel,
as illustrated in the left panel of Figure~\ref{fogs}.

Kaiser's formula~(\ref{Kaiser})
predicts that in the linear regime the angle-averaged
redshift power spectrum $P^{s}(k) \equiv \int P^{s}(\k) do_\k/(4\pi)$
[which is the same thing as the monopole redshift power $P^{s}_0(k)$
discussed in \S\ref{quadrupole}]
is amplified by a constant factor over the unredshifted power spectrum $P(k)$:
\be
\label{PsP}
  {P^{s}(k) \over P(k)} = 1 + \frac{2}{3} \beta + \frac{1}{5} \beta^2
  \ .
\ee
As $\beta$ varies between $0$ and $1$,
the amplification factor
varies between 1 and $28/15$, almost a factor of 2.

Equation~(\ref{PsP}) is valid for any wavenumber $k$ in the linear regime,
so remains valid if the real and redshift space power spectra
are first smoothed over any window $W(k)$ that picks out linear wavenumbers.
That is, if
\be
\label{barP0}
  \smoothP(k) \equiv \int_0^\infty W(k) P(k) \, 4\pi k^2 dk/(2\pi)^3
\ee
and similarly for $\smoothP^{s}(k)$,
then the ratio of the smoothed redshift to real space power spectra is
predicted to be
\be
\label{barPsP}
  {\smoothP^{s} \over \smoothP}
  = 1 + \frac{2}{3} \beta + \frac{1}{5} \beta^2
  \ .
\ee
In particular,
the ratio of the angle-averaged redshift space correlation function
$\xi^{s}(r) \equiv \int \xi^{s}(\r) do_\r/(4\pi)$
[which is the monopole redshift correlation function $\xi^{s}_0(r)$]
to its unredshifted counterpart $\xi(r)$ is,
at linear separations $r$,
\be
  {\xi^{s}(r) \over \xi(r)} = 1 + \frac{2}{3} \beta + \frac{1}{5} \beta^2
  \ ,
\ee
which follows from equation~(\ref{barPsP})
with the smoothing window in~(\ref{barP0})
taken to be the zeroth spherical Bessel function,
$W(k) = \int e^{-i\k.\r} do_\k/(4\pi) = j_0(kr)$.

The unredshifted correlation function $\xi(r)$ and/or
the unredshifted power spectrum $P(k)$
can be measured by deprojecting the angular correlation function of a survey.
This deprojection is liable to be quite noisy, unless the survey is large.
Thus the method of inferring $\beta$ from the ratio of redshift to
real correlation functions or power spectra
is most appropriate if the redshift survey
happens to be a subset of a larger angular survey.
Examples of this are
the Stromlo-APM survey
(Loveday \etal\ 1996b),
which is a 1-in-20 redshift survey
of galaxies brighter than $b_J = 17.15$ in the APM angular survey,
or the QDOT survey
(e.g.\ Saunders \etal\ 1990;
Rowan-Robinson \etal\ 1991;
Saunders, Rowan-Robinson \& Lawrence 1992;
Lawrence \etal\ 1997, in preparation)
which is a 1-in-6 redshift survey of galaxies
brighter than 0.6~Jy from the {\it IRAS\/} Point Source Catalog.

\subsection{Ratio of quadrupole to monopole harmonics of the power spectrum}
\label{quadrupole}

The idea of decomposing the redshift correlation function into harmonics
was proposed by Hamilton (1992),
and the connection to the power spectrum was clarified by
Cole, Fisher \& Weinberg (1994).
Integral expressions for the shape of the redshift correlation function
had previously been given by Kaiser (1987), 
Lilje \& Efstathiou (1989),
and McGill (1990).
This method, like method 1, assumes the plane-parallel approximation.

For plane-parallel redshift distortions,
the redshift power spectrum can be written as a sum of even harmonics
$P^{s}_\el(k)$
\be
\label{Psk}
  P^{s}(\k) = \sum_{\el ~ {\rm even}} {\cal P}_\el(\mu_\k) P^{s}_\el(k)
  \ , \quad
  P^{s}_\el(k) \equiv
    (2\el+1) \int {\cal P_\el(\mu_\k)} P^{s}(\k) \, d o_\k / 4\pi
\ee
where ${\cal P}_\el(\mu_\k)$ are Legendre polynomials,
and $\mu_\k \equiv \hat\z . \hat\k$ is the cosine of the angle between
the wavevector $\k$ and the line of sight $\z$.
The odd harmonics vanish by pair exchange symmetry,
and non-zero azimuthal harmonics
(corresponding to $Y_{\el m}$'s with $m \neq 0$)
vanish by symmetry about the line of sight.

In the linear regime,
Kaiser's formula~(\ref{Kaiser})
shows that the redshift power spectrum reduces to
a sum of monopole ($\el = 0$), quadrupole ($\el = 2$),
and hexadecapole ($\el = 4$) harmonics
\be
  P^{s}(\k) =
    {\cal P}_0(\mu_\k) P^{s}_0(k)
    + {\cal P}_2(\mu_\k) P^{s}_2(k)
    + {\cal P}_4(\mu_\k) P^{s}_4(k)
\ee
where the harmonics $P^{s}_\el$ of the redshift space power spectrum
are related to the unredshifted power spectrum $P(k)$ by
\ba
\label{Psl}
  P^{s}_0(k) &=& \left( 1 + \frac{2}{3} \beta + \frac{1}{5} \beta^2 \right) P(k)
  \nn
  P^{s}_2(k) &=& \left( \frac{4}{3} \beta + \frac{4}{7} \beta^2 \right) P(k)
  \nn
  P^{s}_4(k) &=& \frac{8}{35} \beta^2 P(k)
  \ .
\ea
\begin{figure}[htb]
\begin{center}
\leavevmode
\epsfxsize=2.8in \epsfbox{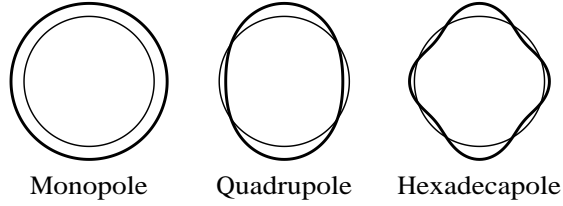}
\end{center}
  \caption[1]{
Shapes of the monopole ($\el = 0$), quadrupole ($\el = 2$),
and hexadecapole ($\el = 4$) harmonics (thick lines),
drawn as perturbations of a sphere (thin lines).
The quadrupole has positive amplitude as drawn,
as appropriate for the redshift power spectrum.
The factor of $i^\el$ between the power spectrum and the correlation function,
equation~(\protect\ref{Pxi}),
introduces a minus sign in the quadrupole correlation function,
which therefore appears squashed, opposite to what is drawn.
\label{lpoles}
}
\end{figure}

Equations~(\ref{Psl}) imply in particular that the ratio $P^{s}_2(k)/P^{s}_0(k)$
of quad\-rupole to monopole harmonics of the redshift power spectrum
in the linear regime is
\be
\label{P2P0}
  {P^{s}_2(k) \over P^{s}_0(k)} =
    {\frac{4}{3} \beta + \frac{4}{7} \beta^2
    \over 1 + \frac{2}{3} \beta + \frac{1}{5} \beta^2}
\ee
which therefore provides a means of measuring $\beta$.
Equations~(\ref{Psl}) predict that
the hexadecapole harmonic $P^{s}_4(k)$ should be quite small,
$\la 0.1$ of the monopole and quadrupole harmonics for $\beta \le 1$,
and it also proves to be noisier, especially at the largest scales,
so in practice the quadrupole-to-monopole ratio has been the statistic
of choice for many authors.

The harmonics $\xi^{s}_\el(r)$ of the redshift correlation function
are defined similarly to those~(\ref{Psk}) of the power spectrum
\be
\label{xisl}
  \xi^{s}(\r) = \sum_{\el ~ {\rm even}} {\cal P}_\el(\mu_\r) \xi^{s}_\el(r)
  \ , \quad
  \xi^{s}_\el(r) \equiv
    (2\el+1) \int {\cal P_\el(\mu_\r)} \xi^{s}(\r) \, d o_\r/(4\pi)
\ee
where $\mu_\r \equiv \hat\z . \hat\r$ is the cosine of the angle
between the separation vector $\r$ and the line of sight $\z$.
The harmonics $P^{s}_\el(k)$ of the redshift power spectrum are related
to the harmonics $\xi^{s}_\el(r)$ of the redshift correlation function by
\be
\label{Pxi}
  P^{s}_\el(k) =
    i^\el 4\pi \!\int_0^\infty\!\!\! j_\el(kr) \xi^{s}_\el(r) \, r^2 d r
  \ , \ \ 
  \xi^{s}_\el(r) =
    i^{-\el} 4\pi \!\int_0^\infty\!\!\! j_\el(kr) P^{s}_\el(k)
    \, k^2 d k/(2\pi)^3
\ee
where $j_\el(kr)$ are spherical Bessel functions.
For the quadrupole harmonic, $\el = 2$,
the $i^\el$ factor in equation~(\ref{Pxi}) introduces a negative sign
between the power spectrum and the correlation function.
Thus the large scale squashing effect of the redshift correlation function,
a negative quadrupole,
corresponds to a stretching of the redshift power spectrum
along the line of sight, a positive quadrupole
(Cole, Fisher \& Weinberg 1994, Fig.~1).

\subsubsection{Smoothed Harmonics of the Redshift Power Spectrum}

In any finite survey,
it is necessary to measure the power spectrum at finite resolution,
given roughly by the inverse scale size of the survey.
Consider then the harmonics of the redshift power spectrum smoothed over
some window $W(k)$
(the following equation generalizes eq.~[\ref{barP0}])
\be
\label{barP}
  \smoothP^{s}_\el \equiv
    \int_0^\infty W(k) P^{s}_\el(k) \, 4\pi k^2 dk/(2\pi)^3
  \ .
\ee
As long as the window $W(k)$ picks out wavenumbers in the linear regime,
the quadrupole-to-monopole ratio of the smoothed harmonics $\smoothP_\el$
will also satisfy the relation~(\ref{P2P0}),
\be
\label{barP2P0}
  {\smoothP^{s}_2 \over \smoothP^{s}_0} =
    {\frac{4}{3} \beta + \frac{4}{7} \beta^2
    \over 1 + \frac{2}{3} \beta + \frac{1}{5} \beta^2}
  \ .
\ee
The relation~(\ref{Pxi}) between the harmonics of the power spectrum
and the correlation function means that the smoothed harmonics
$\smoothP^s_\el$ of the power spectrum, equation~(\ref{barP}),
can also be written as smoothed harmonics of the correlation function:
\be
\label{barPl}
  \smoothP^{s}_\el = \int_0^\infty W_\el(r) \xi^{s}_\el(r) \, 4\pi r^2 dr
\ee
where the real space smoothing window $W_\el(r)$
is related to the Fourier window $W(k)$ by
\be
\label{Wl}
  W_\el(r) = i^\el 4\pi \int_0^\infty \! j_\el(kr) W(k) \, k^2 dk/(2\pi)^3
  \ .
\ee
The point of equation~(\ref{barPl})
is that it shows how smoothed redshift power spectra $\smoothP^{s}_\el$
can be measured from quantities in real (redshift) space,
which is where the data lie.

\subsubsection{Smoothing Windows}

Childers (1978, p.~1) terms the business of choosing smoothing windows $W(k)$
``window carpentry''.
One possibility,
proposed by Hamilton (1995) and used in \S\ref{example} below,
is to use smoothing windows that are power laws times a Gaussian,
$W(k) \sim k^n e^{-k^2}$
with $n$ an even integer,
suitably scaled and normalized.
These windows $W(k)$ have the desirable properties that:
\begin{itemize}
\item[$\circ$]
they are everywhere positive,
so preserving the intrinsic positivity of the power spectrum;
\item[$\circ$]
they vanish at zero wavenumber, $W(k) = 0$ at $k = 0$,
provided that $n \ge 2$, so immunizing the measurement of power against
uncertainty in the mean density
(which makes a delta-function contribution to power at zero wavenumber);
\item[$\circ$]
they are analytically convenient;
\item[$\circ$]
they yield Gaussian convergence as a function of pair separation $r$
in the corresponding real space windows $W_\el(r)$, equation~(\ref{Wl}),
for harmonics $\el \le n$, provided that $n$ is chosen to be an even integer.
\end{itemize}
Suitably scaled, and normalized so $\int W(k) d^3 k/(2\pi)^3 = 1$,
these smoothing windows are
\be
\label{W}
  W(k) {d^3 k \over (2\pi)^3} \equiv
    {2 e^{-q^2} q^{n+2} \, dq \over \Gamma[(n\!+\!3)/2]}
  \ , \quad
  q \equiv {\alpha k \over \smoothk}
  \ .
\ee
The wavenumber $\smoothk$ is an `effective' wavenumber,
the constant $\alpha$ being chosen so that $W(k)$ peaks at approximately
$k \approx \smoothk$.
In Hamilton (1995) and \S\ref{example},
the constant $\alpha$ is chosen so that the smoothed monopole power
at wavenumber $\smoothk$
is equal to the unsmoothed monopole power at the same wavenumber,
$\smoothP^{s}_0(\smoothk) = P^{s}_0(\smoothk)$,
for the particular case where the power spectrum is a power law of index $-1.5$,
that is, for $P^{s}_0(k) \propto k^{-1.5}$.
For the case $n = 2$ in the window~(\ref{W}),
as used in Figure~\ref{xipkoms},
this fixes $\alpha = 1.279$.
The harmonics of the power spectrum smoothed over the window~(\ref{W}) are,
according to equation~(\ref{barPl}),
equal to the harmonics of the correlation function smoothed over
corresponding windows $W_\el(r)$ given by equation~(\ref{Wl}):
\be
\label{Wlag}
  W_\el ( r ) =
  {i^\el [(n\!-\!\el)/2]! \over (3/2)_{(n/2)}} \:
  s^\el e^{-s^2} L_{(n-\el)/2}^{\el+(1/2)} (s^2)
  \ , \quad
  s \equiv {\smoothk r \over 2 \alpha}
\ee
(note that $W_0(0) = 1$)
where $L_{\nu}^{\lambda}$ are Laguerre polynomials
(Abramowitz \& Stegun 1964)
and
\raisebox{0ex}[2ex][0ex]{$(3/2)_{(n/2)} = \Gamma [(n\!+\!3)/2] / \Gamma (3/2)$}
is a Pochhammer symbol.
The smoothing windows~(\ref{Wlag}) happen to be
\raisebox{0ex}[2ex][0ex]{$e^{-s^2/2}$}
times the radial wavefunctions of a 3-dim\-en\-sional simple harmonic
oscillator at energy level $n$ and angular momentum $\el$
(Landau \& Lifshitz 1958, \S33 Problem 4).
The quantization condition that the index $(n\!-\!\el)/2$ be a non-negative
integer ensures Gaussian convergence at large separations $r$.
Despite the forbidding appearance, the smoothing windows $W_\el(r)$
are just polynomials times a Gaussian,
with a rapid stable recurrence relation available to evaluate the polynomials.

It is noteworthy that in order to measure the $\el$'th harmonic with one
of the windows~(\ref{W}), it is necessary to choose $n \ge \el$,
so that the window must go to zero at least as fast as $W(k) \sim k^\el$
as $k \rightarrow 0$.

\subsection{Maximum Likelihood}
\label{ML}

Maximum likelihood (ML) techniques offer an optimal
way to estimate $\beta$ and its uncertainty
from redshift distortions.
The technique is not only optimal,
yielding both the most probable answer
along with a complete probability distribution of answers,
for a given prior,
but is flexible enough to admit complications,
such as the radial (as opposed to plane-parallel)
character of redshift distortions,
which can be difficult to deal with by direct methods.

The ML technique was first applied to measure $\beta$ from linear redshift
distortions by Fisher, Scharf \& Lahav (1994),
and a more thorough implementation was carried through by
Heavens \& Taylor (1995) and then by Ballinger, Heavens \& Taylor (1995).

Likelihood techniques, applied according to Bayesian precepts
(Loredo 1990),
require a prior,
that is, a precise statement of prior assumptions,
including the range of models to be considered,
and the prior probability distributions of the parameters to be measured.
In the usual case,
the prior probability of the parameters is taken to be uniform,
in which case the posterior probability,
the probability distribution of the parameters given the observed data
(which is what you want),
is proportional to the likelihood function,
which is the probability of the observed data given the parameters
(which is what theory tells you).

When estimating $\beta$ from linear redshift distortions,
the prior is liable to include such assumptions as
that the Universe is statistically homogeneous and isotropic,
that the distribution of overdensity on linear scales
is a multivariate Gaussian,
that the power spectrum takes such and such a parametrized form
(e.g.\ Cold Dark Matter (CDM) or one of its cousins),
that observations introduce such and such additional sources of uncertainty
(notably Poisson sampling noise),
that linear redshift distortions conform to the standard model described in
\S\ref{theory},
and that all values of $\beta$ are a priori equally likely
(which is called a uniform prior on the parameter $\beta$).

At first sight the need to state a prior
seems a disadvantage compared to traditional direct approaches
(where for example one measures $\beta$ from some ratio of power spectra,
and one measures power spectra by, well, by measuring power spectra).
With second sight however,
one realizes the advantages of the ML procedure:
\begin{itemize}
\item[$\bullet$]
Being forced to state one's prior assumptions explicitly is a virtue,
not a drawback;
\item[$\bullet$]
If changing your prior (within plausible limits)
makes much difference to the values of the
quantities you are trying to estimate,
then you are not learning much from the data;
\item[$\bullet$]
The mathematical formalism is quite general,
specifying the best way to proceed
even in highly complicated situations.
\end{itemize}

This having been said, it is true that the ML formalism will happily provide
answers for parameters even if the prior is completely wrong.
Thus the prior should, correctly,
always be tested for consistency with the data
(a step that to date has generally not been taken).

If the real overdensity $\bdelta$ forms a multivariate Gaussian field,
as may well be true on linear scales in reality,
then the observed linear redshift overdensity $\bdelta^{s}_\obs$
is also predicted to be a multivariate Gaussian
(any linear combination of a Gaussian field is a Gaussian field),
and the likelihood function ${\cal L}$ is Gaussian
\be
\label{L}
  {\cal L} \propto
    {1 \over |\bC^{s}|^{1/2}}
    \exp \left( - \frac{1}{2}
    \bdelta^{s \dagger}_\obs \bC^{s -1} \bdelta^{s}_\obs
    \right)
\ee
where
$\bC^{s} \equiv \langle \bdelta^{s} \hspace{1pt} \bdelta^{s \dagger} \rangle$
is the expectation value of the survey covariance matrix,
and $|\bC^{s}|$ and $\bC^{s -1}$ are its determinant and inverse.
The $\bdelta^{s}_\obs$ in the likelihood~(\ref{L})
is the observed vector of overdensities,
while the covariance matrix $\bC^{s}$ is the prior.
The likelihood function~(\ref{L}) gives the probability of the observed data,
$\bdelta^{s}_\obs$, given the prior, $\bC^{s}$.
With a uniform prior on the parameters of the covariance matrix $\bC^{s}$,
the likelihood function becomes (up to a normalization factor)
the probability of the parameters given the data.

In general, the prior covariance $\bC^{s}$ is a combination of
signal plus noise terms.
In the simplest case,
one supposes that the only source of noise in a redshift survey of galaxies
is Poisson sampling noise (\S\S\ref{shot}, \ref{redshot}).
The survey covariance $\bC^{s}$ is then a sum of cosmic and Poisson sampling
terms (the following equation is a repeat of eq.~[\ref{Cs}])
\be
\label{Csij}
  C^{s}_{ij} =
    \xi^{s}(\r_i,\r_j) + \delta_D(\r_i-\r_j) [\nbar(\r_i)]^{-1}
\ee
where $\xi^{s}(\r_i,\r_j)$ is the redshift space correlation function,
and $\delta_D(\r_i-\r_j)$ is a Dirac delta function.
For fluctuations in the linear regime,
the redshift correlation function
$\xi^{s}(\r_i,\r_j)$
is predicted to
depend on the linear distortion parameter $\beta$ and the unredshifted
correlation function $\xi(r)$
(or equivalently the unredshifted power spectrum $P(k)$, eq.~[\ref{P}])
in accordance with equation~(\ref{xis}).
The parameters to be measured are then the distortion parameter $\beta$,
and the parameters of some parametrization of the unredshifted power spectrum
$P(k)$.

Actually, the `observed' redshift overdensity $\delta^{s}_\obs(\r)$,
equation~(\ref{deltas}),
depends not only on the observed galaxy density $n^{s}(\r)$ in redshift space,
but also on the selection function $\nbar(\r)$,
which is itself estimated from the survey.
The shape of the selection function
(or the parameters of some parametrization thereof)
would normally be measured as a separate operation,
since that measurement typically involves additional assumptions
about the universality of the luminosity function (see \S\ref{selfn}).
This leaves the overall normalization of the selection function
as an extra parameter to be measured from the likelihood function~(\ref{L}).
In principle, uncertainty in the measurement of the shape of the selection
function should be allowed for by marginalizing over it, that is,
by integrating the likelihood function over the probability distribution
of the parameters of the shape of the selection function.
In practice, the shape of the selection function is normally so accurately
measured, compared to other uncertainties, that it can be treated as a
fixed, known quantity in the likelihood analysis.

In order to proceed numerically,
it is necessary to pixelize the data in some fashion,
so that the covariance matrix $C_{ij}$ becomes a finite matrix.
In principle, the choice of pixelization is irrelevant,
if the pixels are fine and many enough
to represent the survey data accurately.
In practice, the matrix $C_{ij}$ needs to be of tractable size,
and some cunning is needed to pick a pixelization that is small
but nevertheless contains virtually all the information about
the parameters to be measured
(Tegmark, Taylor \& Heavens 1996;
Tegmark \etal\ 1997).
Fisher, Scharf \& Lahav (1994) chose the pixels to be spherical harmonics
to $\el \le 10$ on the sky, with 4 Gaussian pixels in the radial direction.
Heavens \& Taylor (1995) and Ballinger, Heavens \& Taylor (1995)
chose a basis of spherical waves,
that is, spherical harmonics on the sky,
with (weighted) spherical Bessel functions in the radial direction.

Clearly the ML technique, or methods essentially equivalent to it,
is the method of choice for measuring redshift distortions,
certainly on the largest scales where a small number of pixels
contains all relevant information,
perhaps also on smaller scales if clever compression techniques can be
applied
(Vogeley \& Szalay 1996;
Tegmark, Taylor \& Heavens 1996;
Tegmark \etal\ 1997).
Undoubtedly there will be further development of this powerful approach
in the future.

One mysterious aspect of ML results published to date
is that they yield estimates of $\beta$
that appear to be systematically larger than other methods
(see Table~\ref{results}).
One possible reason for the difference is the fact that
the plane-parallel approximation underestimates $\beta$.
However, authors who use the plane-parallel approximation have
generally confined their measurements to opening angles no larger
than $50^\circ$, at which point the $N$-body simulations of
Cole, Fisher \& Weinberg (1994, Fig.~8) indicate that the plane-parallel
approximation underestimates $\beta$ by only 5\%.
Another possibility, suggested by
Ballinger, Heavens \& Taylor (1995),
is that direct methods unfairly penalize larger values of $\beta$,
since larger values of $\beta$ predict a larger variance in the power spectrum,
which is not taken into consideration by direct methods.

The resolution of this mystery remains unclear.

\section{Example of Measuring $\beta$ from Linear Redshift Distortions}
\label{example}
\setcounter{equation}{0}

It is helpful to present an example of the
measurement of $\beta$ from linear redshift distortions,
which is intended not only to illustrate how things work out in practice,
but also
(a) to bring out the difference between {\it IRAS\/} and optically
selected galaxies when they are analysed with (essentially) the same procedure,
and (b) to demonstrate the importance of nonlinearities (fingers-of-god).

The results are given first, in \S\ref{exresults},
and then \S\ref{exanalysis} provides some details of the analysis.

\subsection{Results}
\label{exresults}

\begin{figure}
\epsfxsize=5in \epsfbox{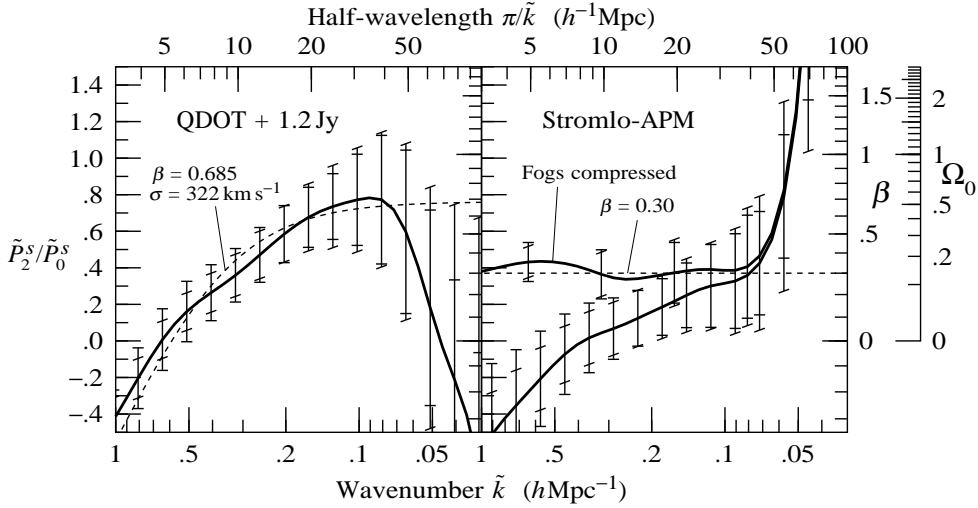}
  \caption[1]{
Ratio $\smoothP^{s}_2(\smoothk)/\smoothP^{s}_0(\smoothk)$
of smoothed quadrupole to smoothed monopole power
as a function of effective wavenumber $\smoothk$ in:
(left) the merged {\it IRAS\/} QDOT plus 1.2~Jy survey;
and (right) the Stromlo-APM survey.
These are the same surveys whose redshift correlation functions are illustrated
in Figure~\protect\ref{xilconts}.
For the QDOT + 1.2~Jy sample in the left panel,
the dashed line is a fit to a model in which the linear redshift distortion
is modulated by a random exponentially distributed pairwise velocity dispersion
(\S\protect\ref{pairdispersion}).
For the Stromlo-APM survey in the right panel,
the upper thick line shows the quadrupole-to-monopole ratio
which results when fingers-of-god are compressed,
and the dashed line is a straight line fit to this line.
The scales on the right show the value of $\beta$ corresponding to the
linear ratio $\smoothP^{s}_2/\smoothP^{s}_0$, equation~(\protect\ref{barP2P0}),
and the inferred value of $\Omega_0$, equation~(\protect\ref{beta}),
in the absence of bias, $b = 1$.
Beware that the smoothing window
\protect\raisebox{0ex}[2ex][0ex]{$
W(k) \propto k^2 \exp[-(1.279 k/\smoothk)^2]$},
equation~(\protect\ref{W}),
is broad, so the points are correlated;
roughly every third error bar is uncorrelated.
The horizontal and slanted bars on the ($1 \sigma$) errors
come from two separate ways of measuring uncertainties.
\label{xipkoms}
}
\end{figure}

Figure~\ref{xipkoms}
shows the ratio $\smoothP^{s}_2(\smoothk)/\smoothP^{s}_0(\smoothk)$
of smoothed quadrupole to smoothed monopole power
measured as a function of effective wavenumber $\smoothk$
in the {\it IRAS\/} QDOT + 1.2~Jy survey
and the Stromlo-APM survey.
These are the same surveys whose redshift correlation functions
are plotted in Figure~\ref{xilconts}.
The result for the QDOT + 1.2~Jy survey is from Hamilton (1995),
who gives additional details of the analysis.
The Stromlo-APM result, new to this review,
follows essentially the same analysis.

On linear scales (to the right of the graphs in Figure~\ref{xipkoms})
the quadrupole-to-monopole ratio is expected to go over to a constant.
While there is some indication, more particularly in the {\it IRAS\/} surveys,
that the ratio is indeed asymptoting to a constant,
it is evident that nonlinearities have some effect even at scales as
large as a half-wavelength\footnote{
The uncertainty principle ensures that 
the correspondence between wavenumber and separation is inevitably uncertain.
Empirically however,
I find that a half-wavelength $\pi/k$ typically
gives a better estimate of the physical scale of separation
than for example a full wavelength $2\pi/k$ or an inverse wavenumber $1/k$.
}
of $\pi/\smoothk \approx 40 h^{-1} \Mpc$.
One should probably not believe the quadrupole-to-monopole ratio
for $\pi/\smoothk \ga 50 h^{-1} \Mpc$,
where the uncertainties are becoming large.
On small scales the nonlinear finger-of-god effect lowers the
quadrupole-to-monopole ratio,
so that the quadrupole power is negative below a half-wavelength of
$\pi/\smoothk \approx 5 h^{-1} \Mpc$ in the QDOT + 1.2~Jy survey,
and below $\pi/\smoothk \approx 8 h^{-1} \Mpc$ in the Stromlo-APM survey.

For the QDOT + 1.2~Jy survey,
nonlinear effects appear to be adequately described by a model in which
the linear distortion is modulated by a random exponentially distributed
pairwise velocity dispersion
(Fisher \etal\ 1994b) (see \S\ref{pairdispersion}).
A least squares fit to the quadrupole-to-monopole ratio over half-wavelengths
$5$--$44 \, h^{-1} \Mpc$ gives a one-dimensional pairwise velocity dispersion
of $322 \, \kms$.
Although this method of measuring the velocity dispersion is crude
compared to Fisher \etal's (1994b) procedure,
the number is nonetheless in good agreement with Fisher \etal's
scale-dependent estimate (\S\ref{fisher}).
As reported by Hamilton (1995),
the best fitting value of the linear redshift distortion parameter
in the QDOT + 1.2~Jy survey is
$\beta = 0.69^{+ 0.21}_{- 0.19}$ ($1\sigma$).

For the Stromlo-APM survey, the nonlinear effects appear large enough
that simple model-fitting to the quadrupole-to-monopole ratio seems
a hazardous procedure.
Certainly there would be a large covariance between $\beta$
and the velocity dispersion, so that neither would be well measured.
One alternative procedure is to deal with nonlinearities
at a more fundamental level,
by identifying individual fingers-of-god in the survey, and compressing them
(Gramann, Cen \& Gott 1994).
Fingers can be identified using a friends-of-friends algorithm
with a window elongated in the radial direction
(e.g.\ Ramella, Geller \& Huchra 1989).
Figure~\ref{xipkoms} shows the quadrupole-to-monopole ratio
in the Stromlo-APM survey
that results from compressing fingers-of-god whose overdensities exceeded 10
when measured through elliptical windows 10 times longer
in the radial than transverse direction
(so the overdensity of the compressed fingers exceeded $10 \times 10 = 100$;
to avoid spurious fingers in distant, sparsely sampled regions of the survey,
the transverse link length was also limited to less than $5 \, h^{-1} \Mpc$).
The result is encouraging in two ways:
first, that the quadrupole-to-monopole ratio is flattened almost to a constant
on all scales (at least until the uncertainties go out of control
at $\pi/\smoothk \ga 50 \, h^{-1} \Mpc$);
and second, that the ratio is hardly changed at half-wavelengths
$\pi/\smoothk \ga 40 \, h^{-1} \Mpc$,
suggesting that the ratio is close to linear at these scales.
I choose to quote the value
$\beta = 0.30^{+0.17}_{-0.15}$ ($1 \sigma$)
measured at a half-wavelength of $\pi/\smoothk = 20 \, h^{-1} \Mpc$.
Even though the statistical uncertainty is smaller at shorter wavelengths,
the systematic uncertainty associated with compressing the fingers-of-god
presumably becomes larger.

That $\beta$ is smaller for optical than {\it IRAS\/} galaxies
is consistent with the standard interpretation
in which $\beta \approx \Omega_0^{0.6}/b$,
and optical galaxies are positively biased relative to {\it IRAS\/} galaxies.
The values of $\beta$ measured here indicate
$b_{\rm optical}/b_{\it IRAS} \sim 2.3$
with an uncertainty of a factor of 2 or 3.

\subsection{Analysis}
\label{exanalysis}

The smoothing window used in Figure~\ref{xipkoms}
is the power law times Gaussian window
$W(k) \propto k^2 \exp[-(1.279 k/\smoothk)^2]$
of equation~(\ref{W}), with index $n = 2$,
the smallest $n$ that allows the quadrupole to be measured.
This choice of $n$ yields the broadest smoothing window of the form~(\ref{W}),
so the errors in the ratio
$\smoothP^{s}_2(\smoothk)/\smoothP^{s}_0(\smoothk)$ of smoothed power spectrum
harmonics are smallest for this $n$,
although the results at different $\smoothk$ are also then most correlated.

A feature of this analysis is that,
although it is the power spectrum that is being measured,
all the calculations are done in real (redshift) space
rather than in Fourier space.
In measuring redshift distortions,
it is important to disentangle the true distortion from the
artificial distortion introduced by a non-uniform survey window.
In real (redshift) space, the observed galaxy density
is the product of the true density and the selection function.
In Fourier (redshift) space, this product becomes a convolution.
Thus the natural place to `deconvolve' observations from the selection function
is real space, where deconvolution reduces to division,
and where the observations exist in the first place.
The deconvolution procedure is `exact', stable in practice,
and admits a near minimum variance pair weighting.

Specifically,
the smoothed harmonics $\smoothP^{s}_\el(\smoothk)$
of the redshift power spectrum are computed,
in accordance with equations~(\ref{barPl}) and (\ref{xisl}),
by taking a suitably weighted sum over galaxy pairs of
$W_\el(r)$, equation~(\ref{Wlag}) for $n = 2$,
times ${\cal P}_\el(\mu_\r)$, the $\el$'th Legendre polynomial:
\be
\label{Psex}
  \smoothP^{s}_\el =
    (2\el+1) \int W_\el(r) {\cal P}_\el(\mu_\r) \xi^{s}(r,\mu_\r) \, d^3 r
\ee
in which
the redshift correlation function $\xi^{s}(r,\mu_\r)$
at separation $r$ and cosine angle $\mu_\r = \hat\z.\hat\r$
to the line of sight $\z$ is estimated by Hamilton's (1993b) estimator
\be
  \xi^{s}(r,\mu_\r) =
    {\langle D D \rangle \langle R R \rangle \over \langle D R \rangle^2} - 1
\ee
where, following the conventional notation of the literature,
$D$ signifies data, and $R$ signifies random background points,
(although in practice all the background integrals here were done as integrals,
not as Monte-Carlo integrals),
and the angle brackets $\langle \, \rangle$ represent
near-minimum-variance weighted
(Hamilton 1993b, eqs.~[60] \& [61])
averages over pairs at separation $r$ and $\mu_\r$.
The line of sight $\z$ is defined separately for each pair
as the angular bisector of the pair,
and to ensure the validity of the plane-parallel approximation,
only pairs closer than $50^\circ$ on the sky are retained.
Poisson sampling noise is removed by excluding self-pairs
(pairs consisting of a galaxy and itself).

Uncertainties in the ratio
$\smoothP^{s}_2(\smoothk)/\smoothP^{s}_0(\smoothk)$
plotted in Figure~\ref{xipkoms}
were estimated by two distinct methods,
distinguished in Figure~\ref{xipkoms}
by horizontal and tilted bars on the errors.
Both methods take account of the correlated character
of the galaxy distribution.

The first method (horizontal bars)
is described and justified in detail by Hamilton (1993b, \S4).
The procedure is to subdivide the survey into a few hundred volume elements,
measure the fluctuation in the power attributable to each volume element,
and form a variance by adding up the covariances of the fluctuations
between the volume elements.
However, there is an integral constraint that implies that
the sum of all covariances between all volume elements is exactly zero,
so it is necessary to truncate the sum at some point.
The procedure adopted is to order the covariances in order of increasing
distance between each pair of volume elements,
and to take the variance to be the maximum value of the cumulative covariance.
The advantage of this method
is that it makes minimal assumptions about the origin and nature of the errors.
The main defect of the method is that it necessarily underestimates
the variance on scales approaching the scale of the survey.

The second method (tilted bars),
inspired by Feldman, Kaiser \& Peacock (1994),
is to take the Poisson (i.e.\ self-pair) variance of pairs $ij$ weighted by
$[ 1 + \nbar^{s}(\r_i) \smoothP^{s}_0 ( \smoothk ) ]
[ 1 + \nbar^{s}(\r_j) \smoothP^{s}_0 ( \smoothk ) ]$.
Properly, this method is valid only for Gaussian fluctuations,
and at wavelengths that are less than some fraction of the scale of the survey;
moreover this estimate fails to take into account the fact that
the covariance of the redshift power spectrum $P^{s}(\k)$
is different in different directions of the wavevector $\k$.
However, the estimate is a useful check on the reliability of the uncertainties
computed by the first method.
The method may be expected to overestimate the variance somewhat on scales
approaching the scale of the survey,
and Figure~\ref{xipkoms} shows that it clearly underestimates the
uncertainties in the nonlinear regime, where fluctuations are non-Gaussian.

\section{Translinear Regime}
\label{translinear}
\setcounter{equation}{0}

Figure~\ref{xipkoms}
indicates that nonlinearities affect the redshift power spectrum even at
half-wavelengths as large as $\pi/k \approx 40 h^{-1} \Mpc$
(although one should bear in mind that the influence of nonlinearity may be
exaggerated in Fig.~\ref{xipkoms} because of the broad smoothing window)
(see also
%Loveday \etal\ 1996a;
Bromley, Warren \& Zurek 1997).
Studies of the ratio of redshift to real space power spectra
(Fisher \etal\ 1993;
Brainerd \etal\ 1996;
see also:
Suto \& Suginohara 1991;
Bahcall, Cen \& Gramann 1993;
Gramann, Cen \& Bahcall 1993;
Brainerd \& Villumsen  1993, 1994)
lead to the same conclusion,
that nonlinearity can affect the redshift power spectrum
even at quite large scales.
Moreover, there is good reason to push to smaller scales,
to improve the statistics.
Thus a good understanding of nonlinearities would appear essential
to a reliable measurement of the linear distortion parameter $\beta$.
With such an understanding, it might even become possible to
break the degeneracy in $\beta \approx \Omega_0^{0.6}/b$
between the cosmological density $\Omega_0$ and bias $b$.

\subsection{Linear or nonlinear?}
\label{nonlin}

Before going any further, it is worth emphasizing that
{\em redshift distortions are more linear than they look}.

This can be seen from Figure~\ref{caustic}
back at the beginning of this review.
For example,
a region that is just turning around in real space,
a mildly nonlinear condition,
appears to be collapsed to a sheet in redshift space.
The edge of the sheet forms a caustic, a line of infinite density,
a thoroughly nonlinear condition
(note that only the boundary of the collapsed sheet appears
at infinite density;
the complete, infinite density, caustic surface of a collapsing overdensity
is the envelope of the finger-of-god in Figure~\ref{fogs}).

This suggests that theories that go a step beyond the linear regime,
for example the Zel'dovich approximation (\S\ref{zeldovich}),
or perturbation theory
(e.g.\ Scoccimarro \& Frieman 1996;
Hivon \etal\ 1995),
may prove quite successful in modelling redshift distortions
in the translinear regime.

\subsection{Random Isotropic Pairwise Velocity Dispersion}
\label{pairdispersion}

The most common approach to nonlinearity has been
phenomenological rather than based on any rigorous theory.
The approach has been
to suppose that the nonlinear redshift correlation function $\xi^{s}$ is
the result of convolving the linear redshift correlation function $\xi^{s}_L$
with the line-of-sight component of
a random isotropic pairwise velocity distribution $f(v)$.
In the plane-parallel approximation,
which should normally be adequate for small, nonlinear scales,
this model gives (with $f(v)$ normalized to
$\int_{-\infty}^\infty f(v) dv = 1$)
\be
  \xi^{s}(r_\para, r_\perp) =
    \int_{-\infty}^\infty \xi^{s}_L(r_\para-v, r_\perp) f(v) d v
  \ .
\ee
Note that this model does not require that the individual galaxies of a pair
should separately have random isotropic velocity distributions,
although some authors have found it convenient also to assume the latter
(e.g.\ Peacock \& Dodds 1994, to allow them to consider cross-correlations
between different populations;
or Heavens \& Taylor 1995, to allow them to deal with radial
rather than plane-parallel redshift distortions).
Convolution in real space becomes multiplication in Fourier space,
so the redshift power spectrum in this model is the product of the
linear redshift power spectrum  $P^{s}_L(\k) = (1+\beta \mu_\k^2)^2 P(k)$
with the line-of-sight Fourier transform
$\FTf(k_\para) = \int_{-\infty}^\infty f(v) e^{i k_\para v} dv$
(note $k_\para = k \mu_\k$)
of the velocity distribution:
\be
\label{Pnl}
  P^{s}(\k) = \FTf(k \mu_\k) P^{s}_L(\k)
            = \FTf(k \mu_\k) (1 + \beta \mu_\k^2)^2 P(k)
  \ .
\ee

Those authors who have invoked this model have generally adopted either a
Gaussian velocity distribution
(Peacock \& Dodds 1994;
Heavens \& Taylor 1995;
Tadros \& Efstathiou 1996),
for which
$f(v) = [(2\pi)^{1/2} \sigma]^{-1} \discretionary{}{}{} \exp[- (v/\sigma)^2/2]$
and
\be
\label{fgau}
  \FTf(k \mu_\k) = \exp \Bigl[- (\sigma k \mu_\k)^2/2 \Bigr]
  \ ;
\ee
or else an exponential pairwise velocity distribution
(Fisher \etal\ 1994b;
Hamilton 1995;
Ballinger, Peacock \& Heavens 1996;
Peacock 1997;
Bromley, Warren \& Zurek 1997;
Ratcliffe \etal\ 1997;
Cole, Fisher \& Weinberg 1995
use an exponential single-particle distribution,
equivalent to a pairwise distribution
that is the convolution of two exponentials),
for which
$f(v) = (2^{1/2} \sigma)^{-1} \exp(- 2^{1/2} |v|/\sigma)$
and
\be
\label{fexp}
  \FTf(k \mu_\k) = {1 \over 1 + \frac{1}{2} (\sigma k \mu_\k)^2}
  \ .
\ee
The velocity dispersions $\sigma$ above are,
after the convention established by
Peebles (1976, 1980 eq.~[76.14]) and
Davis \& Peebles (1983),
1-dim\-en\-sional pairwise velocity dispersions,
which are $2^{1/2}$ larger than a random 1-dim\-en\-sional single-point
(galaxy) dispersion,
and $3^{1/2}$ smaller than a 3-dim\-en\-sional pairwise dispersion.

The exponential pairwise velocity distribution was first proposed
as a fit to observations by Peebles (1976),
and has held up as a good approximation both in observations
(Davis \& Peebles 1983, CfA1;
Fisher \etal\ 1994b, 1.2~Jy survey;
Marzke \etal\ 1995, CfA2 + SSRS2;
Lin 1995, LCRS),
and in $N$-body experiments
(Fisher \etal\ 1994b, Fig.~5;
Zurek \etal\ 1994, Fig.~7),
although the latter reveal appreciable skewness in the distribution
of pairwise infall velocities.

\subsection{Zel'dovich Approximation}
\label{zeldovich}

The Zel'dovich approximation
(e.g.\ Hui \& Bertschinger 1996)
is essentially linear theory expressed in Lagrangian space.
In linear theory, particles (galaxies) move in straight lines from
their initial comoving positions $\r_i$,
with comoving displacements $\Delta\r = D(t) \Delta\r_L$
growing in proportion to the linear growth factor $D(t)$
\be
\label{zeld}
  \r(t) = \r_i + D(t) \Delta\r_L(\r_i)
\ee
where $\Delta\r_L = - \bnabla \nabla^{-2} \delta_L$
is the linear displacement field,
with $\delta_L$ the linear overdensity,
evaluated at some suitably early epoch.
The Zel'dovich approximation supposes that the linear result~(\ref{zeld})
remains true also at nonlinear epochs.
By construction,
the Zel'dovich approximation satisfies the continuity equation,
but it fails to satisfy the Euler equation beyond the linear regime.

Redshift distortions in the Zel'dovich approximation have been studied by
Fisher \& Nusser (1996),
Taylor \& Hamilton (1996),
and Hatton \& Cole (1997).
All three sets of authors tested the predictions of the Zel'dovich
approximations against $N$-body simulations,
and all three adopted the plane-parallel approximation
and chose the ratio of quadrupole to monopole harmonics of the
redshift power spectrum as the test statistic.
Fisher \& Nusser and Hatton \& Cole examined $N$-body simulations
with CDM-like power spectra,
while Taylor \& Hamilton examined simulations with power law power spectra,
$P(k) \propto k^n$ with $n = -1$, $-1.5$, and $-2$.

The conclusions can be summarized as follows.
The quadrupole-to-monopole ratio $P^{s}_2(k)/P^{s}_0(k)$ as a function of $k$
can be characterized by a shape, an overall amplitude,
and an overall scale $k_0$ that is conveniently taken to be
the zero-crossing of the quadrupole power, $P^{s}_2(k_0) = 0$.
Then:
\begin{itemize}
\item[$\bullet$]
The shape of the quadrupole-to-monopole ratio
depends on the shape of the power spectrum,
but is insensitive to the redshift distortion parameter $\beta$.
This shape is accurately predicted by the Zel'dovich approximation
at least down to the zero-crossing of the quadrupole, $k \la k_0$.
\item[$\bullet$]
The overall amplitude of the quadrupole-to-monopole ratio
depends on the linear redshift distortion parameter $\beta$
according to the usual linear formula~(\ref{P2P0}).
\item[$\bullet$]
The zero-crossing scale $k_0$ of the quadrupole
is not reliably predicted by the Zel'dovich approximation.
\end{itemize}

Actually,
Fisher \& Nusser argued that the shape of the quadrupole-to-monopole
ratio is a universal function that is insensitive to the shape of the
power spectrum.
Within the range of CDM-like power spectra this is probably a reasonable
approximation.
However, the investigations of Taylor \& Hamilton showed that
if a broader range of power spectra is considered,
then there is a dependence on the shape of the power spectrum
which, according to the Zel'dovich approximation,
becomes marked as the spectral index $n \rightarrow -3$.
There is general agreement that the shape of the quadrupole-to-monopole ratio
is insensitive to $\beta$,
the latter serving simply to set the overall normalization of the ratio.

The third conclusion, that the Zel'dovich approximation does not
predict the zero-crossing of the quadrupole power reliably,
is based mainly on the results of Taylor \& Hamilton,
but the conclusion seems to be consistent with the results
graphed by Hatton \& Cole
(Fisher \& Nusser offer no information on this point).
It is not entirely clear what is responsible for the discrepancy
in the zero-crossing.

Fisher \& Nusser and Taylor \& Hamilton both
interpreted the agreement between the predicted Zel'dovich
and measured $N$-body shape of the quadrupole-to-monopole ratio
as suggesting that the departure of this ratio from a constant value
in the translinear regime
is caused not by random velocities in nonlinear virialized fingers-of-god,
but rather by coherent infall towards collapsing clusters.
This is encouraging if true, because it lends hope that
redshift distortions can be modelled accurately
with more rigorous theoretical treatments in the translinear regime.

\subsection{Future work}
\label{future}

The shortcomings of modelling nonlinearity in redshift distortions
either with random pairwise velocities,
or with the Zel'dovich approximation,
have been emphasized by Hatton \& Cole (1997).
It is clear that a more thorough and rigorous treatment of redshift distortions
in the translinear regime,
for example using perturbation theory
(e.g.\ Scoccimarro \& Frieman 1996),
is needed.

As usual, bias complicates the problem.
One clear bias is that highly nonlinear fingers-of-god
are more prominent in optical and elliptical populations
than in {\it IRAS\/} and spiral populations
(Fig.~\ref{xilconts}; Guzzo \etal\ 1997).
Long fingers in rich clusters appear stretched to large separations
in redshift space, contaminating the linear and translinear regime.
Perhaps this problem can be side-stepped by identifying and compressing
prominent fingers
(\S\ref{exresults}; Gramann, Cen \& Gott 1994).
Or perhaps estimators can be constructed that are
somehow orthogonal\footnote{
A nice example of what is meant here by orthogonality
is the case of the angular correlation function or angular power spectrum,
which is orthogonal to --- i.e.\ completely unaffected by ---
redshift distortions.
}
to the nonlinear fingers,
as suggested by
Bromley, Warren \& Zurek (1997;
these authors claimed to do just this, but actually their estimator,
based on eq.~[\ref{wBrom}] in \S\ref{bromley} below,
merely corrects for nonlinearity, and is not orthogonal to it).

\section{Measurements of $\beta$ from Linear Redshift Distortions}
\label{measurement}
\setcounter{equation}{0}

\subsection{Compilation}
\label{compilation}

Table~\ref{results} lists
measurements of the distortion parameter $\beta$
from redshift distortions in the linear regime.
The table is intended to be complete up to the first half of 1997.
Compare this compilation to that of
Strauss \& Willick (1995, Table~3, p.~412).
Note that Table~\ref{results}
does not include measurements of $\beta$ from direct measurements of
peculiar velocities, which Strauss \& Willick do include.

\begin{table}[bhtp]
\begin{center}
\caption{Values of $\beta$}
\begin{tabular}{lll}
\hline
Survey & $\qquad\beta$ & Reference \\
\hline
\multicolumn{3}{l}
{1. Ratio of redshift space to real space power spectra (plane-parallel)} \\
\hline 
{\it CfA\/} & $0.53 \pm 0.15$ & Fry \& Gazta\~naga (1994) \\
{\it SSRS\/} & $1.10 \pm 0.16$ & $\cdots$ \\
{\it IRAS\/} & $0.84 \pm 0.45$ & $\cdots$ \\
{\it IRAS\/} & $1.0 \pm 0.2$ & Peacock \& Dodds (1994) \\
Optical & $0.77 \pm 0.15$ & $\cdots$ \\
Stromlo-APM & $0.20^{+0.19}_{-0.22}$ & Baugh (1996) \\
Stromlo-APM & $0.48 \pm 0.12$ & Loveday \etal\ (1996a) \\
Stromlo-APM & $0.20 \pm 0.44$ & Tadros \& Efstathiou (1996) \\
Optical & $0.40 \pm 0.12$ & Peacock (1997) \\
Durham/UKST & $0.52 \pm 0.39$ & Ratcliffe \etal\ (1997) \\
\hline
\multicolumn{3}{l}
{2. Ratio of quadrupole to monopole redshift power (plane-parallel)} \\
\hline
{\it IRAS\/} 2 Jy & $0.69^{+ 0.28}_{- 0.24}$ & Hamilton (1993a) \\
Perseus-Pisces & $0.75 \pm 0.3$ & Bromley (1994) \\
{\it IRAS\/} 1.2 Jy & $0.45^{+ 0.27}_{- 0.18}$ & Fisher \etal\ (1994b) \\
{\it IRAS\/} 1.2 Jy & $0.52 \pm 0.15$ & Cole, Fisher \& Weinberg (1995) \\
QDOT & $0.54 \pm 0.3$ & $\cdots$ \\
LCRS & $0.5 \pm 0.25$ & Lin (1995) \\
QDOT + {\it IRAS\/} 1.2 Jy & $0.69^{+ 0.21}_{- 0.19}$ & Hamilton (1995) \\
{\it IRAS\/} 1.2 Jy & $0.6 \pm 0.2$ & Fisher \& Nusser (1996) \\
QDOT + {\it IRAS\/} 1.2 Jy & $> 0.5$ (95\% confidence) & Taylor \& Hamilton (1996) \\
{\it IRAS\/} 1.2 Jy & $0.8^{+ 0.4}_{- 0.3}$ & Bromley \etal\ (1997) \\
Durham/UKST & $0.48 \pm 0.11$ & Ratcliffe \etal\ (1997) \\
Stromlo-APM & $0.30^{+ 0.17}_{- 0.15}$ & This paper, \S\protect\ref{example} \\
\hline
\multicolumn{3}{l}
{3. Maximum Likelihood (radial)} \\
\hline
{\it IRAS\/} 1.2 Jy & $0.96^{+ 0.20}_{- 0.18}$ & Fisher, Scharf \& Lahav (1994) \\
{\it IRAS\/} 1.2 Jy & $1.1 \pm 0.3$ & Heavens \& Taylor (1995) \\
{\it IRAS\/} 1.2 Jy & $1.04 \pm 0.3$ & Ballinger, Heavens \& Taylor (1995) \\
\hline
\end{tabular}
\end{center}
\label{results}
\end{table}

The Table is subdivided into the three principal methods
described \S\ref{methods}.
However, precise procedures used by different authors differ in many respects,
and the threefold categorization is not as clean as the
headings might suggest.
Comments on the individual measurements are given in the next subsection,
\S\ref{individual}.

The average and standard deviation of values listed in Table~\ref{results}
is $\beta_{\rm optical} = 0.52 \pm 0.26$
for optical galaxies,
and
$\beta_{\it IRAS} = 0.77 \pm 0.22$
for {\it IRAS\/} selected galaxies
(the quoted uncertainty is the deviation of a single measurement;
the lower limit from Taylor \& Hamilton was omitted).
In averaging the numbers,
I felt it was better to assign equal weight to all measurements,
rather than to rely either on the authors' quoted uncertainties
or on my own judgement of their quality.
Of course the data sets --- especially the {\it IRAS\/} data ---
are not all independent of each other, so the standard deviation
reflects differences in methods as much as anything else.
All the same, the dispersion of the measurements turns out to be about the
same as the uncertainties that people quote,
so is perhaps not unreasonable.
If the difference in optical and {\it IRAS\/} $\beta$'s is attributed
to bias, it implies a relative bias of
$b_{\rm optical}/b_{\it IRAS} \approx 1.5$,
consistent with the ratio
$b_{\rm optical}/b_{\it IRAS} \approx 1.3$ measured from
the square root of the ratio of their power spectra
(Peacock \& Dodds 1994; Peacock 1997, Fig.~2).
If optical galaxies are unbiased, $b_{\rm optical} = 1$,
then the inferred value of the cosmological density is
\raisebox{0ex}[2ex][0ex]{$\Omega_0 = 0.33^{+0.32}_{-0.22} \,$}.
If on the other hand {\it IRAS\/} galaxies are unbiased,
$b_{\it IRAS} = 1$, then
\raisebox{0ex}[2ex][0ex]{$\Omega_0 = 0.63^{+0.35}_{-0.27} \,$}.

\subsection{Comments on individual measurements}
\label{individual}

The comments in this subsection are arranged in approximately the
chronological order in which the measurements were published,
except that measurements by the same author or group of authors are
kept together.

\subsubsection{Hamilton \etal}
\label{hamilton}

{\it Hamilton (1993a)}
measured $\beta$ in the {\it IRAS\/} 2~Jy survey
(Strauss \etal\ 1992a)
from a ratio of quadrupole-to-monopole correlation functions,
weigh\-ted with pair separation in the fashion proposed by Hamilton (1992):
\be
\label{xi2xi0}
  {\xi^s_2(r) \over \xi^{s}_0(r) - \bar\xi^{s}_0(r)} =
    {\frac{4}{3} \beta + \frac{4}{7} \beta^2
    \over 1 + \frac{2}{3} \beta + \frac{1}{5} \beta^2}
\ee
where $\bar\xi^{s}_0(r) \equiv 3 r^{-3} \int_0^r \xi^{s}_0(s) s^2 ds$.
Cole, Fisher \& Weinberg (1994, eq. [2.17] and Appendix B)
subsequently pointed out that Hamilton's ratio~(\ref{xi2xi0})
was actually just a ratio of smoothed quadrupole to smoothed monopole power,
equation~(\ref{barP2P0}),
in which the smoothing function is taken to be the second spherical Bessel
function, $W(k) = j_2(kr)$ in equation~(\ref{barP}).
This smoothing window vanishes at zero wavenumber, $W(k) = 0$ at $k = 0$,
immunizing the estimate against uncertainty in the mean density.

{\it Hamilton (1995)} and the present paper,
\S\ref{example} and Figure~\ref{xipkoms},
use essentially the same procedure as Hamilton (1993a),
but with the smoothing function taken to be a power law times a Gaussian,
equation~(\ref{W}) with $n = 2$.

A principal topic of Hamilton (1995) was the presentation of evidence
of a large, by the look of it systematic, difference in
clustering between the near and far regions of the QDOT sample,
which Hamilton suggested might arise from some systematic in the
{\it IRAS\/} PSC fluxes.
To date no satisfactory explanation of this problem has emerged.

{\it Taylor \& Hamilton (1996)}
studied the Zel'dovich approximation as a means of
approximating redshift distortions in the mildly nonlinear regime
(see \S\ref{zeldovich}).
As an application,
they fitted the ratio of quadrupole-to-monopole power measured by
Hamilton (1995) from the {\it IRAS\/} QDOT + 1.2~Jy survey,
the same data as are plotted in Figure~\ref{xipkoms}.
They concluded that $\beta > 0.5$ at the 95\% confidence level,
using a full covariance matrix estimated from the data.

\subsubsection{Fisher \etal}
\label{fisher}

The paper by Fisher \etal\ (1994b) was published after that of
Fisher, Scharf \& Lahav (1994),
but the former seems to be the intellectual predecessor,
so here it appears first.

{\it Fisher \etal's (1994b)}
paper on the {\it IRAS\/} 1.2~Jy redshift survey is especially useful
because of its careful study of nonlinear effects
through Cold Dark Matter (CDM) $N$-body simulations engineered
to the statistical properties of the 1.2~Jy survey.
They pointed out that the redshift correlation function along the line of sight
of sight axis, $\xi^{s}(r_\para, 0)$ with $r_\perp = 0$,
gives a clear signature of the functional form of the
pairwise velocity distribution,
and find that an isotropic exponential pairwise velocity distribution
fits the {\it IRAS\/} data well --- indeed better than a more
detailed anisotropic model that they tried.

Fisher \etal\
measured the linear distortion parameter $\beta$ from the statistic
\be
\label{betaFish}
  \beta = {\partial \ln \bar\xi(r) \over \partial \ln a^2} =
    {3 v_{12}(r) [1+\xi(r)] \over 2 r \bar\xi(r)}
\ee
where $v_{12}(r)$ is the mean pairwise infall velocity at separation $r$,
and $\bar\xi(r) \equiv 3 r^{-3} \int_0^{r} \xi(s) s^2 ds$.
The correlation functions $\xi(r)$ and $\bar\xi(r)$
in equation~(\ref{betaFish}) are in real space, not redshift space,
and Fisher \etal\ adopted the power law fit
$\xi(r) = (r / 3.76 \, h^{-1} \Mpc)^{1.66}$
to the {\it IRAS\/} real space correlation function
estimated by Fisher \etal\ (1994a) from the angular correlation function.
The last equality in~(\ref{betaFish}) is just the pair conservation equation
(Peebles 1980, eq.~[71.7]),
true in both linear and nonlinear regimes,
while the first equality in~(\ref{betaFish}) is true in the linear regime.
Fisher \etal\ showed that the statistic~(\ref{betaFish})
gave a reliable measure of $\beta$ at separations $r \ga 10 \, h^{-1} \Mpc$
in three CDM simulations,
one flat and unbiased, one flat and biased ($b = 1.6$),
and one open and unbiased.

To determine the infall velocity $v_{12}(r)$ in the estimator~(\ref{betaFish}),
they modelled the pairwise velocity distribution as an isotropic exponential
with mean infall velocity $v_{12}(r)$ and dispersion $\sigma(r)$,
the shape (but not amplitude) of these two velocities as a function of
separation $r$ being determined in essence by the CDM simulations,
but scaled and adjusted to match the power law fit to the
{\it IRAS\/} real space correlation function.
To determine the overall amplitudes of the infall velocity $v_{12}(r)$
and dispersion $\sigma(r)$,
they performed least squares fits to two functions
that they measured from the data at separations up to $16 \, h^{-1} \Mpc$:
the quadrupole-to-monopole ratio of correlation functions
$\xi^{s}_2(r)/\xi^{s}_0(r)$;
and the line-of-sight correlation function
$\xi^{s}(r_\para , r_\perp)$ with $r_\perp < 2 \, h^{-1} \Mpc$.
In carrying out the fits,
they used a full covariance matrix estimated from the simulations.
The measured pairwise velocity dispersion
(with scale-dependence fixed by their procedure)
was
$\sigma(r) = 317^{+40}_{-49} \, \kms$
at a separation of $r = 1 \, h^{-1} \Mpc$,
rising to $\approx 400 \, \kms$ at $r = 4 \, h^{-1} \Mpc$,
then declining again to $\approx 350 \, h^{-1} \Mpc$ at $r = 10 \, h^{-1} \Mpc$
(Fisher \etal, Fig.~3).
The final result from the estimator~(\ref{betaFish}) was
$\beta = 0.45^{+0.27}_{-0.18}$.

{\it Fisher, Scharf \& Lahav (1994)}
were the first to study radial redshift distortions
without using the plane-parallel approximation,
and the first to apply a maximum likelihood (ML) formalism to measure $\beta$
(albeit with a rather crude choice of radial density modes).
They expanded the {\it IRAS\/} 1.2~Jy density field in spherical harmonics
about the observer (us),
windowing the density in the radial direction with Gaussian windows
centred at four depths, $38$, $58$, $78$, and $98 \, h^{-1} \Mpc$,
each with a dispersion of $8 \, h^{-1} \Mpc$.
To make spherical harmonics uncorrelated requires full sky coverage,
and the 1.2~Jy survey has the asset that it already covers almost all the sky
at galactic latitudes $|b| > 5^\circ$.
While it is possible to take into account the coupling of harmonics
induced by incomplete sky coverage,
Fisher \etal\ chose instead to complete the sky coverage of the 1.2~Jy survey
by interpolating over the plane of the Milky Way
in a way that smoothly continued structure.
They worked in the Local Group (LG) frame (\S\ref{operatorLG}),
and the equations below do likewise.

In the course of their analysis, Fisher \etal\ did a cute trick,
which is worth pointing out here.
Let $a$ be the redshift space galaxy density $n^{s}(\s)$
in the LG frame, smoothed over some window $f(\s)$
(superscripts LG's on $\s^\LG$ and $n^\LG$ are omitted for brevity):
\ba
  a &\equiv& \int f(\s) n^{s}(\s) \, d^3 s
    = \int f(\s) n(\r) \, d^3 r
  \nn
\label{a}
    &\approx& \int
      \left[ f(\r) + \Bigl( v - \hat\r . \bv^\LG \Bigr)
      {\partial f(\r) \over \partial r} \right]
      \nbar(\r) [ 1 + \delta(\r) ] \, d^3 r
  \ .
\ea
The second equality in equation~(\ref{a})
is true because galaxies are conserved, equation~(\ref{nsn}),
and the last approximation, just a Taylor expansion of $f(\s)$,
is valid for windows $f(\s)$ that are
sufficiently smooth in the radial direction.
Fisher \etal\ emphasized the point that the selection function $\nbar(\r)$
in a flux-limited survey is in fact a function of the true distance $r$
to a galaxy, not of its redshift distance $s$ (see footnote\footnote{
It remains necessary nonetheless to estimate the real space selection function
somehow,
in order to evaluate the quantities~(\protect\ref{wl}) and (\protect\ref{wcl}).
In practice Fisher \etal\ used the selection function $\nbar^{s}(\r)$ estimated
in redshift space,
and discarded all monopole ($\el = 0$) modes.
This is fine for an all-sky survey
--- see the final paragraph of \S\protect\ref{fixnbar}.
}).
Equation~(\ref{a}), which is essentially an integration by parts,
recasts the effect of redshift distortions into the window $f(\r)$
rather than the selection function and the density,
which seems neat.

Actually, the redshift distortion is still there in the last expression
of equation~(\ref{a}),
lurking in the peculiar velocity $v - \hat\r . \bv^\LG$.
In the linear regime,
the velocities $v$ and $\bv^\LG$ are given as usual by equations~(\ref{v})
and (\ref{vLG}), and equation~(\ref{a}) reduces to
\be
\label{alin}
  a = \int \nbar(\r) \left[ f(\r)
    - \beta {\partial f(\r) \over \partial r}
    \, \hat\r \, . \left( {\partial \over \partial\r}
    - \left. {\partial \over \partial\r} \right|_{\r = 0} \right)
    \nabla^{-2} \right] \delta(\r) \, d^3 r
  \ ,
\ee
where it has been assumed that $\int f(\r) \nbar(\r) d^3 r = 0$.
The latter is true for all the windows considered by Fisher \etal,
and is a desirable feature in any case to immunize measurements
against uncertainty in the mean density.

Fisher \etal\ took the spherical transform
of the redshift space density folded through Gaussian radial windows,
which corresponds to choosing windows $f_{\el m}(\r)$
in equation~(\ref{a})
(the subscript $\el m$ is appended to distinguish the windows)
of the form
\be
\label{flm}
  f_{\el m}(\r) \equiv f(r) Y_{\el m}(\hat\r)
\ee
with $f(r) \propto e^{-[(r-r_0)/\sigma]^2/2}$.
They showed from equation~(\ref{alin}) that,
for an all-sky survey,
the resulting spherical transform $a_{\el m}$ could be written
(Fisher \etal\ eq.~[8])
\be
\label{alm}
  a_{\el m} = 
    \int_0^\infty
    \Bigl[ \FTw_\el(k) + \beta \FTw^C_\el(k) \Bigr]^\ast
    \FTdelta_{\el m}(k) \, k^2 dk/(2\pi)^3
\ee
where $\FTdelta_{\el m}(k)$ is the spherical transform~(\ref{drlm})
of the unredshifted overdensity, $\FTw_\el(k)$ is
\be
\label{wl}
  \FTw_\el(k) =
    i^\el 4\pi \int_0^\infty \! j_\el(kr) f(r) \nbar(r) \, r^2 dr
  \ ,
\ee
and $\FTw^C_\el(k)$ is a correction term
embodying the redshift distortions:
\be
\label{wcl}
  \FTw^C_\el(k) =
    i^\el 4\pi \int_0^\infty
    \left[ j'_\el(kr) - \frac{1}{3} \delta^K_{\el 1} \right]
    {d f(r) \over k dr} \nbar(r) \, r^2 dr
  \ ,
\ee
with $j'_\el(x) \equiv dj_\el(x)/dx$ the derivative
of the $\el$'th spherical Bessel function.
The Kronecker delta term $\frac{1}{3} \delta^K_{\el 1}$,
which subtracts from the dipole ($\el = 1$) term
$j'_1(kr)$ its value at $r = 0$,
is the term that arises from the motion of the LG.

It is important in equation~(\ref{alm}) that the survey covers the entire sky.
If the survey were not all-sky,
then the functions $\FTw_\el(k)$ in equation~(\ref{alm})
would be replaced by matrices $\FTw_{\el m}^{\el' m'}(k)$,
which act by matrix multiplication on the spherical transform
$\FTdelta_{\el' m'}(k)$ of the overdensity.
Only in the case of an all-sky survey does the matrix
$\FTw_{\el m}^{\el' m'}(k)$ become diagonal;
the quantities $\FTw_\el(k)$
in equation~(\ref{alm}) are then the eigenvalues of the matrix.

For an all-sky survey,
statistical isotropy implies that the $a_{\el m}$ are uncorrelated,
with variances from equation~(\ref{alm})
\be
  \langle |a_{\el m}|^2 \rangle =
    \int \Bigl| \FTw_\el(k) + \beta \FTw^C_\el(k) \Bigr|^2
    P(k) \, k^2 dk/(2\pi)^3
    + N_{\el m}
\ee
where $N_{\el m}$ is the Poisson sampling noise,
which is just the (expected) self-galaxy contribution to
$\langle |a_{\el m}|^2 \rangle$.
Fisher \etal\ used four separate Gaussian radial windows.
Here there are also covariances
$\langle a^i_{\el m} a^{j \ast}_{\el m} \rangle$
between the different radial windows $i$ and $j$,
but still only covariances with the same $\el m$ are non-zero.
To determine $\beta$,
Fisher \etal\ combined the observed amplitudes $a^i_{\el m}$
measured in the 1.2~Jy survey for $\el = 1$ to 10
into a likelihood function,
adopting as prior an unredshifted power spectrum $P(k)$ with
fixed shape corresponding to a CDM model with $\Gamma = 0.2$
(Efstathiou, Bond \& White 1992),
and fixed amplitude $\sigma_8 = 0.69 \pm 0.04$,
as determined by Fisher \etal\ (1993).
The final result was $\beta = 0.96^{+ 0.20}_{- 0.18} \,$,
which is the value quoted in Table~\ref{results}.

Fisher \etal\ also quote (in proof) a result reported by Fisher (1993),
in which the likelihood analysis is done with the shape parameter $\Gamma$
as well as $\beta$ treated as a free parameter,
with the result $\beta = 0.94 \pm 0.17$ and $\Gamma = 0.17 \pm 0.05$.
The uncertainty here is the conditional $1\sigma$ uncertainty,
which comes from projecting the $\Delta\ln{\cal L} = -0.5$ contour
on to the parameter axes.
If the amplitude $\sigma_8$ is also treated as a free parameter,
then $\beta = 0.47 \pm 0.25$, $\Gamma = 0.15 \pm 0.05$,
and $\sigma_8 = 0.81 \pm 0.06$.
There is a strong anti-correlation between $\beta$ and $\sigma_8$,
and Fisher (1993) argues that the more reliable value of $\beta$ is probably
the higher one obtained when $\sigma_8$ is set to the value measured from the
real space correlation function.

It is instructive to derive a more general version of equation~(\ref{alm}).
Let $w(\r)$ denote the product of a (complex-valued, in general) window $f(\r)$
(not necessarily of the form~[\ref{flm}])
with the selection function $\nbar(\r)$
(not necessarily all-sky)
\be
  w^\ast(\r) \equiv f(\r) \nbar(\r)
  \ .
\ee
Assume, following Fisher \etal, that $\int w(\r) d^3 r = 0$,
always a good idea in any case to protect against uncertainty
in the mean density.
Then the windowed overdensity $a$, equation~(\ref{a}) is,
for linear distortions,
\be
\label{aa}
  a = \int w^\ast(\r) \delta^{s}(\r) \, d^3 r
    = \int w^\ast(\r) \bS^\LG \delta(\r) \, d^3 r
    = \int \Bigl[ \bS^{\LG\dagger} w(\r) \Bigr]^\ast \delta(\r) \, d^3 r
\ee
where $\bS^\LG$ is the linear redshift distortion operator in the LG frame,
equation~(\ref{SLG}),
and $\bS^{\LG\dagger}$ is its Hermitian conjugate,
equation~(\ref{SLGdag}).
It follows that the windowed density $a$ is
\be
\label{aw}
  a = \int \Bigl[ w(\r) + \beta w^C(\r) \Bigr]^\ast \delta(\r) \, d^3 r
\ee
where $w^C(\r)$ is the result of the distortion term
(the part proportional to $\beta$) in $\bS^{\LG\dagger}$,
equation~(\ref{SLGdag}),
acting on the window $w(\r)$:
\be
\label{wc}
  w^C(\r) = 
    \left[
    \nabla^{-2} r^{-2} {\partial \over \partial r}
    \left( {\partial \over \partial r} - {\alpha(\r) \over r} \right) r^2
    - {\hat\r \over r^2} \, .
    \left. {\partial \over \partial\r} \right|_{\r = 0}
    \nabla^{-2} \alpha(\r) r \right] w(\r)
  \ .
\ee
The correction term~(\ref{wc}) can be rewritten in a variety of ways,
including one of the same ilk as Fisher \etal's expression,
equation~(\ref{wcl}),
\be
\label{wcp}
  w^{C\ast}(\r) = 
    \int {\hat\r' \over 4\pi} \, .
    \left( {\r-\r' \over |\r-\r'|^3} - {\r \over r^3} \right)
    {\partial f(\r') \over \partial r'} \nbar(\r') \, d^3 r'
  \ .
\ee
The $\r/r^3$ term in the integrand,
which subtracts from $(\r-\r')/|\r-\r'|^3$ its value at $\r' = 0$,
is the term that arises from the motion of the LG.
Recasting the integral~(\ref{aw}) in spherical transform space,
equations~(\ref{drlm}), (\ref{dkrlm}), and (\ref{aibi}), gives
\be
  a = \int_0^\infty \sum_{\el m}
    \Bigl[ w_{\el m}(k) + \beta w^C_{\el m}(k) \Bigr]^\ast
    \FTdelta_{\el m}(k) \, k^2 dk/(2\pi)^3
\ee
which generalizes equation~(\ref{alm}).
Equation~(\ref{wc}) reveals the nature of the correction term $w^C$,
and it is readily confirmed that Fisher \etal's result~(\ref{wcl})
is regained if the window is chosen to be
$w^\ast(\r) = f(r) Y_{\el m}(\hat\r) \nbar(r)$
in an all-sky survey, where the selection function $\nbar(r)$
is the same in all directions over the sky.

The same derivation can be written compactly in terms of vectors and matrices
in Hilbert space (see \S\ref{hilbert})
\be
\label{ahil}
  a = \w^\dagger \hspace{1pt} \bdelta^{s}
    = \w^\dagger \bS^\LG \hspace{1pt} \bdelta
    = ( \bS^{\LG\dagger} \w )^\dagger \hspace{1pt} \bdelta
    = ( \w + \beta \w^C )^\dagger \hspace{1pt} \bdelta
  \ .
\ee
Here $\w$ is a weighting function, a vector,
$\bS^\LG$ is the linear distortion operator~(\ref{SLG}) in the LG frame,
a matrix,
and $\bdelta$ is the vector of (unredshifted) overdensities.
Equation~(\ref{ahil}) generalizes easily to the case of an array of quantities:
the scalar $a$ in equation~(\ref{ahil}) simply becomes a vector ${\bmia}$,
and the weighting vector $\w$ becomes a matrix.
The covariance matrix of such an array is
\be
  \langle {\bmia} \, {\bmia}^\dagger \rangle =
    ( \w + \beta \w^C )^\dagger \bxi \, ( \w + \beta \w^C ) + \bN
\ee
where $\bxi$ is the (unredshifted) power spectrum matrix,
and $\bN$ is the Poisson sampling noise matrix,
the self-galaxy contribution to
$\langle {\bmia} \, {\bmia}^\dagger \rangle$.

\subsubsection{Bromley \etal}
\label{bromley}

{\it Bromley (1994)}
proposed the idea of measuring $\beta$ from a ratio of variances of
the redshift space density windowed through anisotropic sampling functions.
Consider the statistic $\smoothdelta^{s}$ which is the value of the density
filtered through some window (sampling function) $w(\r)$
randomly positioned in the redshift survey
\be
\label{bard}
  \smoothdelta^{s} \equiv \int w(\r) \delta^{s}(\r) \, d^3 r
  = \int w(\r) {n^{s}(\r) \over \nbar^{s}(\r)} \, d^3 r
\ee
where the last equality is true as long as the window is chosen to have
vanishing volume integral, $\int w(\r) d^3 r = 0$, as was true for the windows
considered by Bromley.
The real space integral in~(\ref{bard}) can be rewritten as a Fourier space
integral
\be
\label{bardk}
  \smoothdelta^{s} = \int \FTw^\ast(\k) \FTdelta^{s}(\k) \, d^3 k/(2\pi)^3
\ee
where $\FTw(\k)$ and $\FTdelta^{s}(\k)$ are the Fourier transforms
of the sampling function and overdensity.
It follows that the expected shot-noise-subtracted variance $\smoothP^{s}$
of the filtered density is equal to
the redshift space power spectrum $P^{s}(\k)$
folded with the power spectrum $|\FTw(\k)|^2$ of the sampling function
\be
\label{varbard}
  \smoothP^{s} \equiv
  \langle (\smoothdelta^{s})^2 \rangle - N =
    \int |\FTw(\k)|^2 P^{s}(\k) \, d^3 k/(2\pi)^3
\ee
where $N$ is the shot noise, the self-pair contribution to
$\langle (\smoothdelta^{s})^2 \rangle$.
Shifting the window $w(\r)$ to another random position in the survey
simply changes the phase of $\FTw(\k)$,
which leaves the variance~(\ref{varbard}) unaltered.
Bromley now proposed to choose windows $w(\r)$ that were the gradient
along some unit direction $\n$ of some spherically symmetric function $R(r)$,
equivalent in Fourier space to $i \n.\k$ times its Fourier transform
$\FTR(k) = \int R(r) e^{i \k.\r} d^3 r$:
\be
\label{w}
  w(\r) = \n . \bnabla R(r)
  \ , \quad
  \FTw(\k) = i \n . \k \FTR(k)
  \ .
\ee
If the unit directions $\n$ are chosen to be
either parallel to or perpendicular to the line of sight to the window,
then the ratio of parallel to perpendicular variances is
(Bromley eq.~[9];
see also Gramann, Cen \& Gott 1994, eq.[10], which however contains an error)
\be
\label{bromstat}
  {\smoothP^{s}_\para \over \smoothP^{s}_\perp} =
  {\int \mu_\k^2 W(k) P^{s}(\k) \, d^3 k \over
    \frac{1}{2} \int (1 - \mu_\k^2) W(k) P^{s}(\k) \, d^3 k} =
  {1 + \frac{6}{5} \beta + \frac{3}{7} \beta^2 \over
    1 + \frac{2}{5} \beta + \frac{3}{35} \beta^2}
\ee
where the last equality follows from Kaiser's equation~(\ref{Kaiser}),
and the smoothing window is $W(k) = k^2 |\FTR(k)|^2$.
If the parallel and perpendicular variances are measured from
many random samplings in a redshift survey, then their ratio
${\smoothP^{s}_\para / \smoothP^{s}_\perp}$
should yield a measure of $\beta$.

Bromley applied this idea to
the Giovanelli \& Haynes
(1991, and references therein;
Wegner, Haynes \& Giovanelli 1993)
redshift survey of the Pisces-Perseus region,
which is complete to blue magnitude $m_B = 15.7$
and contains over 4000 galaxies.
He used a quartic window $R = [ 1 - (r/\lambda)^2 ]^2$,
with $\lambda = 14$--$17 \, h^{-1} \Mpc$.
The value $\beta = 0.75 \pm 0.3$ listed in Table~\ref{results}
is measured from Bromley's Figure~2 at $\lambda = 16 \, h^{-1} \Mpc$,
a length scale that he states gave the best results
in CDM simulations of the procedure.

It is useful to relate Bromley's statistic~(\ref{bromstat})
to the harmonics of the power spectrum.
Both numerator and denominator are sums of monopole and quadrupole
harmonics of the power spectrum filtered through the window
$W(k) = k^2 |\FTR(k)|^2$, equation~(\ref{barP}),
\be
\label{Ppp}
  {\smoothP^{s}_\para \over \smoothP^{s}_\perp} =
  {\smoothP^{s}_0 + \frac{2}{5} \smoothP^{s}_2 \over
    \smoothP^{s}_0 - \frac{1}{5} \smoothP^{s}_2}
  \ .
\ee
Thus the procedure is effectively equivalent to measuring a
quadrupole-to-monopole ratio of smoothed power spectra.

{\it Bromley, Warren \& Zurek (1997)}
developed Bromley's (1994) procedure further,
using a neat trick to correct for nonlinearity.
The paper is notable also for reporting redshift distortions
from two large (17 million particle) high resolution
CDM simulations with $\Omega = 1$ and $H_0 = 50 \,$km/s/Mpc,
one $125 \, h^{-1} \Mpc$ on a side, the other $500 \, h^{-1} \Mpc$ on a side.
The 1-dim\-en\-sional pairwise velocity dispersion of the simulations was
$1100 \, \kms$ and $850 \, \kms$
(at a separation of $1 \, h^{-1} \Mpc$)
respectively for the mass and the haloes
(`galaxies'), considerably higher than observations.
However, Bromley \etal\ were able to engineer a set of haloes
with velocity dispersion $384 \, h^{-1} \Mpc$ and clustering properties
consistent with those of the {\it IRAS\/} 1.2~Jy sample
(Brainerd \etal\ 1996)
by the device of
eliminating all objects that were within 3 Abell radii of the highest
density peaks and that had peculiar velocities greater than $750 \, \kms$.
Such culling is, as they say, `rather contrived',
but it is true that {\it IRAS\/} galaxies are conspicuous by their absence
from rich clusters such as the Coma cluster,
%(Strauss ?)
%Strauss, M. A., Davis, M., Yahil, A., & Huchra, J. P. 1990, ApJ, 361, 49.
%Strauss, M. A., Davis, M., Yahil, A., & Huchra, J. P. 1992a, ApJ, 385, 421.
%Strauss, M. A., Huchra, J. P., Davis, M., Yahil, A., Fisher, K. B.,
%& Tonry, J. 1992b, ApJS, 83, 29.
%Strauss, M. A., Yahil, A., Davis, M., Huchra, J. P., & Fisher, K. B.
%1992c, ApJ, 397, 395.
and it is possible that nature pulls a similar trick.

Bromley \etal's idea for nonlinearity was to modify the anisotropy of
sampling functions in a way designed to cancel nonlinearity.
Suppose that $\FTw_L(\k)$ are a set of `linear' sampling functions,
and suppose that nonlinearity can be modelled with a random isotropic
pairwise velocity dispersion $f(v)$, as in \S\ref{pairdispersion}.
Then the sampling functions can be `corrected' for nonlinearity
by dividing them in Fourier space by the square root of the
line-of-sight Fourier transform $\FTf(k_\para)$ (note $k_\para = k \mu_\k$)
of the velocity dispersion:
\be
\label{wBrom}
  \FTw(\k) = {\FTw_L(\k) \over \FTf(k\mu_\k)^{1/2}}
  \ .
\ee
Bromley \etal\ assumed the exponential model for the pairwise velocity
distribution, equation~(\ref{fexp}), which appears to give a good
description of pairwise velocities in the 1.2~Jy survey
(Fisher \etal\ 1994b).
With this choice,
and with the linear sampling functions chosen as before, equation~(\ref{w}), to
be the parallel and perpendicular gradients of a spherically symmetric function,
the expected shot-noise-subtracted variances~(\ref{varbard})
of the filtered densities
are equal to the redshift space power spectrum folded through the functions
\ba
  |\FTw_\para(\k)|^2 &=&
    \mu_\k^2
    \Bigl( 1 + \frac{1}{2} \sigma^2 k^2 \mu_\k^2 \Bigr)
    k^2 |\FTR(k)|^2
  \nn
  |\FTw_\perp(\k)|^2 &=&
    \frac{1}{2} (1 - \mu_\k^2)
    \Bigl( 1 + \frac{1}{2} \sigma^2 k^2 \mu_\k^2 \Bigr)
    k^2 |\FTR(k)|^2
  \ .
\ea
The factor
$( 1 + \frac{1}{2} \sigma^2 k^2 \mu_\k^2 )$
`corrects' the redshift power spectrum for nonlinearity,
equations~(\ref{Pnl}) and (\ref{fexp}),
so the resulting ratio
$\smoothP^{s}_\para/\smoothP^{s}_\perp$ of parallel to perpendicular
variances should be equal to the linear ratio~(\ref{bromstat}).
Once again, as in~(\ref{Ppp}),
it is instructive to express this ratio in terms of harmonics
of the smoothed power spectrum:
\be
  {\smoothP^{s}_\para \over \smoothP^{s}_\perp} =
  {\smoothP^{s}_0 + \frac{2}{5} \smoothP^{s}_2
    + \sigma^2 ( \frac{3}{10} \smoothP^{sNL}_0
      + \frac{6}{35} \smoothP^{sNL}_2 + \frac{4}{105} \smoothP^{sNL}_4 )
    \over
    \smoothP^{s}_0 - \frac{1}{5} \smoothP^{s}_2
    + \sigma^2 ( \frac{1}{10} \smoothP^{sNL}_0
      + \frac{1}{70} \smoothP^{sNL}_2 - \frac{2}{105} \smoothP^{sNL}_4 )}
\ee
where $\smoothP^{sNL}_\el$ are the harmonics of the power spectrum
smoothed over the window $k^2 W(k) = k^4 |\FTR(k)|^2$
instead of $W(k)$ (cf.\ eq.~[\ref{barP}]),
\be
\label{barbarP}
  \smoothP^{sNL}_\el = \int_0^\infty W(k) P^{s}_\el(k) \, 4\pi k^4 dk/(2\pi)^3
  \ .
\ee
The nonlinear correction is equivalent to replacing `linear' estimators of
the monopole and quadrupole harmonics of the redshift power spectrum
by `nonlinear' estimators
\ba
  \smoothP^{s}_0 & \rightarrow &
    \smoothP^{s}_0
    + \sigma^2 \left( \frac{1}{6} \smoothP^{sNL}_0
      + \frac{1}{15} \smoothP^{sNL}_2 \right)
  \nn
  \smoothP^{s}_2 & \rightarrow &
    \smoothP^{s}_2
    + \sigma^2 \left( \frac{1}{3} \smoothP^{sNL}_0
      + \frac{11}{42} \smoothP^{sNL}_2
      + \frac{2}{21} \smoothP^{sNL}_4 \right)
  \ .
\ea
Two comments can be made about the nonlinear correction.
Firstly, it involves incorporating judicious quantities of
monopole, quadrupole, and hexadecapole terms into the linear estimators
of the parallel and perpendicular (or monopole and quadrupole) power,
which seems like a simple and stable procedure.
Secondly, the nonlinear correction terms involve
the power spectrum multiplied by an extra factor of $k^2$,
equation~(\ref{barbarP}),
so that the correction becomes larger further into the nonlinear regime,
at larger $k$, as one might expect.

Bromley \etal\ applied their procedure both to their $N$-body simulations,
and to the {\it IRAS\/} 1.2~Jy survey.
For the spherical function $R(r)$,
they adopted a Gaussian $R(r) = e^{-(r/\lambda)^2/2}$,
in place of Bromley's (1994) quartic.
As it happens, the resulting (linear) smoothing window
$W(k) = k^2 |\FTR(k)|^2 \propto k^2 e^{-(k \lambda)^2}$
coincides with that of Hamilton (1995),
which is the same smoothing window used in Figure~\ref{xipkoms},
and the nonlinear smoothing window $k^2 W(k) \propto k^4 e^{-(k\lambda)^2}$
also belongs to the class of windows specified in equation~(\ref{W}).

The simulations gave good agreement with expectation.
For the 1.2~Jy survey, Bromley \etal\ assumed a velocity dispersion of
$\sigma = 320 \, \kms$,
which is a reasonable approximation to the (scale-dependent)
velocity dispersion measured by Fisher \etal\ (1994b).
The final result was $\beta = 0.8^{+0.4}_{-0.3} \,$.

\subsubsection{Cole, Fisher \etal}
\label{cole}

{\it Cole, Fisher \& Weinberg (1994, 1995)}
wrote a pair of papers,
in the first of which they described and tested with $N$-body simulations a
procedure to measure the redshift power spectrum in the plane-parallel
approximation, and in the second of which they applied their procedure
to measure $\beta$ from the quadrupole-to-monopole ratio of power
in the {\it IRAS\/} 1.2~Jy and QDOT surveys.

To ensure the validity of the plane-parallel approximation,
they first windowed the surveys through sets of spherical bowler hat windows
$w(|\r-\r_c|)$
(a bowler is a top hat convolved with a Gaussian),
centred at random positions $\r_c$ sufficiently far from the observer
that the windows subtended an opening angle no greater than $50^\circ$.
Cole \etal\ (1994, Fig.~8)
showed that at this opening angle and using their procedure,
the plane-parallel approximation causes $\beta$
to be underestimated by approximately 5\%.
They defined the line of sight to each window
as the direction to its centre $\r_c$.
In the plane-parallel approximation,
the shot-noise-subtracted power spectrum $\coleP^{s}(k, \mu_\k)$
measured through a window $w(|\r-\r_c|)$
(normalized to $\int w(r) d^3 r = 1$)
is the convolution of the
true redshift power spectrum $P^{s}(k, \mu_\k)$
with the power spectrum $|\FTw(k)|^2$ of the window
(Cole \etal\ 1994, eq.~[3.4]),
\be
\label{hatP}
  \coleP^{s}(k, \mu_\k) =
    \int \left|\FTw(|\k-\k'|)\right|^2 P^{s}(k', \mu_{\k'}) \, d^3 k'/(2\pi)^3
  \ .
\ee
It follows
(Cole \etal\ 1994, eq.~[3.5])
that the harmonics $\coleP^{s}_\el(k)$ of the power spectrum measured
through the window
are equal to the harmonics $P^{s}_\el(k)$ of the actual power spectrum
convolved with the harmonics $W_\el(k,k')$ of the power spectrum of the window:
\be
  \coleP^{s}_\el(k) =
    \int_0^\infty W_\el(k,k') P^{s}_\el(k')
    \, 4\pi k'^2 d k'/(2\pi)^3
\ee
where
\be
  \left|\FTw(|\k'-\k|)\right|^2 =
    \sum_\el (2\el+1) {\cal P}_\el(\hat\k.\hat\k') W_\el(k,k')
  \ .
\ee
The spherical symmetry of the window $w(r)$ is essential here to ensure that
the $\el$'th harmonic of the true power spectrum maps only to the $\el$'th
harmonic of the observed power spectrum.
The fact that the kernels $W_\el(k,k')$ are different for different
harmonics $\el$ means that the ratio $\coleP^{s}_2(k)/\coleP^{s}_0(k)$
of measured quadrupole-to-monopole harmonics
no longer satisfies the simple formula~(\ref{P2P0}).
However, Cole \etal\ (1994, Table~1 and Fig.~5) showed that
the measured ratio $\coleP^{s}_2(k)/\coleP^{s}_0(k)$ is equal to
the true ratio $P^{s}_2(k)/P^{s}_0(k)$ multiplied by a correction factor
$\approx 0.7$ which, although appreciably different from unity,
is nevertheless insensitive to the shape of the power spectrum.
In practice they adopted correction factors appropriate for the
$\Gamma = 0.25$ CDM spectrum of Efstathiou, Bond \& White (1992).

The Fourier transform $\FTw(k)$ of the bowler hat window
is the product of the first spherical Bessel function $j_1(k R_{\rm sph})$
with a Gaussian, where $R_{\rm sph}$ is the radius of the top hat.
Cole \etal\ used only wavenumbers $k$ for which $k R_{\rm sph}$
are zeros of $j_1(k R_{\rm sph})$
(in practice they used only the first two zeros,
at $k R_{\rm sph} = 4.49$ and $7.72$).
With this choice, the integrand in equation~(\ref{hatP}) vanishes
at $\k' = 0$,
which eliminates leakage from power at zero wavevector,
so immunizing the measurement of power against uncertainty in the
mean density
(which makes a delta-function contribution to power at zero wavevector).
This clever trick was used first by Fisher \etal\ (1993).

Cole \etal\ (1995) combined their measurements of the redshift power spectrum
$\coleP^{s}(k, \mu_\k)$
from bowler windows at different depths using an inverse-variance weighting,
the variance of the power from each window being estimated in the manner of
Feldman, Kaiser \& Peacock (1994).
They fitted the averaged $\coleP^{s}(k, \mu_\k)$ to a sum of harmonics,
from which they formed the quadrupole-to-monopole ratio
$\coleP^{s}_2(k)/\coleP^{s}_0(k)$,
which they divided by the aforesaid correction factor $\approx 0.7$
to obtain the true ratio $P^{s}_2(k)/P^{s}_0(k)$.
To allow for nonlinearity,
they adopted the model in which the linear redshift distortion
is modulated by a random one-point (not pairwise) exponential velocity
dispersion.
To arrive at a final value of $\beta$ and the velocity dispersion $\sigma$
in each of the 1.2~Jy and QDOT surveys,
they carried out a least squares fit to the quadrupole-to-monopole ratios
as a function of wavenumber $k$,
using a full covariance matrix
estimated from an ensemble of mock catalogues
constructed from $N$-body simulations.

{\it Fisher \& Nusser (1996)}
studied the Zel'dovich approximation as a means of
approximating redshift distortions in the mildly nonlinear regime
(see \S\ref{zeldovich}).
As an application,
they fitted the ratio of quadrupole-to-monopole power measured by
Cole, Fisher \& Weinberg (1995) from the {\it IRAS\/} 1.2~Jy survey,
finding $\beta = 0.6 \pm 0.2$.

\subsubsection{Fry \& Gazta\~naga}
\label{fry}

{\it Fry \& Gazta\~naga (1994)}
compared redshift to real space correlation functions in each
of three redshift surveys,
the first Center for Astrophysics (CfA1) survey
(Huchra \etal\ 1983),
the Southern Sky Redshift Survey (SSRS)
(da Costa \etal\ 1991),
and the {\it IRAS\/} 2~Jy survey
(Strauss \etal\ 1992a).
Their procedure was:
measure the volume-averaged correlation functions
$\bar\xi = V^{-2} \int_V \xi(r_{12}) d^3 r_1 d^3 r_2$,
in spherical volumes $V$ for redshift space,
and conical volumes $V$ for real space;
find the best power law fits to $\xi^{s}(r)$ and $\xi(r)$
that reproduce the behaviour of the volume-averaged $\bar\xi$
in each case;
infer $\beta$ from the ratio $\xi^{s}(r)/\xi(r)$
of the fitted power law correlation functions
at the largest separations probed, $r \sim 5$--$10 \, h^{-1} \Mpc$;
repeat this for several volume-limited subsamples,
and adopt an average $\beta$.
The resulting values of $\beta$ seem surprisingly large for such modest
separations.
However,
given the indirectness of the procedure for measuring large scale power,
and some lack of rigour in the error analysis,
one might be inclined to take the numbers with a pinch of salt.

\subsubsection{Peacock \etal}
\label{peacock}

{\it Peacock \& Dodds' (1994)}
principal goal was not so much to measure $\beta$
as to reconstruct the linear power spectrum of mass fluctuations
from a compilation of available observational evidence.
A feature of the paper was the derivation
of empirical analytic formulae
relating linear and nonlinear power spectra,
using a procedure inspired by one proposed by Hamilton \etal\ (1991).
Updated versions of Peacock \& Dodds' formulae are given by
Peacock \& Dodds (1996).

The value of $\beta$ measured by Peacock \& Dodds (1994)
followed from a comparison of redshift space to real space power spectra.
For the real space power spectrum
they adopted the
APM power spectrum of Baugh \& Efstathiou (1993),
while for redshift space power spectra they considered power spectra
from the {\it IRAS\/} QDOT survey
(Feldman, Kaiser \& Peacock 1994),
the Stromlo-APM survey
(Loveday \etal\ 1992),
and the CfA2 survey
(Vogeley \etal\ 1992).
To combine the various observed power spectra into a single canonical
linear power spectrum, they
introduced linear bias factors $b$,
equation~(\ref{b}),
corrected the galaxy power spectra for nonlinear evolution,
and modelled nonlinearity in the redshift distortions
with a random Gaussianly distributed velocity dispersion.
To break the degeneracy between $\Omega_0$ and bias $b$
that occurs in $\beta \approx \Omega_0^{0.6}/b$,
they brought into consideration the power spectra of Abell clusters
(Peacock \& West 1992)
and of radio galaxies
(Peacock \& Nicholson 1991),
which they assumed to be biased
but (unlike galaxies) unaffected by nonlinear evolution.
With these assumptions they found
$b_{\rm optical}/b_{IRAS} = 1.3$ and
$\beta_{IRAS} = 1.0 \pm 0.2$,
which imply also
$\beta_{\rm optical} = 0.77 \pm 0.15$.
The absolute values of the bias factors were less well determined
than their relative values;
the best models had $b_{IRAS} \approx 0.8$.

{\it Peacock (1997)} carried out an improved version of
Peacock \& Dodds' (1994) analysis,
arriving at the notably lower value of
$\beta_{\rm optical} = 0.40 \pm 0.12$
from optical data alone.
As in the earlier paper,
for optical data
Peacock used the real space APM power spectrum from
Baugh \& Efstathiou (1993; also 1994),
and redshift space power spectra from
the Stromlo-APM survey
(Loveday \etal\ 1992)
and the CfA2 survey
(Vogeley \etal\ 1992).
The reason for the lower value of $\beta$ was that
Peacock increased the real space APM power spectrum from
Baugh \& Efstathiou (1993)
by a factor of 1.25 in order to match it to the amplitude of
the real space correlation function of Stromlo-APM, as measured
by Loveday \etal\ (1995)
using the cross-correlation technique of Saunders \etal\ (1992).
This adjustment reduces
$\beta_{\rm optical}$ from $0.77$ to $0.40$,
in accordance with equation~(\ref{PsP}).

Peacock carried out a separate analysis for {\it IRAS\/} data,
but here the results were less conclusive.
He used a real space {\it IRAS\/} power spectrum
obtained by transforming the cross-correlation function between
the 1-in-6 QDOT survey and its parent QIGC catalogue, measured by
Saunders, Rowan-Robinson \& Lawrence (1992),
while for the redshift power spectrum he adopted
Tadros \& Efstathiou's (1995) power spectrum of
the combined {\it IRAS\/} 1.2~Jy and QDOT surveys,
in place of the Feldman \etal\ (1994) QDOT power spectrum.
The redshift power spectrum of the combined {\it IRAS\/} 1.2~Jy and QDOT
surveys is lower than that of the QDOT survey thanks in large part to a region
in Hercules, which Tadros \& Efstathiou showed produces an anomalously large
upward excursion in the QDOT power spectrum
(cf.\ Hamilton 1995; Oliver \etal\ 1996).
Tadros \& Efstathiou's downwardly revised redshift power spectrum lay below
Saunders \etal's real space power spectrum at wavenumbers
$k < 0.05 \,h\, \Mpc^{-1}$,
nominally suggesting a negative $\beta$,
and causing Peacock to discard those points.
Such machinations do not inspire confidence,
and Peacock avoided drawing any definitive conclusion.

Besides improved data,
Peacock applied various theoretical improvements:
he used an improved treatment of nonlinear evolution
(Peacock \& Dodds 1996);
the ratio of optical and {\it IRAS\/} real space power spectra
indicated some scale dependence in bias,
which he fitted with a two-parameter power law model
(Mann, Peacock \& Heavens 1997)
in place of the one-parameter linear bias model;
he abandoned the use of Abell cluster and radio galaxy power spectra;
and he used an exponential rather than Gaussian pairwise velocity distribution
to model nonlinearity in redshift distortions.

In the latter half of the paper,
Peacock went on to discuss
bias, $\Omega_0$, and the shape of the reconstructed, linearized power spectrum,
adducing a variety of arguments to conclude
that a low bias, low $\Omega_0$ model is preferred
over a high bias, high $\Omega_0$ model.

\subsubsection{Heavens, Taylor, \etal}
\label{heavens}

{\it Heavens \& Taylor (1995)}
were the first to apply a full-blown maximum likelihood procedure
to measure $\beta$,
in a manner designed to retain as much information as possible
about $\beta$ in the linear regime.
They took the {\it IRAS\/} 1.2~Jy survey as the data set,
conservatively cut to galactic latitude $|b| > 10^\circ$.
They chose to work in a basis of spherical waves,
and in the frame of the Local Group (LG).
First, they defined the unredshifted overdensity modes $\delta_{n\el m}$
by the discrete spherical transform
(the definition below is the complex conjugate of Heavens \& Taylor's
definition, but it conforms to the convention of the present review,
eq.~[\ref{dklm}], which follows that of Peebles 1980, \S46)
\be
\label{dnlm}
  \delta_{n\el m} \equiv
    c_{n\el} \int_{r < r_{\max}} \!\!\!\!\!\!\!\!\!\!
    j_\el(k_{n\el}r) Y_{\el m}(\hat\r) \delta(\r) \, d^3 r
  \ , \ \ 
  \delta(\r) = \sum_{n\el m} c_{n\el}
    j_\el(k_{n\el}r) Y^\ast_{\el m}(\hat\r) \delta_{n\el m}
\ee
where the integration is over a finite sphere of radius
$r_{\max} = 200 \, h^{-1} \Mpc$,
the $c_{n\el}$ are normalization constants,
and the wavenumbers $k_{n\el}$ are chosen to satisfy the boundary condition
$j'_\el(k_{n\el} r_{\max}) = 0$, which ensures that the predicted
linear peculiar velocity field vanishes on the boundary of the sphere
(see Fisher \etal\ 1995b for a discussion of boundary conditions).
Next, they defined observed, redshifted modes $D_{n\el m}$ by
\be
  D_{n\el m} \equiv
    c_{n\el} \int_{r < r_{\max}} \!\!\!\!\!\!\!\!\!\!
    j_\el(k_{n\el}r) Y_{\el m}(\hat\r) w(r) \nbar(\r) \delta^{s}(\r) \, d^3 r
\ee
which, besides being in redshift space,
involve an additional factor $w(r) \nbar(\r)$ in the integrand
compared to~(\ref{dnlm}).
The radial weighting function $w(r)$
(which in practice depended also on $n\el$)
is a near-minimum-variance weighting
which helps to reduce the noise in each mode,
hence to increase the information content of each mode,
and hence to reduce the number of modes required.
They derived approximate expressions for the weightings $w_P(r)$
and $w_\beta(r)$ that optimize each mode $D_{n\el m}$ for measuring
respectively the normalization of the power spectrum $P$
and the distortion parameter $\beta$.
What it means to optimize modes for the measurement of particular parameters
is clarified by
Tegmark, Taylor \& Heavens (1997)
and Tegmark \etal\ (1997).
Heavens \& Taylor
showed that their modes $D_\mu$
(abbreviating $n\el m = \mu$)
are linearly related to the unredshifted overdensity modes $\delta_\mu$ by
\be
\label{Dmu}
  D_\mu = \sum_\nu ( \Phi_{\mu\nu} + \beta V_{\mu\nu} ) \delta_{\nu}
  \ .
\ee
The matrix $\Phi_{\mu\nu} + \beta V_{\mu\nu}$ can be recognized as
the representation in spherical transform space of the operator
$w(r) \nbar(\r) \bS^\LG$,
where $\bS^\LG$ is the redshift distortion operator~(\ref{SLG})
in the LG frame\footnote{
Actually they used the distortion operator $\bS$ appropriate to the CMB frame,
equation~(\ref{S}),
rather than the distortion operator $\bS^\LG$ in the LG frame,
equation~(\ref{SLG}),
and they dropped the dipole ($\el = 1$) modes from the likelihood analysis.
This is not quite correct,
although the error should be minor for a near all-sky survey.
They used a selection function, their equation~(54),
from Fisher, private communication;
whether this was a real or redshift selection function is unclear.
If a redshift selection function,
then strictly the distortion operator $\bS^{s\,\LG}$
relative to the redshift space selection function should be used,
as discussed in \S\protect\ref{operatornbar}.
Again, Heavens \& Taylor dropped the monopole ($\el = 0$) modes from
the likelihood analysis, so whatever the case,
any error should be minor for this near all-sky survey.
}.
Heavens \& Taylor introduced one further refinement,
to model nonlinearity as a random Gaussian velocity field
(cf.\ \S\ref{pairdispersion}),
which has the effect of modifying the matrix
$\Phi_{\mu\nu} + \beta V_{\mu\nu}$ in equation~(\ref{Dmu})
by pre-multiplying it by another matrix.
The expected covariance matrix of the modes is then
\be
\label{Cmunu}
  C_{\mu\nu} \equiv \langle D_\mu \hspace{1pt} D^\ast_\nu \rangle =
    \sum_{\alpha} ( \Phi_{\mu\alpha} + \beta V_{\mu\alpha} )
                  ( \Phi_{\nu\alpha} + \beta V_{\nu\alpha} )^\ast P(k_\alpha)
    + N_{\mu\nu}
\ee
where $N_{\mu\nu}$ is the Poisson sampling noise,
which is just the (expected) self-galaxy contribution to
$\langle D_\mu D^\ast_\nu \rangle$.
In all,
Heavens and Taylor's covariance matrix $C_{\mu\nu}$
took into account
finite sky coverage,
linear redshift distortions,
nonlinearity modelled as a random Gaussian velocity distribution,
and Poisson sampling noise.

The prior covariance $C_{\mu\nu}$, equation~(\ref{Cmunu}),
depends
on the unredshifted power spectrum $P(k)$
and on the distortion parameter $\beta$.
Heavens \& Taylor
adopted as prior the shape of the power spectrum measured by
Peacock \& Dodds (1994),
but retained its overall amplitude $\sigma_8^2$
and the distortion parameter $\beta$
as parameters to be measured from the likelihood analysis.
They assumed a Gaussian likelihood function, equation~(\ref{L}),
and included observed modes $D_{n \el m}$
with $k_{n\el} < 0.1 \, h \Mpc^{-1}$,
a total of 604 modes, with maximum values $\el = 17$ and $n = 6$.
The prior covariance $C_{\mu\nu}$, equation~(\ref{Cmunu}),
was computed including modes $\delta_{n\el m}$ up to $\el = 30$ and $n = 20$.
The final result of the likelihood analysis was
$\beta = 1.1 \pm 0.3 \,$.
The uncertainty here is the conditional $1\sigma$ uncertainty,
which comes from projecting the $\Delta\ln{\cal L} = -0.5$ contour
on to the parameter axes.
Heavens (1997, private communication) states that
a more conservative estimate of the uncertainty in $\beta$ is the marginal
uncertainty, $\pm 0.5$ ($1\sigma$),
which results from integrating the likelihood over the
probability distribution of the amplitude $\sigma_8$.
Heavens \& Taylor commented that the maximum likelihood value of $\beta$
dropped to $0.5$ when the maximum $k_{n\el}$ was increased
from $0.1$ to $0.12 \, h \Mpc^{-1}$,
which increases the number of modes from 604 to 1048.
They attributed the change in $\beta$ to the onset of nonlinearity,
but it seems likely that statistical fluctuation rather than nonlinearity
would cause so dramatic change in $\beta$
for so small a change in the cutoff wavenumber.

{\it Ballinger, Heavens \& Taylor (1995)}
took Heavens \& Taylor's procedure a step further by
parametrizing the power spectrum by its amplitude in six
logarithmically spaced bins,
rather than just an overall normalization.
Otherwise the analysis proceeded essentially exactly as in Heavens \& Taylor,
with a similar result, $\beta = 1.04 \pm 0.3$.

Tegmark, Taylor \& Heavens (1997)
report that the 604 modes used by Heavens \& Taylor (1995)
can be compressed to as few as 60 signal-to-noise eigenmodes
(so called Karhunen-Lo\`eve modes)
of $\beta$,
without appreciable loss of accuracy in $\beta$.
This compression step,
which involves solving an eigenvalue equation in the covariance matrix
$C_{\mu\nu}$,
does not obviate the necessity to compute the covariance matrix at full size,
but it can reduce the work involved in computing likelihood contours
for the parameters.

\subsubsection{Lin}
\label{lin}

{\it Lin (1995)}
examined redshift distortions, and measured the pairwise velocity dispersion,
in the Las Campanas Redshift Survey (LCRS)
(Shectman \etal\ 1996),
which contains 23697 galaxies in six narrow slices
each $1.\!\!^\circ 5$ in declination and $80^\circ$ in right ascension,
three each in the north and south galactic caps.
Lin measured $\beta = 0.5 \pm 0.25$
using Hamilton's (1992) estimator
of the quadrupole-to-monopole power, equation~(\ref{xi2xi0}).

\subsubsection{Baugh}
\label{baugh}

{\it Baugh (1996)}
compared the real space correlation function $\xi(r)$ of the APM survey
obtained by Fourier transforming the power spectrum derived by
Baugh \& Efstathiou (1993),
to the redshift space correlation function $\xi^{s}(r)$
of the Stromlo-APM survey from Loveday \etal\ (1995).
Uncertainty in the APM real space power spectrum
arises from uncertainty in cosmological geometry (i.e.\ in $\Omega_0$),
and in the evolution of the power spectrum with redshift, parametrized as
$P(k) \propto (1+z)^{-\alpha}$ at fixed comoving wavenumber $k$
(so $\alpha = 2$ for linear growth in an $\Omega = 1$ Universe).
Baugh (his Table~2) quoted three values of $\beta$,
two assuming no evolution, $\alpha = 0$ and $\Omega_0 = 1$ or $0.2$,
giving respectively
\raisebox{0ex}[2ex][0ex]{$\beta = 0.61^{+0.20}_{-0.23}$}
or
\raisebox{0ex}[2ex][0ex]{$0.45^{+0.21}_{-0.23} \,$},
and one value with evolution, $\alpha = 2$ and $\Omega_0 = 1$, giving
\raisebox{0ex}[2ex][0ex]{$\beta = 0.20^{+0.19}_{-0.22} \,$}.
Evidently the value of $\beta$ is quite sensitive to the
assumed geometry and evolution.
Observations suggest quite strong evolution.
For example,
Le F\`evre \etal\ (1996)
find from the Canada-France Redshift Survey (CFRS)
an evolutionary index\footnote{
Le F\`evre \etal\ define evolution with respect to proper,
not comoving coordinates:
their index $\epsilon$ is related to Baugh's index $\alpha$ by
$3 + \epsilon = \alpha + \gamma$,
where $\gamma \approx 1.7$ is the logarithmic slope of the correlation function.
}
$\epsilon = 0.4 \pm 1.1$, corresponding to $\alpha = 1.7 \pm 1.1$.
Table~\ref{results} in this review therefore lists Baugh's value
\raisebox{0ex}[2ex][0ex]{$\beta = 0.20^{+0.19}_{-0.22}$}
with evolution.
Presumably the uncertainty arising from evolution could be resolved
in a manner similar to Peacock (1997) ---
the APM power spectrum could be normalized so as to agree with the
amplitude of the real space correlation function of Stromlo-APM deduced by
Loveday \etal\ (1995).

\subsubsection{Loveday \etal}
\label{loveday}

{\it Loveday \etal\ (1996a)}
examined the Stromlo-APM redshift survey
(Loveday \etal\ 1996b),
a 1-in-20 survey of galaxies 
brighter $b_J = 17.15$
in the Automatic Plate Measuring (APM) survey
(Maddox \etal\ 1990a,b, 1996).
They considered both of the first two methods described in \S\ref{methods}.
They measured the real space correlation function
$\xi(r)$ from the projected
cross-correlation between the Stromlo-APM survey and its parent APM survey,
using the procedure of
Saunders, Rowan-Robinson \& Lawrence (1992),
as described by Loveday \etal\ (1995).
They measured the monopole, quadrupole, and hexadecapole harmonics
$\xi^{s}_\el(r)$ of the redshift space correlation function
using the procedure of Hamilton (1993b).

They tested the different methods with mock catalogues constructed from
two ensembles of $N$-body simulations
described by Croft \& Efstathiou (1994),
10 flat CDM simulations with a cosmological constant,
$\Omega_M = 0.2$ and $\Omega_\Lambda = 0.8$,
and 9 flat Mixed Dark Matter simulations,
$\Omega_{\rm CDM} = 0.7$ and $\Omega_\nu = 0.3$
($\nu$ for neutrinos denotes Hot Dark Matter).
Loveday \etal\ found from the simulations that nonlinearities affect
the angle-averaged (monopole) correlation function $\xi^{s}(r)$
up to separations
$r \sim 10 \, h^{-1} \Mpc$,
its volume integral
$J^{s}(r) \equiv \int_0^r \xi^{s}(s) s^2 ds$
to separations
$r \sim 20 \, h^{-1} \Mpc$,
and the quadrupole and hexadecapole harmonics $\xi^{s}_2(r)$ and $\xi^{s}_4(r)$
to scales $r \sim 30 \, h^{-1} \Mpc$.
They concluded that the first method, the ratio of redshift to real space
correlation functions (or its integrals) is likely to yield more reliable
measures of $\beta$.
In practice the ratio $J^{s}(r)/J(r)$ of integrated correlation functions
proved considerably less noisy than $\xi^{s}(r)/\xi(r)$,
so the former was their statistic of choice.

For the Stromlo-APM survey,
Loveday \etal\ found $\beta = 0.48 \pm 0.12$ from
the ratio $J^{s}(r)/J(r)$
at its most accurately measured point,
a separation of $r = 17.8 \, h^{-1} \Mpc$.
This is also the point where the observed ratio $J^{s}(r)/J(r)$
reached its maximum value,
so they also concluded more conservatively that the data
favoured $\beta \la 0.6$.
They also considered Hamilton's (1992) quadrupole-to-monopole ratio,
equation~(\ref{xi2xi0}),
but were unable to draw any conclusions from this estimator.
The conclusion that nonlinearities have a large effect on the
quadrupole-to-monopole ratio in the Stromlo-APM survey agrees with
the analysis of the present review, \S\ref{example}.

\subsubsection{Tadros \& Efstathiou}
\label{tadros}

{\it Tadros \& Efstathiou (1996)}
carried out an analysis similar to Baugh (1996), \S\ref{baugh},
except that they worked with the power spectrum
instead of the correlation function.
Tadros \& Efstathiou
computed the redshift space power spectrum $P^{s}(k)$ of the Stromlo-APM survey,
and compared this to the real space power spectrum $P(k)$ of the APM survey
computed by Baugh \& Efstathiou (1993).
Table~\ref{results} quotes the `Flux Limited' (misprinted `Flux Weighted')
result from Tadros \& Efstathiou's Table~2,
since this weighting is expected to (and in fact did) give lower variance
than the `Volume Limited' weighting.
As with Baugh (1996),
systematic uncertainty in the real space APM power spectrum arises from
uncertainty in its rate of evolution with redshift,
and again Tadros \& Efstathiou gave values both without evolution,
$\beta = 0.74 \pm 0.48$ for $\alpha = 0$,
and with evolution, $\beta = 0.20 \pm 0.44$ for $\alpha = 1.3$.
Table~\ref{results} quotes the value with evolution.

\subsubsection{Ratcliffe \etal}
\label{ratcliffe}

{\it Ratcliffe \etal\ (1997)}
studied the Durham/UKST (UK Schmidt Telescope) redshift survey,
which is a $> 75\%$ complete 1-in-3 redshift survey of galaxies
brighter than $b_J \approx 17$ in the
Edinburgh/Durham Southern Galaxy Catalogue
(Collins, Heydon-Dumbleton \& MacGillivray 1988;
Collins, Nichol \& Lumsden 1992).
The survey contains $\approx 2500$ galaxies over a region
$\sim 20^\circ \times 70^\circ$ centred on the South Galactic Pole.

Ratcliffe \etal\ used both of the first two methods
described in \S\ref{methods}.
For the first method,
they measured both the angle-averaged redshift correlation function $\xi^{s}(r)$
and the real space correlation function $\xi(r)$ from the survey,
inferring the latter from the projected redshift correlation function
using Saunders \etal's (1992) procedure.
The resulting ratio $\xi^{s}(r)/\xi(r)$ of redshift to real correlation
functions was quite noisy,
so they considered instead the ratio $J^{s}(r)/J(r)$
of the integrated correlation functions
$J(r) \equiv \int_0^r \xi(s) s^2 ds$,
finding $\beta = 0.52 \pm 0.39$ at a separation $r = 20 \, h^{-1} \Mpc$.

For the second method,
Ratcliffe \etal\ used Hamilton's (1992) estimator
of the quadrupole-to-monopole power, equation~(\ref{xi2xi0}).
Combining results over separations $r \approx 10$--$20 \, h^{-1} \Mpc$
gave $\beta = 0.48 \pm 0.11$.

Ratcliffe \etal\ tested their procedures extensively with
two sets of CDM $N$-body simulations,
one set an $\Omega_M = 1$ biased $b = 1.6$ model,
the other a spatially flat $\Omega_M = 0.2$, $\Omega_\Lambda = 0.8$
unbiased model with a cosmological constant.
These simulations were not matched in every respect to the observations
--- in particular the simulations had a considerably larger small scale
velocity dispersion.
On the assumption of an exponential pairwise velocity distribution,
the 1-dim\-en\-sional pairwise velocity dispersions
were measured to be
$980 \pm 22 \, \kms$ and $835 \pm 60 \, \kms$
respectively for the $\Omega_M = 1$ and $0.2$ models,
compared to
$416 \pm 36 \, \kms$ for the Durham/UKST survey.

\section{Cosmological Redshift Distortions}
\label{cosmological}
\setcounter{equation}{0}

\begin{figure}
\begin{center}
\leavevmode
\epsfxsize=3.2in \epsfbox{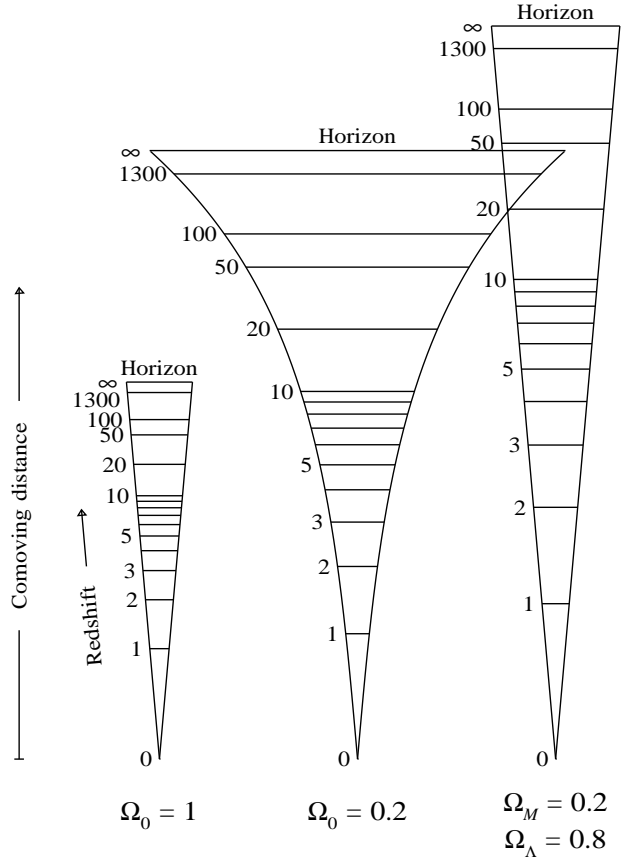}
\end{center}
  \caption[1]{
In this diagram, each wedge represents a cone of fixed opening angle,
with the observer (us) at the point of the cone, at zero redshift.
The wedges show the relation between physical sizes,
namely the comoving distances in the radial (vertical)
and transverse (horizontal) directions,
and observable quantities,
namely redshift and angular separation,
in three different cosmological models:
(left) flat matter-dominated Universe;
(middle) open matter-dominated Universe with $\Omega_0 = 0.2$;
(right) flat Universe with a cosmological constant, $\Omega_\Lambda = 0.8$.
\label{z}
}
\end{figure}

The relation between redshift and radial comoving distance,
and between angle and transverse comoving distance,
is different for different cosmological models,
as illustrated in Figure~\ref{z}.
The differences produce a cosmological redshift distortion
that is zero at zero redshift,
but that becomes more marked at higher redshift.

The idea of using cosmological redshift distortions to measure cosmological
parameters, notably the cosmological constant, was first proposed by
Alcock \& Paczy\'nski (1979).
More recently, several authors have considered
the combined effect of distortions from peculiar velocities and from cosmology,
both to see how cosmological distortion affects the measurement of $\beta$,
and to assess the prospect of measuring cosmological parameters from
upcoming redshift surveys
(Ballinger, Peacock \& Heavens 1996;
Matsubara \& Suto 1996;
%Suto \& Matsubara 1996;
Nakamura, Matsubara \& Suto 1997;
de Laix \& Starkman 1997).

Alcock \& Paczy\'nski (1979) argued, and subsequent authors have concurred,
that the largest cosmological distortion is a $\Lambda$-squashing,
a squashing in the radial direction that occurs if there is a large
cosmological constant.
By comparison,
the differences in redshift distortions between low and high $\Omega_M$ models
in the absence of a cosmological constant
are relatively small.

The differences between redshift distortions
predicted by models with and without a cosmological constant
at first increase with redshift,
but then saturate at $z \approx 1$
for physically interesting models, those with $\Omega_M \ga 0.1$
(Ballinger, Peacock \& Heavens 1996;
Matsubara \& Suto 1996).
Even at these redshifts,
the cosmological distortion is not easy to distinguish from
redshift distortions caused by peculiar velocities.
This degeneracy could in principle be resolved
because the cosmological and peculiar velocity signals evolve differently
with redshift,
but in practice the uncertain evolution of bias muddies the issue
(Ballinger, Peacock \& Heavens 1996).
Nakamura, Matsubara \& Suto (1997)
emphasize that cosmological redshift distortions
will affect the linear distortion parameter $\beta$ at the 10--20\% level
in the Sloan Digital Sky Survey (SDSS),
which aims to go to a median depth of $z \approx 0.1$.
de Laix \& Starkman (1997)
conclude that the SDSS will not provide a clean signal of
cosmological parameters from redshift distortions in the linear regime.
%
%\section{Prospects}
%
%Published redshift surveys of galaxies have reached the point
%where it is possible to measure the linear distortion parameter $\beta$,
%at the level of a small number of standard deviations above zero.
%Upcoming large redshift surveys
%such as the Two Degree Field (2dF) survey
%(Colless?)
%and the Sloan Digital Sky Survey
%(?;
%http://)
%should greatly improve the accuracy with which redshift distortions
%can be measured, both on linear and nonlinear scales.
%
%On the theoretical front,
%there is a clear need for a more rigorous understanding of redshift distortions
%in the translinear linear.
%
%And an aphorism to end with:
%Quality data deserve quality statistical analysis.

\section*{Acknowledgements}

This work was supported by
NSF grant AST93-19977
and by
NASA Astrophysical Theory Grant NAG 5-2797.
I thank Jon Loveday and George Efstathiou for providing a copy of the
Stromlo-APM survey in advance of publication.

%\end{sloppypar}

\end{document}